\documentclass[10pt,journal,compsoc]{IEEEtran}

\ifCLASSOPTIONcompsoc
  \usepackage[nocompress]{cite}
\else
  \usepackage{cite}
\fi

\ifCLASSINFOpdf
   \usepackage[pdftex]{graphicx}
\else
\fi

\usepackage{underscore}
\usepackage[document]{ragged2e}
\usepackage{dblfloatfix}
\usepackage{caption}
\usepackage{hyperref}
\usepackage[all]{hypcap}
\DeclareUnicodeCharacter{0301}{\'{e}}

\begin{document}
\title{An NFC-Enabled Anti-Counterfeiting System \\ for Wine Industry}
\author{Neo~C.K.~Yiu,~\IEEEmembership{Member,~IEEE\\ Department~of~Industrial~and~Manufacturing~Systems~Engineering\\ The~University~of~Hong~Kong}
\IEEEcompsocitemizethanks{\IEEEcompsocthanksitem Neo C.K. Yiu was with Department of Industrial and Manufacturing Systems Engineering, The University of Hong Kong, Hong Kong}
\thanks{Manuscript first submitted on May 2, 2014; revised on Nov 12, 2020 (v5).}}

\IEEEtitleabstractindextext{
\justify
\begin{abstract}
Wine counterfeiting has been posing significant challenges to the wine industry, and has undermined the international wine trading market and even the global economy hugely. The situation of counterfeiting has even been exacerbating in wine industry and its global supply chain. There has been a number of anti-counterfeiting approaches which have been proposed and adopted utilizing different authentication technologies, in response to growing threats of counterfeiting to the wine industry. The proposed NFC-Enabled Anti-Counterfeiting System (NAS) is developed for luxury-good industry such as wine industry, aiming at upholding provenance and authenticity of the wine products from counterfeits via the product pedigree, transaction records and supply chain integrity maintained along the supply chain. Consumers can therefore safeguard their stake by authenticating a specific wine product with their NFC-enabled smartphones before purchasing at the retail points. NAS utilizes Near-field Communication (NFC), which has emerged as a promising technology and communication protocol for developing innovative alternatives, to facilitate the wine record processing of wine products and in turn combat wine and spirit counterfeiting. The integrated NAS is consisted of a wide range of hardware and software components, and the best combination of settings, parameters and deployments will therefore be identified. Other possible implementation issues, such as tag selection, tag programming and encryption, setup of back-end database servers and the design of NFC mobile application will also be discussed in this project. The critical design of NAS is vital not only to the key of product anti-counterfeiting of wine industry, but also to the strong foundation for other innovative supply chain solutions, such as the NFC-enabled purchasing system, later built on top of NAS with improved and integrated anti-counterfeiting functionalities. The system demonstration of NAS is available at \url{http://youtu.be/qOWPibxESL4}.
\end{abstract}

\begin{IEEEkeywords}
NFC-enabled Anti-counterfeiting System, Near-field Communication, Anti-counterfeiting, Wireless Communication Technologies, Wine Industry, Internet-of-Things, Supply Chain Integrity, Track-and-Trace System, End-to-End Traceability.
\end{IEEEkeywords}}

\maketitle
\IEEEdisplaynontitleabstractindextext
\justifying
\IEEEpeerreviewmaketitle

\ifCLASSOPTIONcompsoc
\IEEEraisesectionheading{\section{Introduction}\label{sec:introduction}}
\else
\section{Introduction}
\label{sec:introduction}
\fi

\IEEEPARstart{I}{nternational} wine trade has been facing rampant and exacerbating problems of counterfeiting. This chapter explains the reasons why wine industry and the wine product itself have been targeted by counterfeiters, the statistical evidence of the growing wine counterfeit market, and the current anti-counterfeiting techniques adopted by different supply chain nodes of the wine industry.

\subsection{The Susceptibility of Wine Industry to Counterfeiting}
One may question why the wine industry and the wine product itself would be an attractive target of counterfeiting activities. Some have suggested that luxurious wine products are relatively valued but lacking anti-counterfeiting features on product packaging. Some reviews believed that the recent emergence of those spectacular vintages, like 2000 and 2005 Bordeaux, has already fueled the situation of wine counterfeiting as the market of those extravagant and rare wines with the wine bottles, has become super-heated, making the wine industry a hotbed for product counterfeiting. As such, the more valued and rarer the wine product is, the more likely the wine product and its winemakers are targeted to wine counterfeiting activities \cite{1}. 

There are various reasons why the wine product itself with its attributes, are susceptible to wine frauds and counterfeits. For instance, the market value of fine-wine product and the wine industry as a whole have been soaring in recent years owing to the limited supply, diminishing supply of older vintages and the increasing wealth creating more potential buyers and demands in those developing wine markets and countries, where both demand of wine products and counterfeiting activities in wine industry has been rocketed even after Hong Kong manifested itself as a wine trading and distribution center with all taxes on wine abolished in 2008 \cite{2}. The aforementioned reasons have already created vast and attractive opportunities for those counterfeiting activities with profit yielded, and posed negative impacts to wine industry and international wine trade. Furthermore, it is not surprising to indicate that many wine consumers are not really familiar with the knowledge and technique required for wine appreciation and recognition. It turns out that there have been other reasons why wine products are attractive for being counterfeited, such as the difficulty in proving provenance of those vintage wine bottles and the variability of how the content of those old vintages taste.

The most common technique of wine counterfeiting, as suggested in \cite{wayofcounterfeit} is printing a fake label with a subtly misspelled brand name or a slightly different logo in hopes of fooling wine consumers, which is also applicable to other luxury-goods industries prone to counterfeiting. More ambitious counterfeiters might remove an authentic label and place it on a bottle of similar shape, usually from the same vineyard, containing a cheaper wine. Savvy buyers could identify if the cork stopper does not match the label, but how many ordinary wine consumers could manage to identify the fake merely with the naked eye? The wine counterfeiting has been getting more aggressive and sophisticated, particularly due to the rising bottle price contributed by the huge demand in consumer market. With empty bottles of valued wine products gathered and even available for sales in consumer market at a attractive cost, problem of re-bottling wine product with cheaper wine content or chemicals refilled is rampant in the wider wine industry. 

Furthermore, wine bottles have been seen copying with the original label, artwork and trademarked name, with slight change on the name and logo of the originals. It has then been different enough to be classified as ordinary wine product instead of a product counterfeit, and it could still trick numerous industry participants and wine consumers along the supply chain. The bottles of wine products have always been regarded as originals, since some new or unsuspecting wine consumers and some supply chain participants are not familiar with the real label due to their lack of wine knowledge, and their illiteracy in French or English making the situation even worse. Some counterfeiters in developing countries and wine markets also purchase empty wine bottles with cheaper wine or unknown liquor simply refilled using a syringe, re-cork the bottles, replace the capsule packaging materials, and sell the bottles as new to the next node along the supply chain. These types of wine counterfeiting techniques are much easier to pass off as the original than bottles produced by cloning the original labels. 

\subsection{Statistical Evidence of Growing Wine Counterfeit Market}
The recent growth of publicity on the counterfeited wine has given a large body of evidence that counterfeiting exists on a variety of different levels, and even in a spiraling trend throughout the international wine market. Specifically, there is growing evidence of fine wine counterfeit market and even the whole counterfeit industry.

\begin{figure}[h]
    \centering
    \captionsetup{justification=centering}
    \includegraphics[width=0.5\textwidth]{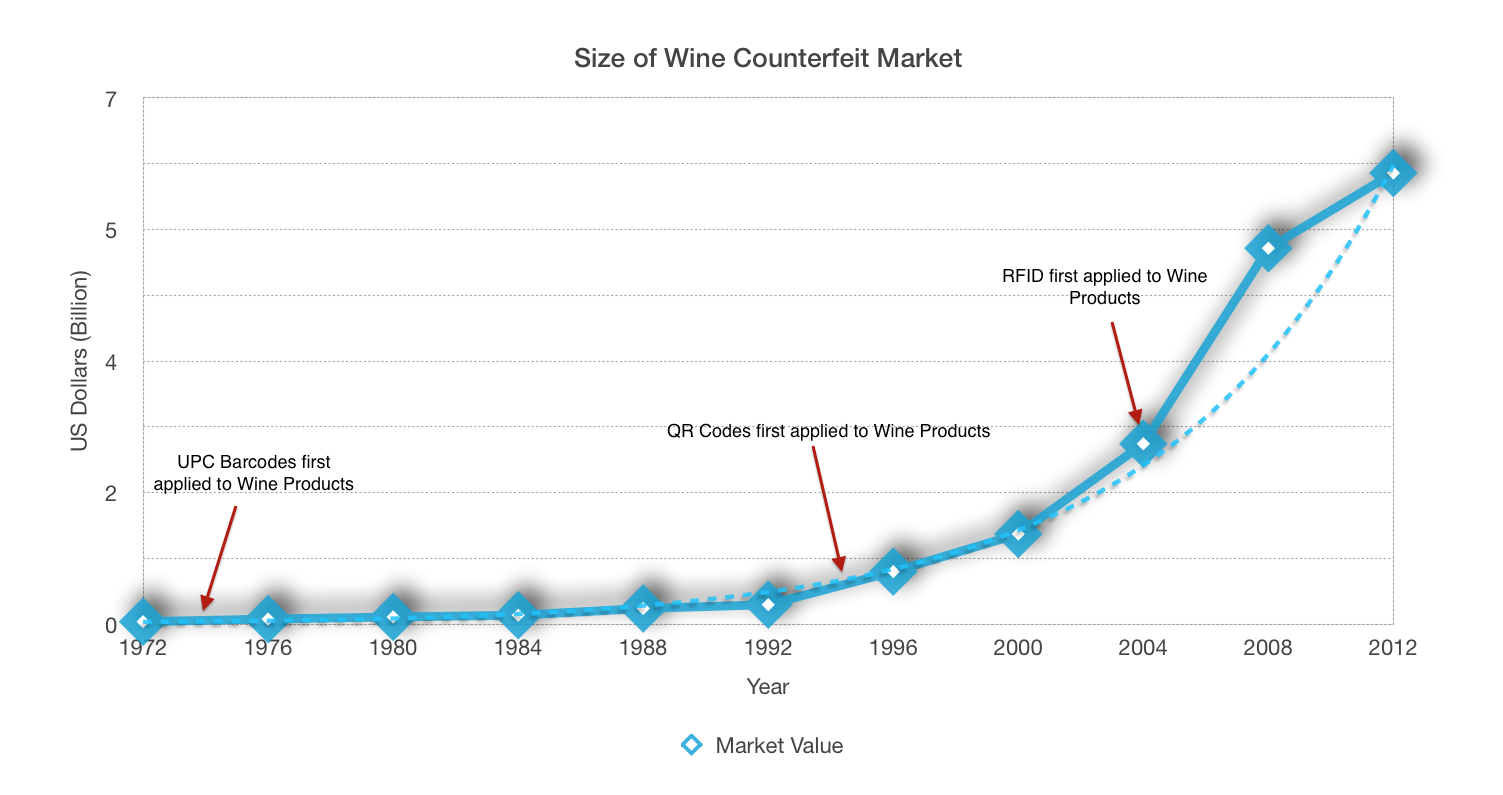}
    \caption{\textit{The size of wine counterfeit market between 1972 - 2012}}
    \label{fig:1}
\end{figure}

Indeed, the US government have estimated the global wine counterfeit industry, with quite different definition of wine counterfeiting, at \$5 billion (USD) with an upsurging rate of 1,700\% over the past 10 years in 2008 as well. Similarly back in 2012, the International AntiCounterfeiting Coalition (IACC) reported that wine counterfeiting is a problem costing \$6 billion (USD) a year, and it was in fact that the problem has grown over 10,000\% in the past two decades, in part fueled by consumer demand and the limited effectiveness of those labeling technologies applied to the wider wine industry. \textit{Fig.~\ref{fig:1}} demonstrates the estimates on the size of the wine counterfeit market performed by the two organizations, as well as the previous estimates, from 1972, based on the trend as specified by both organizations.

Concerning the wine counterfeiting market, there has been a pile of reported cases of counterfeit wine over recent years all around the world. Some of the more famous cases including 1990 Penfolds Grange, 1994 and 1995 Sassicaia worth over £1 million and the fake Rioja estimated close to one million bottles under which a famous wine collector, named William Koch, issued lawsuits over the counterfeited wine claiming his magnums of 1921 Chateau Pétrus, valued for about \$4 million (USD), was filled with cheap California Cabernet \cite{3}. Some even happened to the vintage that was never produced by the original wine producers. Burgundy winemaker Laurent Ponsot discovered that 106 out of 107 bottles of his wine at auction were faked including a sale of Clos Saint Denis 1945 and other old vintages they did not even begin producing this particular appellation until 1982 \cite{4}.

\subsection{Current Anti-counterfeiting Techniques Adopted in Wine Industry}
As the wine counterfeit has become a more challenging issue to wine industry, a bundle of anti-counterfeiting companies have developed and marketed their innovations and technologies in wine industry. This section is to provide a summary of the techniques utilized to combat wine counterfeit. 

There are many anti-counterfeiting techniques to confirm the authenticity of wine products; for example, examining the vintage corks has also been one of them due to the fact that vintage corks have appeared to be really tough to clone, compared with any other aspects of a wine product, which could easily be copied. It appears even more secured that corks are branded, and the original branding ink might be paler when the time goes by, but often the year of origin would still be clearly visible for viewing. Nonetheless, it is possible to detect the anomalies on the cork with the bottle sealed, but only limited to those experts of wine products not the ordinary wine consumers. Given the fact that this is one of the primary ways to verify the wine product, the wine product itself has to be uncorked by the ordinary wine consumers so as to confirm that wine product is authentic \cite{1}.

It seems that authenticating a wine through examining the cork is not common and has transferred low effectiveness to the industry as wine consumers will be incapable of authenticating a wine utilizing such technique. Is there any other techniques could be adopted easily by ordinary wine consumers without the need of uncorking a wine product at retail points before making purchase? Yes, a series of labeling technologies applied to the wine products could help anti-counterfeiting effort for the wider wine industry. So as to relieve the expanding wine counterfeit market, the UPC Barcode was first applied to the industry in 1974, followed by the QR codes and RFID which were adopted in different wine products in 1994 and 2004 respectively as shown in \textit{Fig.~\ref{fig:1}}. 

Notwithstanding, there have been rapid changes in the technology applied to counterfeit prevention and even for the anti-counterfeiting purposes. Counterfeiters have likely been only a step behind the introduction, and so the longevity and the trustworthiness of any existing labeling technology have been questioned, though there is a series of labeling technologies available for adoption, in order to combat the rampant wine counterfeiting, uncertain authenticity of wine products with labels counterfeited, such as barcodes, QR codes and even RFID tags. While the wine industry participants are becoming increasingly aware of the potential negative ramifications of counterfeit parts with altered serial numbers, barcodes and QR codes sourced and distributed in the supplier networks or along the supply chain. In fact, wine industry participants have regarded RFID tags as the preferred labeling solutions which can ultimately protect their brands and the products; however, doubts and threats such as self-replicating RFID virus, cloning of RFID tags and replay attacks could still be performed during the RFID application to the wine products at different nodes along the supply chain \cite{6} \cite{7}. Despite the fact that there are more innovations based on a variety of emerging technologies such as the Applied DNA Sciences botanical test, Jean-Charles Cazes of Chateau Lynch-Bages also stated that: “For us right now there is no technology in which we are sure would be viable on the market in 20 or 40 years.” \cite{5}

The incentives of building the one-of-a-kind NAS were then supported by the above findings covered and visualized. In response to the urgent need from the wine industry, an open, rapid and affordable anti-counterfeiting system adopted with newer communication technology – NFC, in which the technology has never been adopted in the wine industry or the supply chain industry as a whole for such purpose, should therefore be developed in this project.

\section{Background}
\subsection{Overview of The Proposed NFC-enabled Anti-Counterfeiting System (NAS)}
The NFC-enabled Anti-Counterfeiting System (NAS) is an integrated and “open” system tailor-made for winemakers, supply chain participants and wine consumers, aiming at operating two mobile applications to direct a NFC-enabled smartphone to connect with a back-end database system, managed solely by the winemaker. Through scanning the NFC tags on packaging of wine bottles using NFC technology of the smartphone, the proposed NAS has been developed and setup for the purpose of prompting consumers to authenticate wine products at retailer points.

The whole NAS is consisted of FIVE major components, which are (1) a back-end system for wine data management performed by the winemakers, (2) a mobile application, \emph{ScanWINE}, for tag-reading purpose of wine products at retailer points performed by wine consumers and supply chain participants before accepting a wine product, (3) another mobile application, \emph{TagWINE}, performing tag-writing purpose for wine products at wine bottling stage performed by the winemakers, (4) the NFC tags attached on the bottleneck for the anti-counterfeiting purposes and actions, and (5) the NFC-enabled smartphones or tablets utilized in any NAS process. 

NAS enables supply chain participants and wine consumers to confirm authenticity and provenance of wine products, and to preserve the supply chain integrity anytime and anywhere, utilizing the \emph{ScanWINE} mobile application, NFC-enabled smartphones and its underlying technology - NFC, at different nodes along the supply chain or at retail points before making purchase for the case of wine consumers. While the winemakers could utilize the \emph{TagWINE} mobile application with their NFC-enabled handheld devices (or simply their NFC-enabled smartphones registered in the integrated NAS) so as to provide enhanced security using NFC technology, in addition to its counterparts based on other existent labeling technologies such as Barcode technology or communication technology such as Radio-frequency identification (RFID), which have already been applied to the bottling and packaging stages of wine products, so as to deliver enhanced anti-counterfeiting features.

NAS has been addressing anti-counterfeiting aspects and authenticity of a wine product since its production stage with NFC-enabled labels or tags included in the product package. NAS is an open-and-break, recycling-prevented, and clone-prevented system. It offers convenience and simplicity to winemakers, supply chain participants and wine consumers without any sophisticated tool or sound knowledge and experience in wine products and the industry required. NAS is applicable to any type of wine bottle, due to the specific design of NFC tags working with metallic packaging materials such as foil package. NAS is also aimed at serving as a multi-lingual verification platform promoting a concept of seamless value chain for the international wine industry in which original winemakers, supply chain participants and wine consumers all around the world could be participated in any NAS process without any language barrier.

The anti-counterfeiting concept of NAS is designed to be based on the frangible labels for which once the labels are broken with the wine uncorked, the respective wine record of the wine product is destroyed so as to also prevent counterfeiters from recycling the wine bottle. NAS could be a more secured and pragmatic solution for different industry participants to protect their own assets and interests. The NFC hardware setup costs are low compared with the precious value of wine products, and hence the overall expenses of system implementation will be justifiable, as merely those invaluable vintage wine products already substantiate enough incentives for developing and implementing NAS in wine industry.

\section{Selection, Deployment and Preparation of NAS System Components}
A combination of software and hardware components are selected and prepared, so as to develop and implement NAS in wine industry. For instance, open-source software modules and tools are needed for developing both the web-based database application and the mobile applications for authentication purpose. Hardware components, such as NFC tags and NFC-enabled smartphones, should also be narrowed down so that the web-based database application and mobile applications could be connected to the hardware components, and make the whole integrated NAS and its anti-counterfeiting features running as well as offering enhanced anti-counterfeiting capabilities to wine industry in reality.

\subsection{Selection of NFC-Enabled Smartphone}
In today’s market of 2013, there is only Android as an operating system found to be compatible with NFC technology. It implies that all the NFC-enabled smartphones are actually running the operating system of Android, instead of Apple’s iOS or any other operating systems. As such, what we actually need so as to simulate the integrated NAS, is to source an NFC-enabled smartphone such that mobile applications utilizing NFC technology of the hardware, could therefore be developed and functioned in the NFC-enabled smartphone. 

There are in fact 114 NFC-enabled smartphones in the market as of December 2013, but how should we determine which smartphone is the most suitable one for the development of NAS? What should be the selection criteria of the desired NFC-enabled smartphone of the NAS? There are mainly FIVE of the selection criteria defined, namely (1) the smartphone should be implanted with the updated model of NFC chips and controllers, (2) the NFC-enabled chip should be found compatible with as many NFC tags in the market as possible, (3) the smartphone must be released after Fall 2011, which was the released season of the one-of-the-kind universal NFC tag – NTAG203, (4) the selected NFC-enabled smartphone must be compatible with the universal NFC tag – NTAG 203 and other Type 2 NFC tags, and (5) the smartphone must be equipped with basic technical requirement, such as implanting with quad-core processor, running on the Android 4.3 (Jelly Bean) at least, built with WiFi and Bluetooth, etc.

The latest Samsung release in 2013 – Samsung Galaxy Note 3 (Model Number SM-N9005) was then narrowed down as the main NFC-enabled device and the desired smartphone for which we built NFC-enabled mobile applications on, as the Note 3 does meet all the criteria required for developing NAS. Samsung Galaxy Note 3 is the latest model of the Galaxy Note Series in 2013, for which it is NFC-enabled and its NFC controller is also compatible with the NXP NTAG203, fulfilling all the remaining aforementioned requirements as well. 

The NFC controller chip of the desired NFC-enabled smartphone (Samsung Note 3) is Broadcom BCM20793S. Regarding the compatibility with NFC tags, the NFC controller chips of the desired NFC-enabled smartphone supports communication with Type 2 NFC Tags. There are of course some other NFC-enabled smartphones which also fulfil the criteria of NAS, but not as more suitable as the desired Samsung Galaxy Note 3 given the fact that it has checked all the five aforementioned selection criteria. There is a full list of all the NFC-enabled smartphones or devices in the market as specified in \textit{Appendix~\ref{a1}}.

\subsection{Selection of NFC Tags}
\begin{figure}[h]
    \centering
    \captionsetup{justification=centering}
    \includegraphics[width=0.35\textwidth]{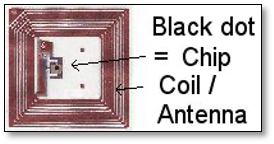}
    \caption{\textit{A formation of an NFC Tag}}
    \label{fig:2}
\end{figure}

An NFC Tag is an non-powered passive target consisting of an NFC PCB chip and an antenna in a substrate as described in \textit{Fig.~\ref{fig:2}}. NFC tags have relatively low radio frequency of 13.56 MHz, which is too weak for intense interaction. When an NFC tag is exposed to an “initiator” (NFC reader, NFC writer or NFC-enabled smartphone) with a more powerful radio-frequency field, the radio-frequency field of the NFC tag is then strengthened to facilitate tag-reading and tag-writing processes. Most NFC tags contain standardized NDEF messages, which are parsed and formatted to readable content. NDEF, as detailed in \textit{Appendix~\ref{a4}}, is strictly a message format and a common data format for NFC forum-compliant tags and devices, which is essential to any tag communication steps such as tag-reading and tag-writing.

Similarly, there is also a group of FIVE selection criteria set so as to narrow down the most suitable NFC tags for developing the integrated NAS, namely (1) the compatibility between the NFC tag itself and the NFC controller chip of the selected NFC-enabled smartphone, (2) the surface material onto which the NFC tag is applied, (3) the size of the NFC tag, (4) the memory capacity of the NFC tag, and (5) the write endurance of the NFC tag. There are also some minor criteria, which are the unit cost and the souring availability.

\begin{figure}[h]
    \centering
    \captionsetup{justification=centering}
    \includegraphics[width=0.2\textwidth]{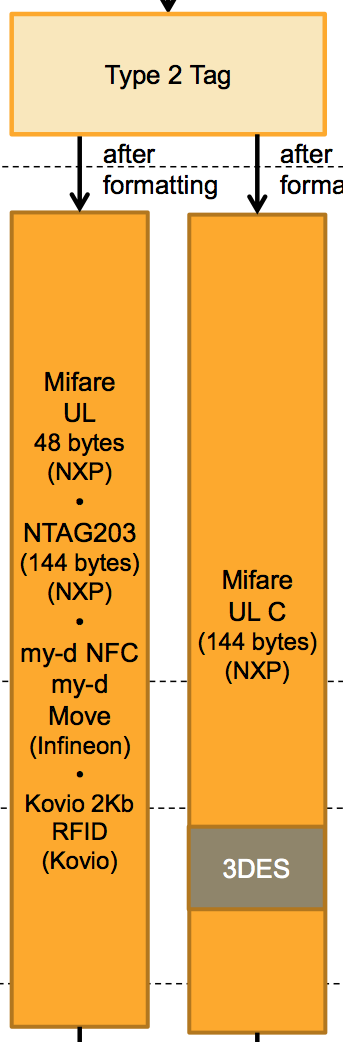}
    \caption{\textit{The Type 2 NFC Tag Group}}
    \label{fig:3}
\end{figure}

Regarding the compatibility between the chosen NFC tag and the preferred NFC-enabled smartphone, Samsung Galaxy Note 3 was selected and sourced as the preferred NFC-enabled smartphone in which its NFC controller chip is the aforementioned Broadcom BCM20793S. According to the NFC Forum Type Tag Platform, NFC Standards, Products existed with its specifications, as detailed in \textit{Appendix~\ref{a2}}, it has proved that those NFC tags with the NFC Forum Type Tag Platform – Type 2 Tags would be the most suitable and compatible NFC tags with the specific NFC controller chip implanted in the Samsung Galaxy Note 3. In short, our targeted NFC tags should ONLY be those under the group of “Type 2 Tag” as specified in \textit{Fig.~\ref{fig:3}}. Based on the first criteria, it could be concluded with the fact that only Type 2 Tag will be chosen for further screening of the NFC tag to be adopted for the development of NAS, which could either be any Type 2 tags, such as Mifare Ultralight (UL), NTAG203, Mifare Ultralight C, Kovlo 2Kb RFID, etc. 

\begin{figure}[h]
    \centering
    \captionsetup{justification=centering}
    \includegraphics[width=0.45\textwidth]{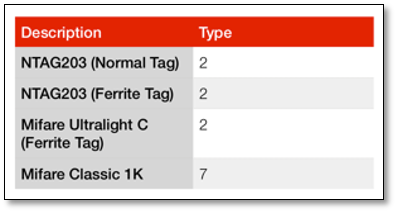}
    \caption{\textit{Different NFC tag models and its type}}
    \label{fig:table}
\end{figure}

For the tag physical size, it would be optimal if the selected NFC tags come with a shape of circle and around 25-35 mm of diameter with at most 0.5 mm in thickness for packaging a wine product of 0.75 liter, while the preferred write-tag memory capacity should be 144 Bytes. (For Type 2 Tag, there is only 48-byte or 144-byte available). The write endurance should be at least 10000 times. Through grouping the outcomes of all the aforementioned selection criteria, there are FOUR NFC tags selected and sourced for further comparison, which are all produced by NXP Semiconductors under which three are Type-2 in nature and one is Type-7 in nature, as compared in \textit{Fig.~\ref{fig:table}}. For those Type-2 NFC tags, two of them are actually NFC Ferrite Tag with which this type of NFC tag could enable the integrated NAS to work and operate under the metallic environment, such as foil packaging around most of the bottlenecks at wine products. Adoption of the special protective layers on NFC tags will be discussed and these special tags would be introduced in the improvement stage so as to further enhance the anti-counterfeiting capability of NAS based on the results obtained from future usability research activities.

\begin{figure}[h]
    \centering
    \captionsetup{justification=centering}
    \includegraphics[width=0.26\textwidth]{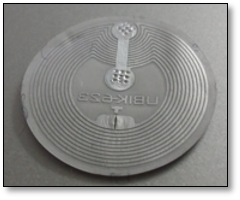}
    \caption{\textit{The Sample NFC Tag - NTAG203}}
    \label{fig:4}
\end{figure}

NTAG203, as demonstrated in \textit{Fig.~\ref{fig:4}}, is the most universal NFC tag in the world. Given the fact NTAG203 is compatible with and supported by most of the models of NFC-enabled smartphone and device, it will in all likelihood be suitable for the integrated NAS developed. While Mifare Ultralight (UL) C is developed and designed for limited-use applications, such as authentication and NFC Forum Tag Type 2 applications, for which it is a more secured version of Mifare Ultralight (UL) integrated with 3DES encryption.

\begin{figure}[h]
    \centering
    \captionsetup{justification=centering}
    \includegraphics[width=0.43\textwidth]{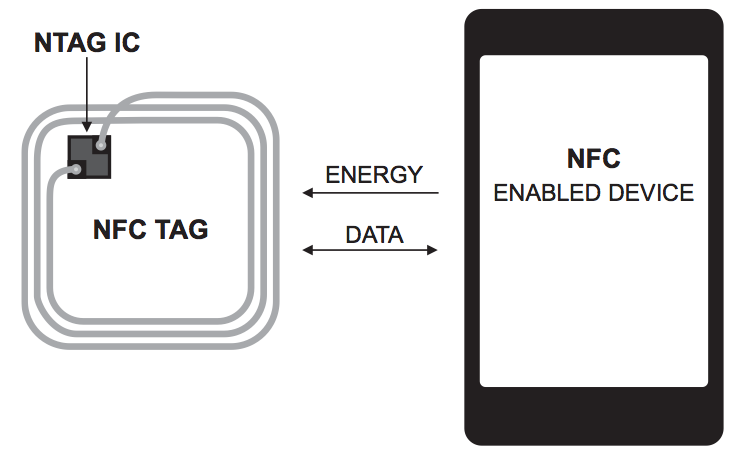}
    \caption{\textit{The Near-Field communication procedure between an NFC tag and an NFC-enabled device}}
    \label{fig:interaction}
\end{figure}

An intelligent anti-collision function, according to ISO/IEC 14443, allows operating more than one card in a common field simultaneously. The anti-collision algorithm selects each card individually and ensures that the execution of a transaction with a selected card is performed correctly without data corruption resulting from other cards in the same common field, as demonstrated in \textit{Fig.~\ref{fig:interaction}}. As such, there will not be 2 NFC tags being written or read at the same time.

\begin{figure}[h]
    \centering
    \captionsetup{justification=centering}
    \includegraphics[width=0.30\textwidth]{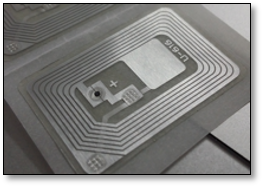}
    \caption{\textit{The Sample NFC Tag - Mifare Class 1K}}
    \label{fig:m1k}
\end{figure}

Regarding Mifare Class 1K in \textit{Fig.~\ref{fig:m1k}}, the MF1ICS50 is designed for simple integration and user convenience, which could allow complete ticketing transactions to be handled in less than 100 ms. The most suitable NFC tag for the development of NAS will further be determined once the NAS is functional and ready later for a series of tests for different situations, user cases and categories of pre-determined NFC tags. Mifare Class 1K was also sourced in this project for the purpose of control experiment to determine whether the NFC controller chip of the preferred NFC-enabled smartphone could be compatible with this Type-7 tag or not. All details and specifications of the NFC tags and NFC-enabled smartphone are included in \textit{Appendix~\ref{a3}}.

\subsection{Introduction of NFC Ferrite Tag (Anti-Metal Tag)}
After the initial phase of NFC tag selection process, the NFC tags sourced were proved not available for working under metallic environment. A set of NFC ferrite tags, which were 1) NTAG 203 NFC Ferrite Tag and 2) Mifare Ultralight (UL) C NFC Ferrite Tag, were then sourced so as to tackle the communication problem attributed by metallic interference. 
\begin{figure}[h]
    \centering
    \captionsetup{justification=centering}
    \includegraphics[width=0.41\textwidth]{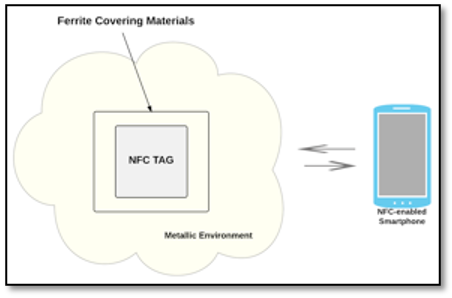}
    \caption{\textit{The NFC Ferrite Tag Working in Metallic Environment}}
    \label{fig:metalic}
\end{figure}

\begin{figure}[h]
    \centering
    \captionsetup{justification=centering}
    \includegraphics[width=0.41\textwidth,height=1.9in]{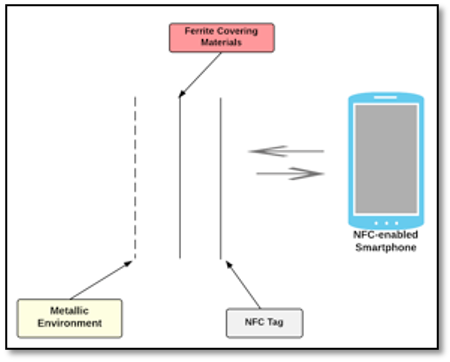}
    \caption{\textit{The Ferrite-Covering Material Shields NFC Antenna from the Metal Surface}}
    \label{fig:metalic1}
\end{figure}

The NFC Ferrite Tag (Anti-metal tags) is proved to be working on metal surfaces because the anti-metal, Ferrite-covering materials shield the NFC antennae from the metal surface. For instance, when there is a metal object present in the communication field, such as the foil packaging material surrounding the wine bottleneck, near the antenna of that NFC tag , the Ferrite sheet, which is like a thin magnetic sheet physically, is required on the NFC tag to avoid communication failure caused by the metallic interference. It is believed that the NFC Ferrite Tag can change and optimize the magnetic flux path to avoid interference to the NFC tag, and result in high surface resistance and effectiveness in preventing resonance and suppressing coupling, as depicted in \textit{Fig.~\ref{fig:metalic}} and \textit{Fig.~\ref{fig:metalic1}}. 

\section{The Proposed NFC-Enabled Anti-Counterfeiting System}
\subsection{System Overview of NAS}
So as to relieve the current rampant situation of wine counterfeiting in emerging markets and developing countries, this project proposes an anti-counterfeiting system aimed to provide an up-to-dated wine pedigree of every wine product throughout its supply chain where both supply chain participants and wine consumers can share, update and retrieve while utilizing the NAS for the purpose of authenticating wine products. The anti-counterfeiting mechanism was designed and developed based on a winemaker, and according to the automated approach to update the wine pedigree of a wine product whenever there is an NFC-based scanning process performed by any supply chain participant with their NFC-enabled device for updating the wine pedigree itself, such as the record of wine acceptance by the next node along the supply chain of specific wine product. NAS also enables wine consumers to authenticate the wine product via their own NFC-enabled smartphones at retail points before purchasing it.

NAS requires various participants along the supply chain to record wine transactions and update the states of specific wine products, through applying the NFC technology enabling NAS. As such, supply chain integrity on the wine products is preserved by forming a chain of custody based on wine transaction records stored in a central database owned merely by the winemakers, in which only winemakers will be allowed to modify the data stored in the back-end database for a specific wine product. All the supply chain participants are expected to verify incoming wine product and reject the suspicious ones, for which no data would be returned to the NFC-enabled mobile applications, during the NAS scanning process, at their node along the supply chain. The system primarily targets at high-end fine wine products, and the NFC hardware costs are relatively lower compared to the value of these wine products, and hence the system implementation cost will be justifiable. More importantly, the valuable wine product itself already provides enough incentives for wine consumers to verify them, using their own NFC-enabled smartphone, instead of the NFC-enabled handheld device provided by retailers, before purchasing the wine products. 

The NAS is consisted of mainly three parts to actualize the aforementioned anti-counterfeiting functionalities, namely the (1) back-end database servers, (2) the NFC tags compatible with the NFC-enabled handset selected, and (3) two mobile applications running on the NFC-enabled smartphone. Wine consumers could simply tap on the smartphone to check an authentication certificate of a wine product to confirm the originality of a luxury wine with full wine pedigree, wine details and even the wine pictures retrieved and shown on the user interface of the mobile application - \emph{ScanWINE}, and winemakers could simply tap to edit and add data of a specific wine product while performing the bottling process of their manufacturing steps, utilizing another mobile application - \emph{TagWINE}.

\subsection{The Design Methodology of NAS}
The anti-counterfeiting system can be accessed through the Internet. It is divided into FOUR layers – Application Controller Layer, Wine Database Layer, User Execution Layer and Internet Protocol Communication Layer. Wine Database Layer is comprised of the database itself and its database management system owned by winemakers. Application Controller Layer actually provides users and those NFC-enabled devices with access to server controllers, corresponding user interfaces, as well as sending request to wine databases. Data Interchange Output Layer and Internet Communication Protocol provide communication standard for both Application Controller Layer and User Execution Layer to communicate with each other. User Execution Layer can be accessed through Hypertext Markup Language (HTML) or via JavaScript Object Notation (JSON) using Hypertext Transfer Protocol (HTTP) for sending requests to Application Controller Layer. Both Wine Database Layer and Application Controller layer are connected over a Local Area Network (LAN) owned by the winemakers, while those Application Controllers interface with the components of User Execution Layer through the Internet. Wine Products Database, LAN Wireless Router and Application Controller Layer must be separated with firewalls so as to prevent from any attempted external attacks.
\begin{figure*}[h]
    \centering
    \captionsetup{justification=centering}
    \includegraphics[width=1\textwidth]{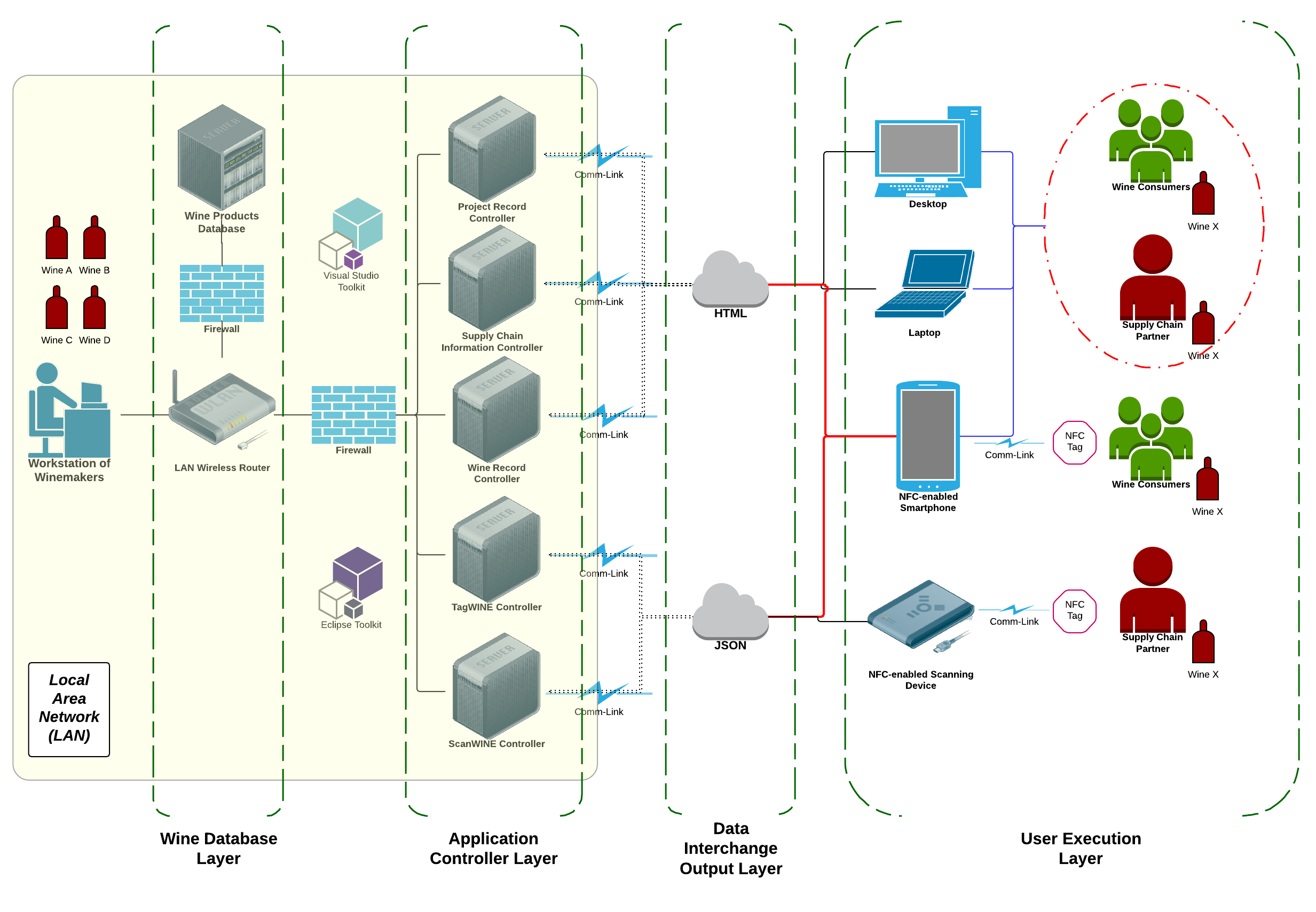}
    \caption{\textit{The Design Methodology of NAS}}
    \label{fig:method}
\end{figure*}

Regarding Wine Database Management Layer, the wine database is acting as a heart of the whole NAS and it is supposed to be solely owned by winemakers. It consists of tables that record wine pedigree information, including wine product details, vintages, transaction records, and any related project along with supply chain information regarding specific wine products. The wine database system updates, retrieves and transmits the required information from or to those tables involved according to instructions given by Application Controller Layer, thus integrating the supply chain data with transaction record systematically, logically and automatically to form wine pedigree based on the request sent from User Execution Layer. In short, this layer stores wine records produced in the workstation of related winemakers.

Application Controller Layer is where the operating logic and control of user interface for NAS are developed and executed. The operating logic is mainly composed of C\# and JavaScript syntax to perform operations of NAS with five application controllers, namely (1) Project Record Controller containing and operating related wine records for specific projects, (2) Supply Chain Information Controller containing and operating data of any related supply chain participant, (3) Wine Record Controller containing logic for managing those wine records which are responded from requests with accurate wine record and performing control of user interface for the wine management database, (4) the application controllers of the \emph{TagWINE} (only accessible to winemakers) and (5) \emph{ScanWINE} (Wine Consumers and Supply Chain Participants) in which they contain the operating logic of wine record requests and retrievals communication mechanism between User Execution Layer and Wine Database Layer. A set of design patterns, written in JavaScript of application controller and those written in C++/C\# set in the “AppController” of with the back-end Microsoft SQL database are used to access the data stored in wine database of Wine Database Layer, with data returned the required back to the mobile applications running on NFC-enabled smartphones. In addition, the Visual Studio and Eclipse Toolkits will be needed if there are interactions between Wine Database Layer with both Microsoft SQL database of those supply chain participants, and \emph{TagWINE} applications under User Execution Layer.

While the Data Interchange Output Layer is where the user interface of database web application and JSON are utilized to publish the content generated via operating on NAS, and the content will then be output and transferred to User Execution Layer for users to execute business transactions and decisions on acceptance at different nodes along the supply chain. The NAS users could make instructions or requests, such as agreeing the partnership, to the winemaker’s back-end database using the “Participants” version of web-based database with their desktop, laptop or smartphone device through the user interface of the database web application of NAS. The instructions, such as wine acceptance, updating transaction records and requesting for specific wine records while at any retail point, can also be made by scanning an NFC tag attached on the wine bottle and transmitted back to Wine Database Layer in JSON format. The JSON could ensure the machine-to-machine interaction without human intervention to suit the fast-moving nature of supply chain through using NFC tags, as the mobile applications will send data output in JSON format to Application Controller Layer, which is also the reason of employing NFC in the proposed track-and-trace system in which associated decisions will be made automatically by the mobile applications and controllers of the backend database. The communication path, highlighted in red as specified in the design methodology detailed in \textit{Fig.~\ref{fig:method}}, depict the fact that NFC-enabled smartphones could send requests to Wine Database Layer through in both HTML and JSON formats as the smartphone itself can access mobile version of web-based databases deployed for those supply chain participants and perform those NFC-related functions with the mobile applications of NAS running on it respectively.

For User Execution Layer, where an user loads an interface from Application Controller Layer to operate a web application of the database and to make purchasing and business decisions on specific wine products, HTML pages are enhanced with the use of Cascading Style Sheets (CSS) and JavaScript to improve the user-friendliness of the web application database system. The NAS user instructs the corresponding application controller to perform the business logic accordingly through the User Execution Layer. The results are then transferred to and displayed on the user interfaces at the User Execution Layer. The cluster, highlighted in red dotted line as specified in \textit{Fig.~\ref{fig:method}}, implies both supply chain participants and wine consumers could utilize the database web application running on desktop, laptop and smartphone, highlighted in blue as specified in \textit{Fig.~\ref{fig:method}}, to perform functions required for different industrial operations.

\subsection{System Architecture of NAS}
\begin{figure}[h]
    \centering
    \captionsetup{justification=centering}
    \includegraphics[width=0.5\textwidth]{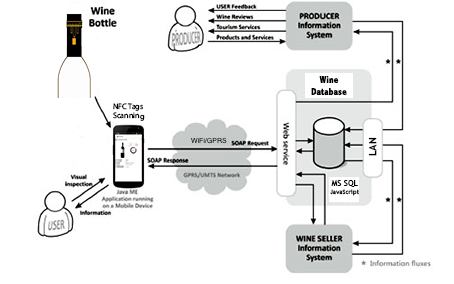}
    \caption{\textit{The General Operational Structure of NAS}}
    \label{fig:systemarchi}
\end{figure}

 The \textit{Fig.~\ref{fig:systemarchi}} has demonstrated a general operational structure of NAS. For web application of the wine database owned solely by the winemakers, there are mainly two modules of information system, which are (1) the Producer Information System and (2) the Wine Seller Information System. The former is actually a system consisted of wine product details, consumers' feedback, reviews, etc. It only stores information regarding specific wine products, under which winemakers will key in wine data to update states of the wine products on the back-end database via the web application providing user interface, and in return, the NAS could therefore provide winemakers with data about the wine products for further improvement on the wine product itself and better practice in product management. While the latter is consisted of transaction records of specific wine products, the details of the supply chain participants, which could be any distributor, wholesaler and re-sellers of wine products produced by winemakers, the details of project involving sales of a combination of different wine products between winemakers and specific supply chain participants. Both modules will then be connected and integrated together through the Local Area Network (LAN) and form a basic operating layer of the wine database web application, with the back-end database developed in Microsoft SQL, owned by specific winemakers. The database web application is actually served as a platform for the wine database to communicate with those mobile applications running on NFC-enabled smartphones of NAS, with those application controllers deployed to both the mobile applications and the web application of the back-end database.

While at the users’ end, wine consumers perform steps to authenticate the specific wine product at a retail point or when the wine product is being accepted by the next nodes along the supply chain of such wine products, using \emph{ScanWINE}, running on any designated NFC-enabled smartphones of NAS. Supply chain participants could also use \emph{ScanWINE} running on their NFC-enabled smartphones, to scan the NFC tag attached on any wine bottleneck. In the meantime, the NFC tag acts as a bridge to link both the mobile applications and the back-end database based on the WID (Wine identifier which is a basic element of tag value stored in the NFC tag. With WID, the specific wine record could be located and referred by the web application and both mobile applications, for which the communication, such as the request sent from the application for the information of specific wine record based on the WID scanned and the wine record returned from the back-end wine database, could be transmitted in JSON format via HTTP with the application controllers under the web service platform, using WiFi or General Packet Radio Service (GPRS) provided by those internet service providers in the region.

\begin{figure}[h]
    \centering
    \captionsetup{justification=centering}
    \includegraphics[width=0.5\textwidth]{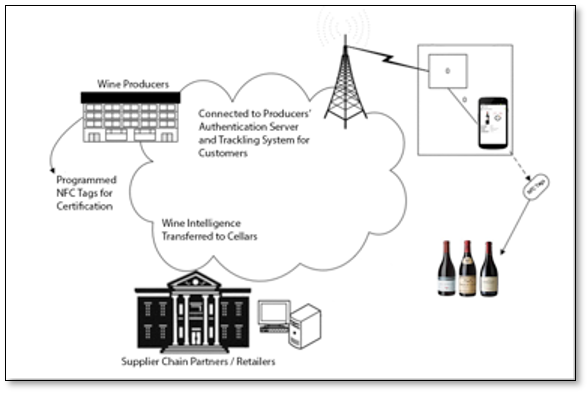}
    \caption{\textit{An Example of NAS System Flow}}
    \label{fig:systemflow1}
\end{figure}

As described in \textit{Fig.~\ref{fig:systemflow1}}, when a specific wine produced and being transferred hand-in-hand between different nodes along the supply chain, the wine record such as the transaction record of that wine product will be automatically updated simultaneously and accordingly once the tag affixed on the wine bottleneck is scanned with an NFC-enabled device owned by any supply chain participants at every node along the supply chain. For instance, when a supply chain participant adopted NAS and utilized the NFC-enabled smartphone to scan the NFC tag while they accepted the wine intelligence shared by a specific winemaker, the \emph{ScanWINE} mobile application running on the NFC-enabled device will connect to the Authentication Controller of the back-end wine database owned by that winemaker, and so the transaction record will be updated automatically.

\begin{figure*}[h]
    \centering
    \captionsetup{justification=centering}
    \includegraphics[width=1\textwidth]{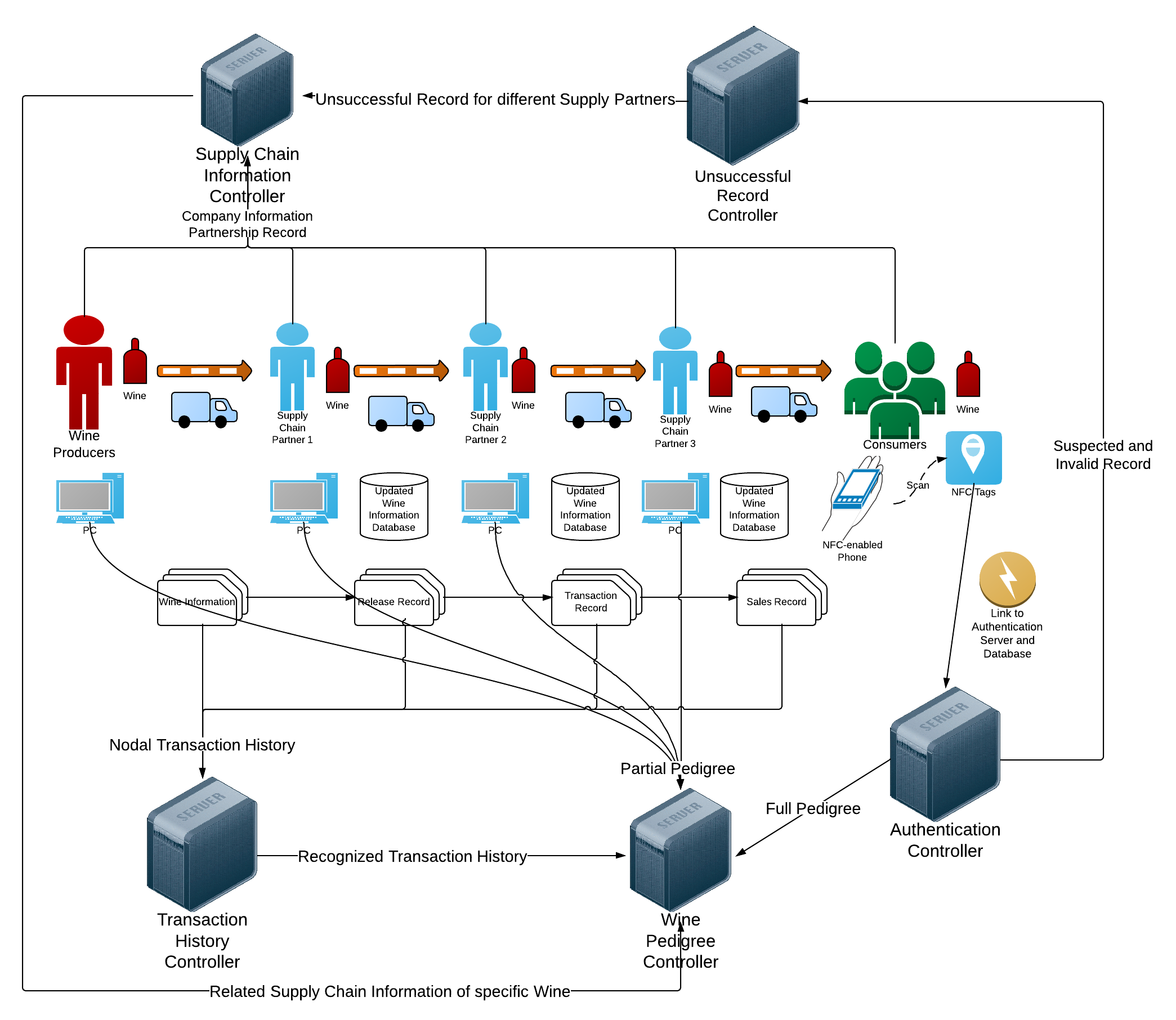}
    \caption{\textit{The System Structure Flow of NAS}}
    \label{fig:systemflow2}
\end{figure*}

According to the system structure flow chart of NAS along the supply chain of specific wine product, as shown in \textit{Fig.~\ref{fig:systemflow2}}. In fact, NAS comprises FIVE important controllers to perform different functionalities, namely (1) Supply Chain Information Controller, (2) Transaction History Controller, (3) Authentication Controller, (4) Wine Pedigree Controller, and (5) Unsuccessful Record Controller, which were all grouped under the controller-layer logic of the back-end database.

The Supply Chain Information Controller is responsible for collecting and managing company information of other related supply chain participants, record of past collaborative projects with the winemaker and the wine products which have been held by the supply chain participants. For example, the data, such as the combination of wine products involved, in the collaborative project could be updated simultaneously when those wine products were being sent by the winemaker along the supply chain and being accepted by the subsequent supply chain participants through scanning the NFC tag attached, with \emph{ScanWINE} running on their NFC-enabled smartphones. The data of the supply chain information should be registered beforehand by the supply chain participants and verified by the host company of the infrastructure, which could be the winemaker, before the first transaction record was sent to Wine Pedigree Server. 

The Transaction History Controller is crucial for the wine pedigree of different wine products processed on NAS because they form an entire picture of the product history of wine products at specific nodes along the supply chain. Transaction History Controller will collect and manage all the nodal transaction history data including wine information data and release records provided by winemakers, and other transaction records such as the sales records provided by subsequent supply chain participants. This controller also forms the basis of tracing product authenticity when suspicious counterfeits emerged. As wine products move along the supply chain, each nodal supply chain participant should scan the NFC tag affixed on wine bottleneck using their NFC-enabled smartphones, supposedly connected to the internet, with metadata exchanged in JSON, and via HTTP with the Transaction History Controller, so that each nodal record could therefore be updated. Eventually, the aggregated data of related transaction history stored in the winemaker’s wine database could be updated, automatically and accordingly, under which no one, but only the winemaker, is able to modify the transaction record manually, which is as well a security consideration of NAS.

The Authentication Controller verifies those nodal transaction history data and the updated supply chain data retrieved from Transaction History Controller and Supply Chain Information Controller respectively. Authentication Controller could filter suspicious activities along the supply chain, such as cloned NFC tag on any wine products being detected and identified. The filtered wine records are then sent to Wine Pedigree Controller for storage and being processed on Wine Database Layer. For the wine records which could not pass through Authentication Controller, with its wine products being labeled as suspected wine counterfeit during the authenticating process specified in Authentication Controller, the metadata - “Wine Status” of that wine product would then be turned from “Valid” to "\emph{Invalid}". When the "\emph{WID}" (a component of tag value) stored in NFC tags scanned by the supply chain participants or wine consumers, was found unmatched with that in the back-end database maintained by winemakers, the related wine record will be sent to Unsuccessful Record Controller for storage, further reviewed and future reference by other supply chain participants sharing the related and corresponding winemaker. The supply chain participants will also be responsible for reporting such a case of attempted counterfeiting to the winemaker of that wine products and returning the “Invalid” wine product back to the original winemaker for fear that the suspected wine counterfeit will still be circulated in the wine market. Nonetheless, the “Invalid” wine record stored in Unsuccessful Record Controller will anyway be shared to Supply Chain Information Controller to update the state of affected wine products according to the anti-counterfeiting mechanism of the NAS-driven supply chain of wine industry, responsively reporting any suspected case of wine counterfeit, for further documentation and reference for every winemaker.

The supply chain participants can also verify partial wine pedigrees from the point of manufacturing to their previous node along the supply chain which are indeed previous owners of the wine products, by sending requests to Wine Pedigree Controller storing the legitimate wine records, which in turn retrieved transaction records from Transaction History Controller as well as company information data from Supply Chain Information Controller to generate the required pedigree in Wine Pedigree Controller. Several validation mechanisms are also embedded in these controllers, rejecting any wine product with suspicious partial wine pedigrees detected and found. Besides, some validation mechanisms also involve both Wine Pedigree Controller and Authentication Controller, in which the former storing partial legitimate wine records. The latter, which is comprised of a series of authenticating logic for verifying wine pedigree records generated and passed by the former, is aimed at retrieving requests from those NAS users and responding with resultant metadata of wine pedigree data in JSON format, eventually demonstrated on the user interface of \emph{ScanWINE} running on their NFC-enabled smartphones, along with the aforementioned anti-counterfeiting functionalities delivered. When a wine consumer is satisfied with the genuine wine product utilizing NAS and in turn purchased the wine product with confidence, NAS will automatically update the sales record to Transaction History Controller, which will subsequently updated the state of the wine pedigree record in Wine Pedigree Controller. The tag would no longer be workable in any further scanning processes detected with the same wine product after the sale record updated as the wine product will expect to be consumed. Any further state transitioned at any post-purchase point along the supply chain will therefore be deemed suspicious and should be rejected immediately as NAS is assumed to be non-applicable to any post-purchase steps after a wine product being purchased at any retail point.

\subsection{Use Case Analysis of NAS}
Considering the whole NAS, there will be mainly three application systems contributed to actualize the anti-counterfeiting functionalities enabled by NAS, which are the web-based wine database storing those important wine details, the mobile application - \emph{TagWINE} in which winemakers could utilize it to write the specific wine record into the NFC tag while performing bottling during the wine production process, and another mobile application - \emph{ScanWINE} in which both the supply chain participants and wine consumers could use it to authenticate a wine product before accepting a wine product sent from the winemaker or their previous nodes along the supply chain by other supply chain participants or buying the wine product by wine consumers at retail points.

\begin{figure*}[h]
    \centering
    \captionsetup{justification=centering}
    \includegraphics[width=1\textwidth]{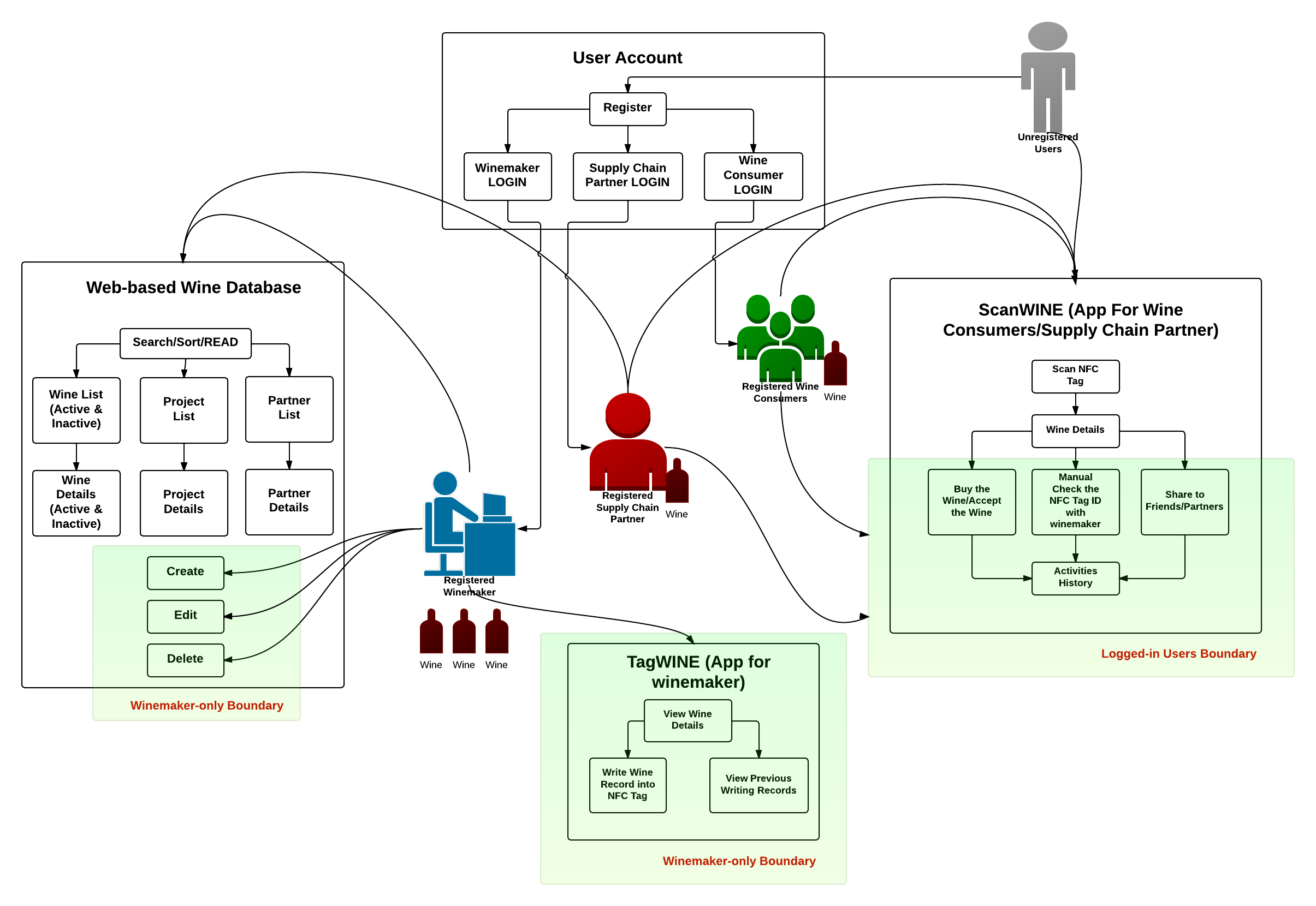}
    \caption{\textit{Use Case Analysis of NAS}}
    \label{fig:usecaseanalysis}
\end{figure*}

There are FOUR users will well be involved in NAS, which are winemakers, supply chain participants, wine consumers and unregistered users. According to the use case diagram demonstrated in \textit{Fig.~\ref{fig:usecaseanalysis}}, the web-based wine database could only be accessed if both the winemakers and supply chain participants are registered so that they could access to the database system for usage; however, only winemakers, which are also administrators of the database system, could perform functionalities such as create, edit or delete on every wine product, project and partnership records stored in the wine database (even those owned by the registered supply chain participants), implying that the wine database owned by the supply chain participants is just an extension of that owned by winemakers. The registered supply chain participants could only read, search or even sort those wine records related to specific winemakers, and they have no right for modifying any wine record owned by respective winemakers. Regarding \emph{TagWINE}, the area, highlighted in green as specified in \textit{Fig.~\ref{fig:usecaseanalysis}}, states that only winemakers will be eligible for accessing the mobile application to perform tag-writing and other functionalities, such as viewing wine details and viewing previous writing records.

Regarding \emph{ScanWINE}, any type of user could essentially utilize the mobile application on their purpose, including those unregistered users, other supply chain participants and wine consumers to perform functionalities of NAS to authenticate a wine product along its supply chain, distribution channels and even at retail points with required wine records returned. However, the area of registered and logged users, highlighted in green as specified in \textit{Fig.~\ref{fig:usecaseanalysis}}, states that functionalities, such as buying, accepting and sharing specific wine products, or even manually checking NFC tag identifiers against that stored in winemaker’s database will only be bound for the registered supply chain participants and wine consumers to use, under which the activity history for individual accounts should be accessed only if corresponding registered users are logged so that those activity records could be viewed by those account users.

\section{The System Architecture of Wine Database Web Application}
The Anti-counterfeiting model of NAS is assumed and designed to be based on a winemaker, in this project, which is also the only entity hosting technical infrastructure of NAS including the back-end wine database instead of a cluster of winemakers. Being the heart of NAS, both wine database itself and its web application, owned merely by the winemaker, are where the wine records of all wine products would be stored. The state changes involved in any operation on wine records, such as creating, appending and even sharing them with other registered supply chain participants, will be defined and recorded in this system component. Regarding the wine database web application, in addition to those wine records stored for the purpose of being queried and requested, at any node along the supply chain, by those supply chain participants and wine consumers while they are in the process of accepting or purchasing the wine products. Such state-transitioned requests on specific wine records are firstly processed in Wine Pedigree Controller which would further update states of specific wine records stored in the wine database accordingly. The wine database also stores invalid wine records of any suspicious wine products found along the supply chain accordingly with state-transitioned request processed in Unsuccessful Record Controller. Similarly, data of trusted supply chain participant with their associated projects involving a combination of wine products produced by the same winemaker, would also be updated in the wine database, whenever there is any state-transitioned request relating to supply chain participant data processed in Supply Chain Controller.

\subsection{The Sitemap Design of Wine Database Web Application}
Regarding the Site Map of the wine database demonstrated in \textit{Appendix~\ref{b1}}, all the functionalities that would well be available on the user interface of wine database web application, are categorized mainly into four tabs on the web application are (1) Active Wine, (2) Inactive Wine, (3) Project Details and (4) Supply Chain Participant data respectively. The nature of these functionalities are mainly in searching for targeted wine records, sorting the lists with a variety of attributes in different sequential preferences. The functionality of sharing designated wine records could be achieved under the viewing page of wine records such as the web page of “View Details” or “Edit”. Besides, a login page will be designed and required for winemakers’ login so that only the winemaker could get access into the wine database. The registration page of winemakers and other supply chain participants will also designed and developed to adopt NAS.

\begin{figure}[h]
    \centering
    \captionsetup{justification=centering}
    \includegraphics[width=0.5\textwidth]{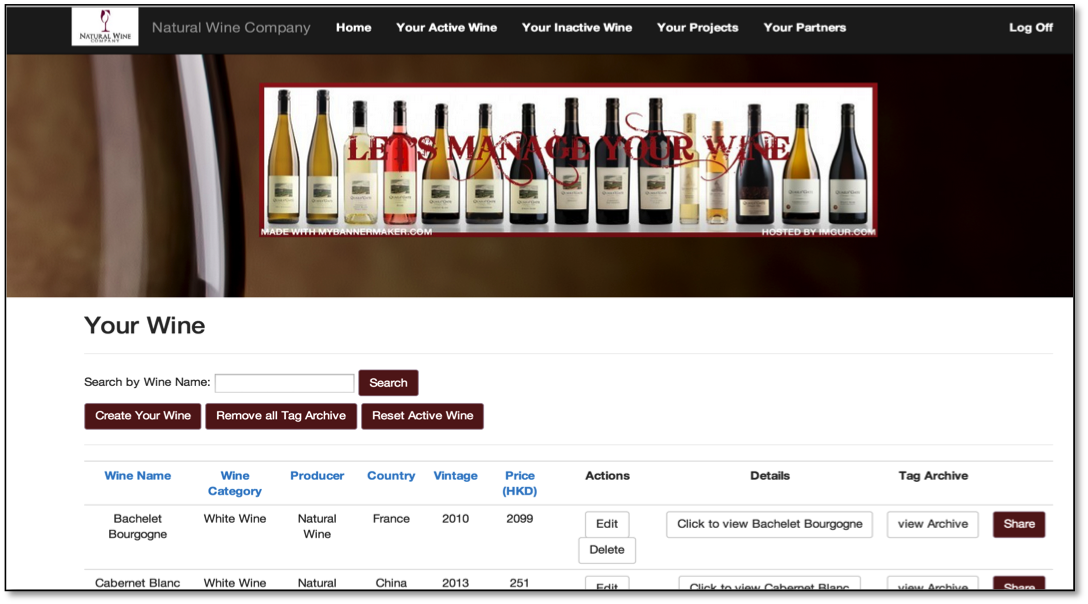}
    \caption{\textit{The Wine Database Web Application}}
    \label{fig:databasewebapp}
\end{figure}

Regarding the tab of Active Wine, it allows winemakers to view the wine details of wine products the winemakers produced and bottled, and to performing functionalities, such as searching for specific wine records required, sorting to customize the sequence of the list of wine products, and even viewing the full pedigree of specific wine products, so that the winemakers will have a better understanding on status of the wine products they produced, also via comparing wine records of different wine products whenever there will be new versions of wine records required to be updated by the winemaker. With the wine database web application demonstrated as shown in \textit{Fig.~\ref{fig:databasewebapp}}, it follows that the winemaker could therefore have a better planning of their wine products and better management on those associated wine records, projects, as well as enhancing the effectiveness on the projects they will be working alongside with their trusted supply chain participants, since the product details of every individual wine involving in certain project could be viewed and committed easily with a specific supply chain participant. All wine records listed under this tab are actually extracted from the wine database based on the operating logic set in Wine Pedigree Controller.

Regarding the tab of Inactive Wine, it allows winemakers to identify and list all the “invalid” wine records of the corresponding suspicious wine counterfeits being identified along the supply chain, under which the respective wine pedigree will be tagged and updated accordingly in the Authentication Controller. The same functionalities designed under the tab of Active Wine could also be performed for specific operations, such as searching wine records based on specific requirements, sorting to customize the sequence of wine records listed, and viewing details of the wine pedigree of invalid wine products, so that the better wine reviewing processes could be performed by the winemaker with suspicious wine products returned by that supply chain participant, at the node along the supply chain, identifying the issue. The invalid wine record is actually retrieved from the wine database based on the operating logic defined in the Unsuccessful Record Controller. Wine records will be deemed “invalid” only if the wine status of that wine product is changed from “valid” to “invalid”, in case the matching process against the data stored in specific NFC tag and the counterpart stored in the back-end database is failed.

Regarding the Project List, winemakers can create a project including a combination of different wine products produced by them or a certain quantity of a specific type of wine product, into it, alongside the details of which supply chain participants with their group such project belongs to. The design of the project list can allow winemakers to have a better planning on their resource involving different supply chain participants at different levels, instead of only being able to get access to wine records of the wine products they produced only, which is only in an aspect of managing wine records. The wine records are spanned into a project with supply chain participants involved so as to attain a well-rounded analytical management on those wine records and transaction records of their wine products. For instance, if winemakers want to create a project spanned by a group of supply chain participants, with different wine products of certain quantity, it will be much more convenient for them to manage all wine products in one go in NAS under which they can freely perform those functionalities, such as creating wine records, putting it to the tab of inactive wine record or sharing a specified wine product to targeted participants, simply by clicking the required buttons on the web application.

While for the tab of participant information management, winemakers can create the contacts of their trusted supply chain participants who should also be registering with the winemakers, so as to be shared with associated wine records while the wine products will be accepted by them along the supply chain. With the trusted supply chain participants registered with the winemakers as their contacts in their database system, it will then be possible to choose wine products produced by the winemakers and share it with their trusted and targeted supply chain participants. It follows that the targeted participants could retrieve the shared wine products once they logged into their account of the wine database web application again, after the wine product itself is physically transferred ownership from the winemaker to that supply chain participant. How winemakers share the wine product with their trusted participants is simply by clicking the sharing button and selecting the wine products they owned and to be shared, on the user interfaces such as “Create” or “Edit” under the tab of active wine record in the sitemap aforementioned in \textit{Appendix~\ref{b1}}. The sharing operation will be completed, and the list view of wine products will also be updated accordingly and automatically when they logged in to their accounts of the wine database web application again. All the details about the projects and the supply chain participants are actually stored in the wine database with data processing logic defined in the Supply Chain Controller. 

\subsection{The Design of NAS Wine Database Database System}
All the functionalities of the wine database web application explained above are depending the development and architecture of the back-end wine database itself. The detailed design of the wine database system is demonstrated and substantiated with use cases in both Entity and Attribute Diagram and Entities Relationship Diagram, as demonstrated in \textit{Appendix~\ref{b2}} and \textit{Appendix~\ref{b3}} respectively.

For the design of database architecture, there will be around five steps we need to take so as to ensure the wine database web application designed is problem-free and efficient enough, namely (1) identifying entities with its attributes, (2) defining the relationships, (3) assigning keys, (4) determining the corresponding data type on each column of data stored in the entities table, and (5) performing normalization taken for specific entities. The processes of defining database architecture require structuring and organizing content of the wine database web application, in order to enable winemakers and supply chain participants to better capture the data they require. The content architecture is vital to a well-designed and scaled database system for winemakers and supply chain participants who will have their own account accessing and maintaining the database system connected with the counterparts of other NAS users for sharing specific wine records of wine products with ownership being transferred along the supply chain. Capturing accurate system requirements of developing the NAS as well as its wine database web application would be key as different users would have varying needs and requirements, and the wine database web application developed will need to accommodate these needs and requirements accordingly. The database design for NAS is detailed and demonstrated in \textit{Appendix~\ref{b4}}.

\section{The Android Application Development of NAS}
The Cordova provides a set of uniform JavaScript libraries that could be referred during the development of both mobile applications, with device-specific native back-end code for those JavaScript libraries. Cordova is available for the following platforms: iOS, Android, Blackberry, Windows Phone, Palm WebOS, Bada, and Symbian. As such, the Cordova Native Library, Android Framework and the SDKs could therefore be adopted and called for developing both \emph{ScanWINE} and \emph{TagWINE} of NAS, which are both Android applications and running on Android 4.3 OS. Most importantly, the mobile applications built with the application programming interface (API) of PhoneGap could enable the NFC features by including the NFC Plug-ins so that both mobile applications could be NFC-enabled and utilized the NFC technology of the selected smartphones. The details of adopting the PhoneGap framework and how the PhoneGap plug-ins are interacting with the web applications and the Android Mobile applications are explained in \textit{Appendix~\ref{b5}}.

\subsection{The Introduction of NFC-Plugins and Its APIs}
Through including NFC-Plugins into the ADT with PhoneGap APIs with the source files placed under the "\emph{src}" as "\emph{NfcPlugin.java}", there is a wide range of functions, written in Java, available to be invoked depending on the use cases. Another API, named "\emph{phonegap-nfc}" will also be included in the "\emph{src}" in which it includes all necessary functions of formatting NDEF tags. The phonegap-nfc plugin actually allows functions of reading and writing NFC tags from a PhoneGap application, written in JavaScript, under which the NFC-plugins will be activated once it is imported in "config.xml" of both mobile applications.

\begin{figure}[h]
    \centering
    \captionsetup{justification=centering}
    \includegraphics[width=0.5\textwidth]{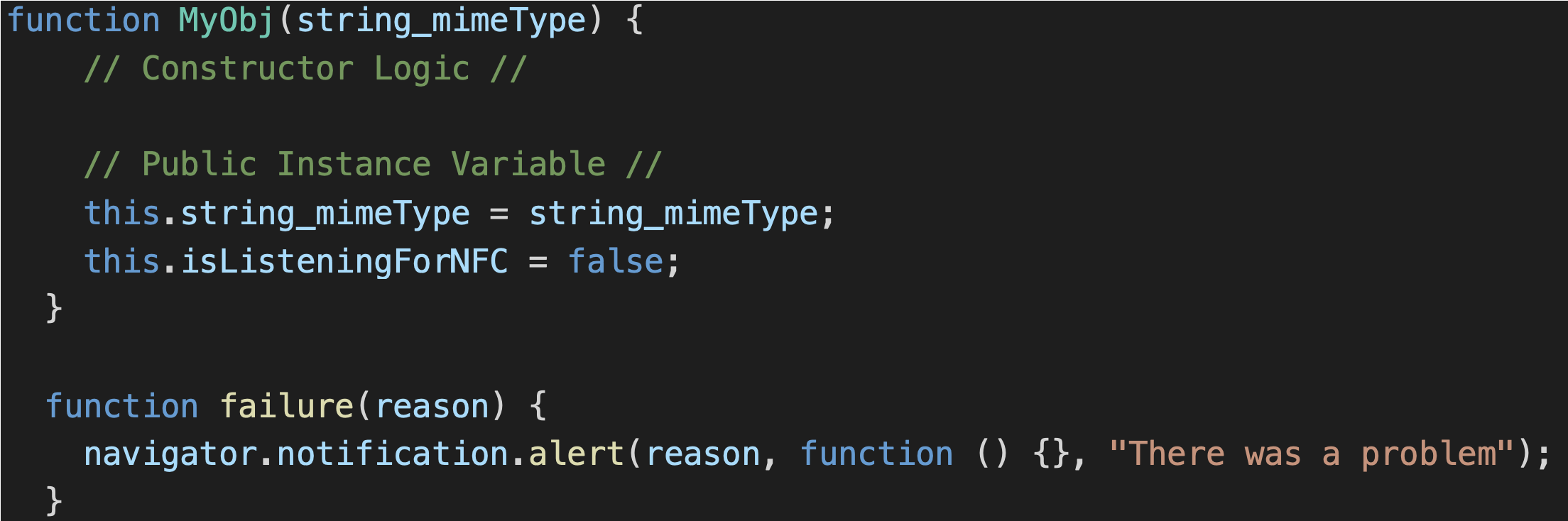}
    \caption{\textit{Functional Definition of "failure" in "phonegap-nfc-simplenfc"}}
    \label{fig:mime}
\end{figure}

\begin{figure}[h]
    \centering
    \captionsetup{justification=centering}
    \includegraphics[width=0.5\textwidth]{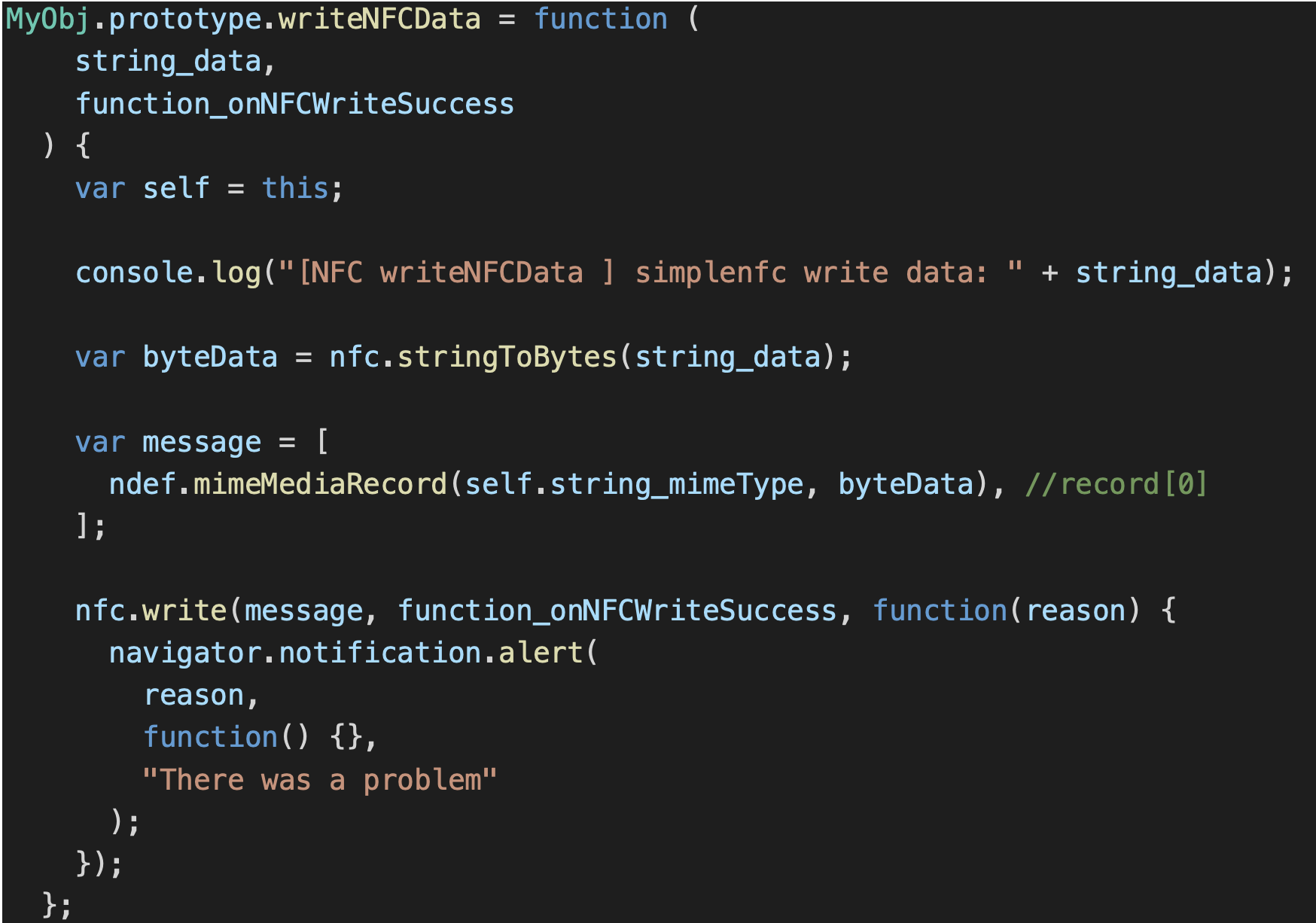}
    \caption{\textit{Object Definition of "writeNFCData" in "phonegap-nfc-simplenfc"}}
    \label{fig:writenfc}
\end{figure}

\begin{figure}[h]
    \centering
    \captionsetup{justification=centering}
    \includegraphics[width=0.5\textwidth]{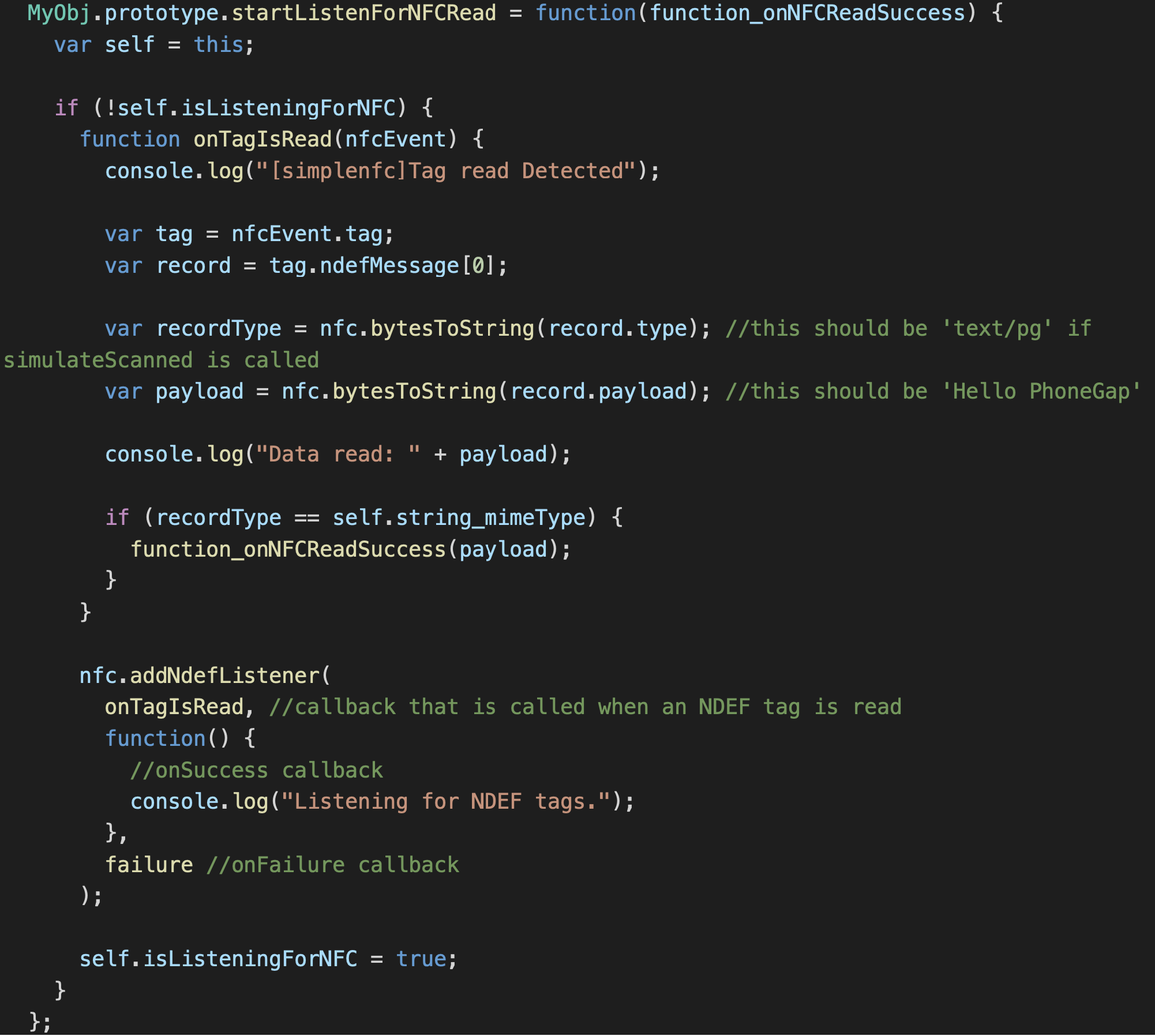}
    \caption{\textit{Object Definition of "startListenForNFCRead" in "phonegap-nfc-simplenfc"}}
    \label{fig:objstructure}
\end{figure}

The "\emph{phonegap-nfc-simplenfc}" service is developed based on "\emph{phonegap-nfc}", and there are only two functions and two objects constructed in this service. As demonstrated in \textit{Fig.~\ref{fig:mime}}, "\emph{MyObj(string_mimeType)}" is actually defining a public instance variable and specifies to be type of MIME, while the "\emph{failure(reason)}" is deployed for fear that there is a failure case during any tag-interacting process. While for the two objects, namely "\emph{writeNFCData}" and "\emph{startListenForNFCRead}", as demonstrated in \textit{Fig.~\ref{fig:writenfc}} and \textit{Fig.~\ref{fig:objstructure}} respectively, the former is actually for \emph{TagWINE} with which a specific string data (tag value) will be written into any NFC tag, while the latter is for \emph{ScanWINE} with which there should be string of \emph{record.type} and \emph{record.payload} (NDEF message/tag value) returned during any NFC-reading process. The conceptual notions on the details of NDEF message, payload, and the formatting structure of any NFC tag are all included and shown in \textit{Appendix~\ref{a4}}.

The reason why "\emph{phonegap-nfc-simplenfc}" is developed based on the "\emph{phonegap-nfc}" is owing to the fact it could provide numerous formatting functions (events) on interacting NFC tags. The objects developed in "\emph{phonegap-nfc-simplenfc}" are defined to invoke the related low-level functions defined in "\emph{phonegap-nfc}", namely "\emph{nfc.NdefListener}" and "\emph{nfc.write}". Regarding the "\emph{nfc.NdefListener}" as invoked in \textit{Fig.~\ref{fig:objstructure}}, it is indeed a callback function for any NDEF event, under which the logic patterns embedded in this callback function could be invoked only if an NDEF event is triggered which could be a NFC tag is scanned. For the "\emph{nfc.write}" as invoked in \textit{Fig.~\ref{fig:writenfc}}, it is actually a function that writes an NDEF message (tag value), which is an array of one or more NDEF records, as explained in \textit{Appendix~\ref{a4}}, to a NFC tag.

Furthermore, given the fact that callback function in "\emph{nfc.NdefListener}" would only be invoked if there is a NDEF event to trigger such invocation, an event handler is needed to be in place for such event-driven design pattern. For instance, the "\emph{onTagIsRead(nfcEvent)}" in the object definition of "\emph{startListenForNFCRead}", as demonstrated in \textit{Fig.~\ref{fig:objstructure}}, is actually the event handler for processing the required NDEF events which are fired when any NFC tag is being read provided that such events are "\emph{passive}" for which another function is required for waking up the "\emph{events}". However, sometimes the event handler is not required due to the fact the event itself is "\emph{active}", such as the design patterns adopted in "\emph{nfc.write}" as described in \textit{Fig.~\ref{fig:writenfc}}, where an event handler is not needed and the functional object is about performing NFC-writing processes to write any "\emph{byteData}" into "\emph{messages}". In short, the event-handling function is to handle "\emph{passive}" NFC-related events. With the advent of NFC-plugins and its associated APIs, objects and NDEF events defined in "\emph{phonegap-nfc-simplenfc}", both \emph{ScanWINE} and \emph{TagWINE} could therefore perform tag-reading and tag-writing functionalities with the NFC-enabled smartphones, via invoking related functional objects in the "\emph{mainLogic.js}" of both mobile applications. 

\subsection{The Communication between Server API and Application JavaScript APIs}
It is normal that there are questions on how the required tag value stored in any NFC tag can be returned to the user interface of both mobile applications, and how the communication between the database server and both mobile applications could be performed. Indeed, such communication between the server and both mobile applications is actually enabled by two APIs in which one from \emph{AppController} generating the server API and another from the “\emph{wineappclient}” generating Application JavaScript API.

There are nine functions and an object designed and included in the class of the server-end "\emph{AppController}". It contains logical patterns of showing a specific array of wine records with some functions such as "\emph{getAllWine}" or "\emph{getWine}" so that requests sent mainly from the \emph{TagWINE} could be processed with required wine records retrieved from the database based on the object-relational mapping. There are also functions to generate random hashed tag values, such as "\emph{genTagHash}" and "\emph{createHash}", defined in AppController so as to preserve the privacy and security on the wine data stored in the NFC tags, and NAS as a whole.

\begin{figure}[h]
    \centering
    \captionsetup{justification=centering}
    \includegraphics[width=0.5\textwidth]{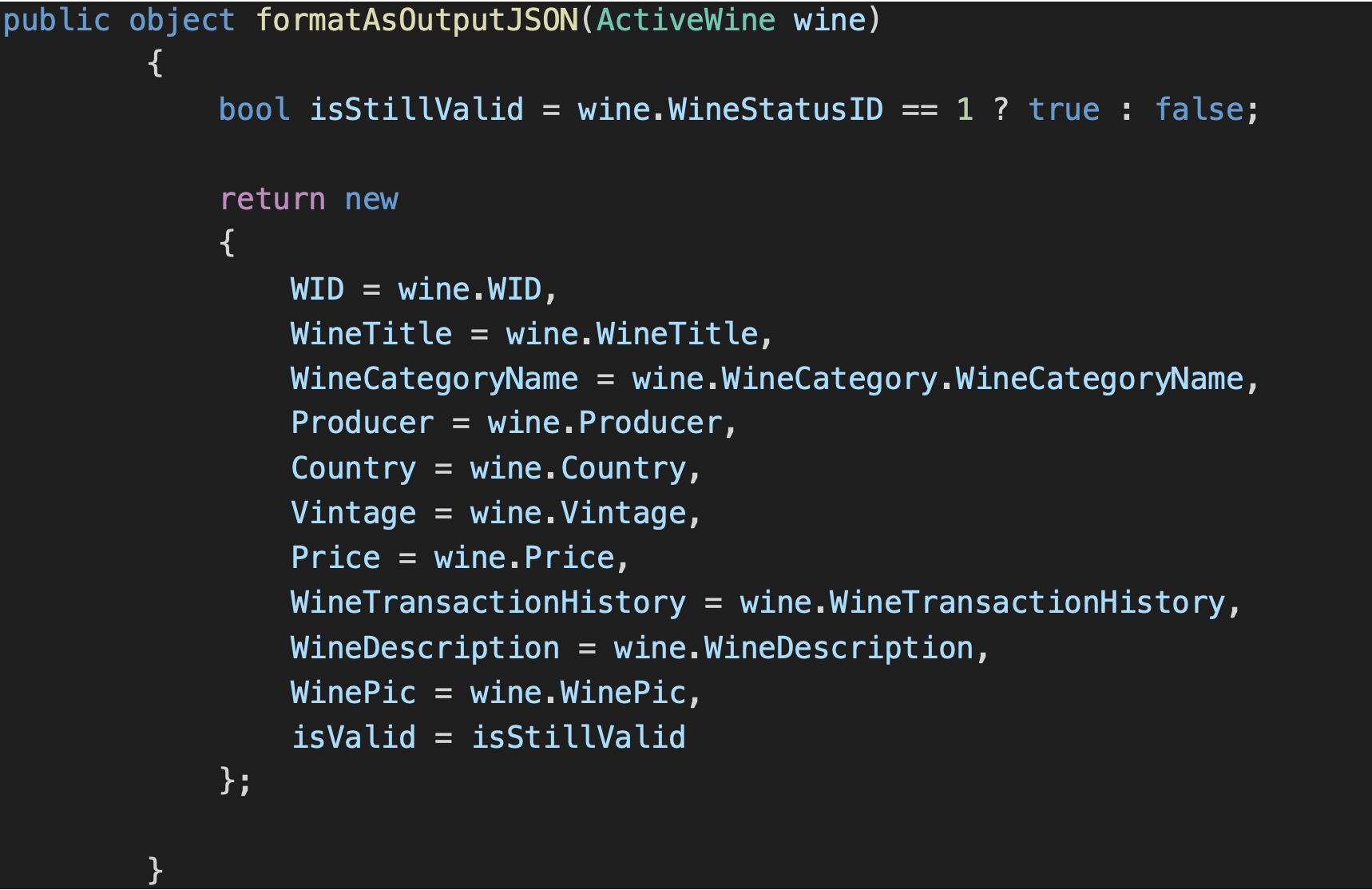}
    \caption{\textit{Object Definition of "formatAsOutputJSON" in AppController}}
    \label{fig:formatjson}
\end{figure}

Some required wine record data processed in both mobile applications could be formatted as JSON outputs in AppController so that these JSON records can be invoked by those corresponding objects of the Application JavaScript API and being returned on the user interface of both \emph{ScanWINE} and \emph{TagWINE}. Through applying method of "formatAsOutputJSON" as demonstrated in \textit{Fig.~\ref{fig:formatjson}}, an array of metadata of the required wine record could be returned in JSON format, in case the wine product itself is still with a wine status of “valid”. The returned data object will maintain no references on the order of metadata returned in the data objects, and once such data object returned to Application Javascript API (\emph{wineappclient}), it will then be iterated over its metadata elements in the collection, storing each object reference to the next consecutive element of the data object, starting from the first metadata element.

\begin{figure}[h]
    \centering
    \captionsetup{justification=centering}
    \includegraphics[width=0.5\textwidth]{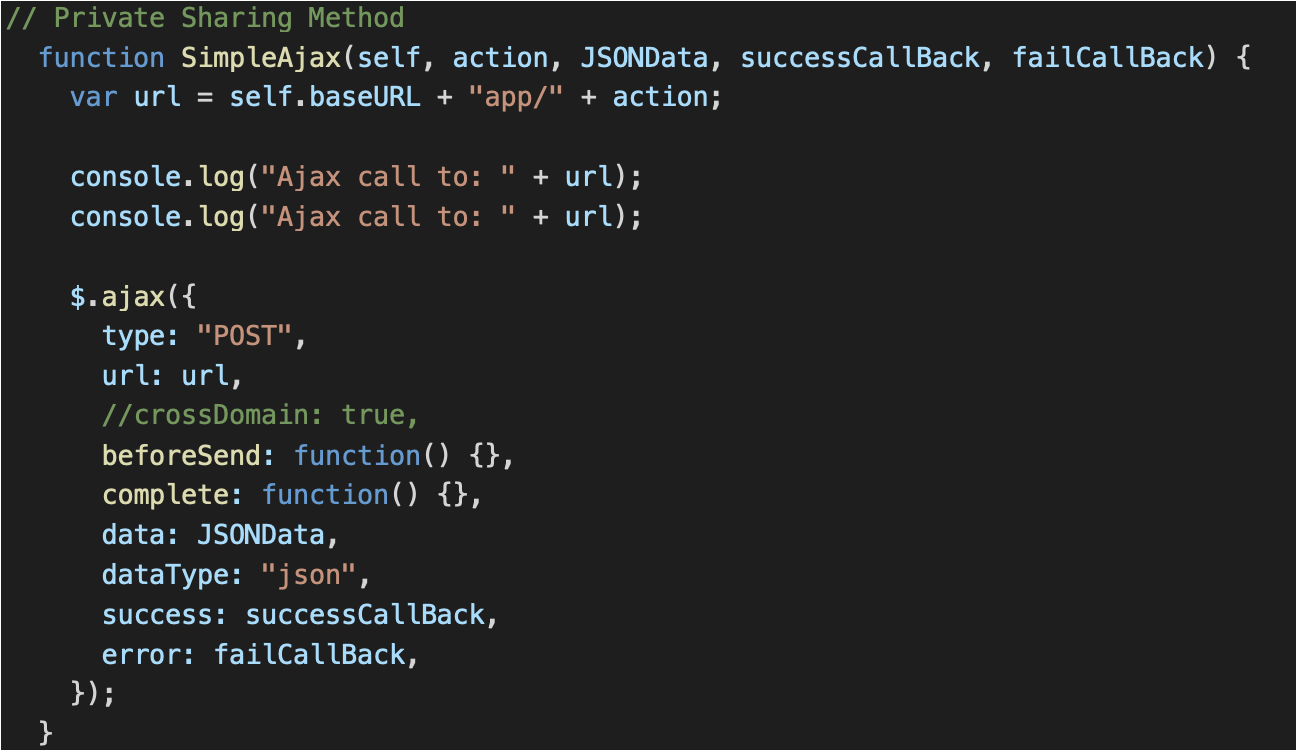}
    \caption{\textit{Functional Definition of "SimpleAjax" in wineappclient}}
    \label{fig:simpleAjax}
\end{figure}

Application JavaScript API is indeed referred as "\emph{wineappclient}", and a public instance variable was defined, in \emph{wineappclient}, so that the URL of the wine database web application could therefore be referred, as \emph{baseURL}, for a series of subsequent callback activities processed in \emph{wineappclient}. In addition, there is also a private method named "\emph{SimpleAjax()}", as demonstrated in \textit{Fig.~\ref{fig:simpleAjax}}, to define a standard Ajax callback method for any callback activities, processed in "\emph{wineappclient}" with payload of the related wine record, URL and preferred data type specified. The function is indeed a private sharing method under which it can only be invoked within its respective class boundary.

Once the private sharing method is set and applied to every “public” sharing object, which directly called back wine data from the respective methods in AppController (Server API), the callbacks from those public sharing objects will request callback of the required wine data via the Ajax call method specified in "\emph{SimpleAjax}". For instance, all the required wine data will be called back with the data type of JSON, and all the subsequent events handled depending if the callback will be succeeded or not, are all specified in the method, namely "\emph{successCallBack}" and "\emph{failCallBack}" respectively.

\begin{figure}[h]
    \centering
    \captionsetup{justification=centering}
    \includegraphics[width=0.5\textwidth]{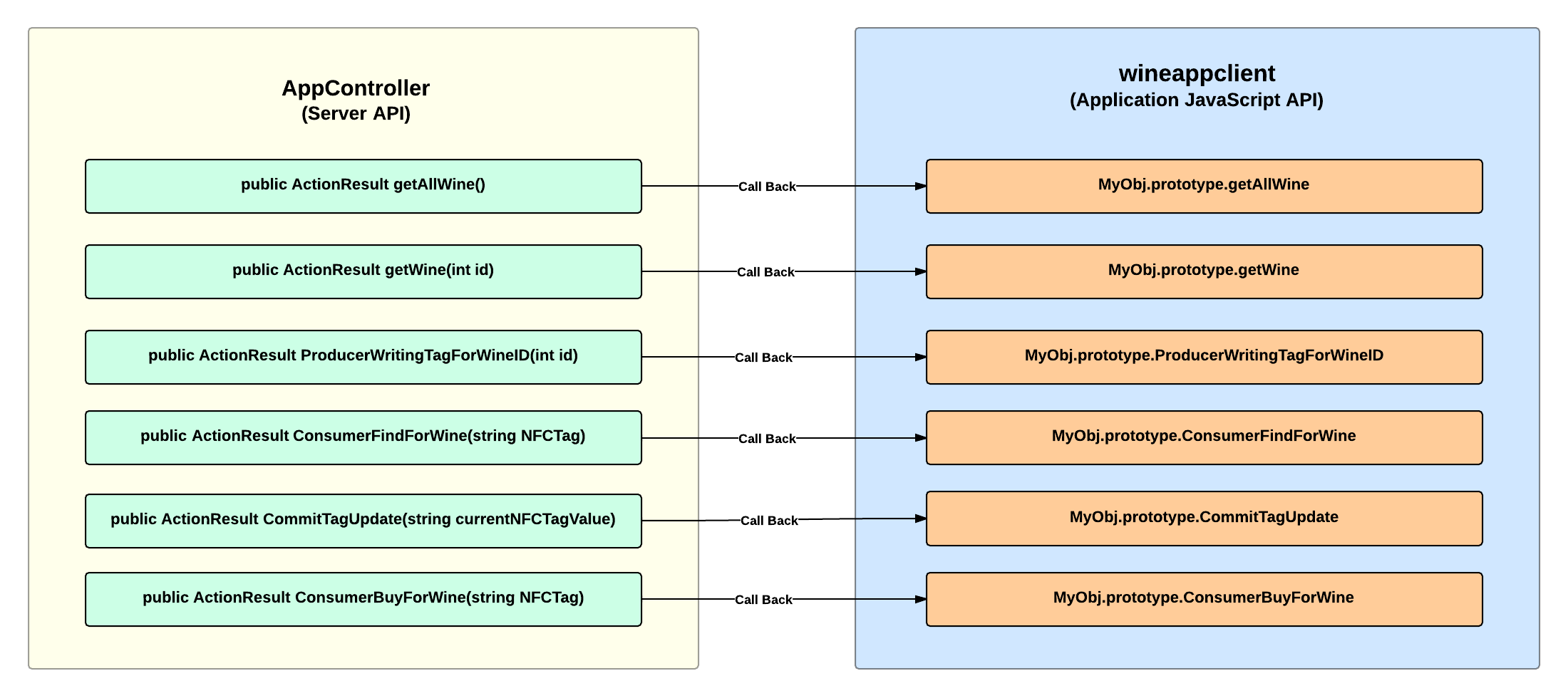}
    \caption{\textit{Callback Relationship Between AppController and wineappclient}}
    \label{fig:callbackrelation}
\end{figure}

Every object included in the \emph{wineappclient} (Application JavaScript API) is indeed set as “public” under which it can be invoked beyond the local class boundary, and therefore those could be invoked by the corresponding methods defined in the "\emph{mainLogic.js}" resulting in the defined anti-counterfeiting functionalities of the mobile applications could thus be demonstrated. There are actually six objects in the \emph{wineappclient}, which are deployed with callback events so that the required wine data, generated from those corresponding methods in the \emph{AppController} (Server API) with data objects returned in JSON format. The corresponding callback relationship between \emph{wineappclient} and \emph{AppController} is depicted in \textit{Fig.~\ref{fig:callbackrelation}}.

Generally speaking, the reason why an object of \emph{wineappclient} could call back wine data from the corresponding methods defined in "\emph{AppController}" is actually based on the synchronization of variables and actions set defined in \emph{wineappclient}, which in turns must be referred to same function at the server end (\emph{AppController}). For instance, the object of "\emph{ConsumerBuyForWine}" defined in \emph{wineappclient} with the action defined, would invoke its counterpart method - "\emph{ConsumerBuyForWine}" defined in \emph{AppController} at the server end. The callback could therefore be reached, and the required wine data, such as "\emph{isBuySuccess}", "\emph{reason}" and "\emph{wine}", could then be returned in JSON as specified in "\emph{formatAsOutputJSON}", and be passed in as payload during the invocation on those public objects defined in "\emph{mainLogic.js}" subsequently.

There is another example, such as the object of "\emph{getWine}" defined in \emph{wineappclient} with an action defined with “\emph{getWine}", which is exactly referring to the corresponding method of "\emph{getWine()}" included in "\emph{AppController}" at the server end. There is also a situation that "\emph{getWine()}" might return nothing back to the corresponding objects of "\emph{AppController}" in case the wine record is actually equal to "\emph{null}". The callback would be triggered and the required wine record will therefore be returned back to \emph{wineappclient} for further external invocations with methods defined in "\emph{mainLogic.js}" of both \emph{ScanWINE} and \emph{TagWINE}.

\subsection{The Tag-Writing Android Application – TagWINE}
\emph{TagWINE} is a component of NAS, which enables winemakers to write or update the data of wine records into NFC tags before the bottling operation of wine production processes at the winery. Apart from the functions of writing and editing specific data of wine records stored in NFC tags, winemakers could also view the details of wine products produced in the winery while using \emph{TagWINE}, and check the previous state-transitioned activities on the wine records, such as the writing and editing history of specific wine records.

The approach of developing both the \emph{TagWINE} and \emph{ScanWINE} is actually Object-oriented Programming (OOP), in which objects (data and code) are building blocks on different layers constituting different APIs of the mobile applications. The object itself is actually a modular and reusable element which could be invoked within the class specified, from methods of other APIs of the same application project, to even the "\emph{mainLogic.js}" of the application. Through developing the applications using the approach of OOP, the structure of different APIs built in the application project and even the code base could be more easily understood, maintained, debugged and even be further improved in the future in case there is any new functionality required to be added to the mobile applications of NAS.

\begin{figure}[h]
    \centering
    \captionsetup{justification=centering}
    \includegraphics[width=0.5\textwidth]{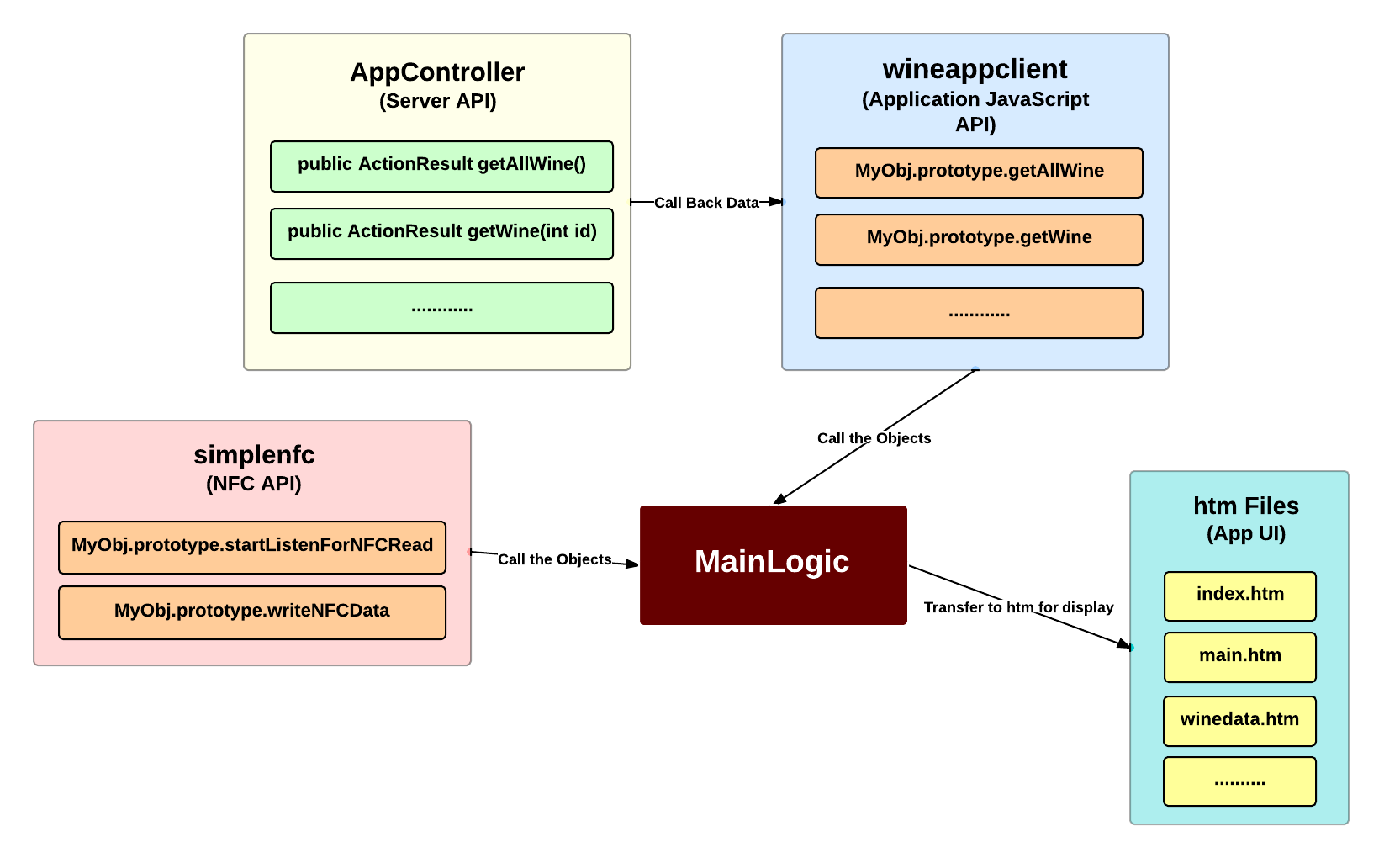}
    \caption{\textit{The Invocation Relationship of mainLogic.js with Different APIs}}
    \label{fig:mainlogic1}
\end{figure}

Based on the concept of OOP, both mobile applications are actually developed mainly based on the "\emph{mainLogic.js}" invoking different objects of different APIs. For instance, there are actually two core APIs, which are vital on whether both mobile applications could invoke required wine data from the Server APIs and adopt NFC-related methods from the NFC API (\emph{simplenfc}). As such, objects in "\emph{wineappclient}" were defined so as to enable the "\emph{callback}" relationship between those methods in the "\emph{AppController}" (Server End) and the "\emph{wineappclient}" (Application End) as specified in \textit{Fig.~\ref{fig:mainlogic1}}, while those objects in the "\emph{simplenfc}" (NFC API) were developed so that the events in the original NFC API – "\emph{phonegap-nfc}" could be included and enabling the NFC-reading and NFC-writing functions of the mobile applications running on NFC-enabled smartphone.

\begin{figure}[h]
    \centering
    \captionsetup{justification=centering}
    \includegraphics[width=0.4\textwidth]{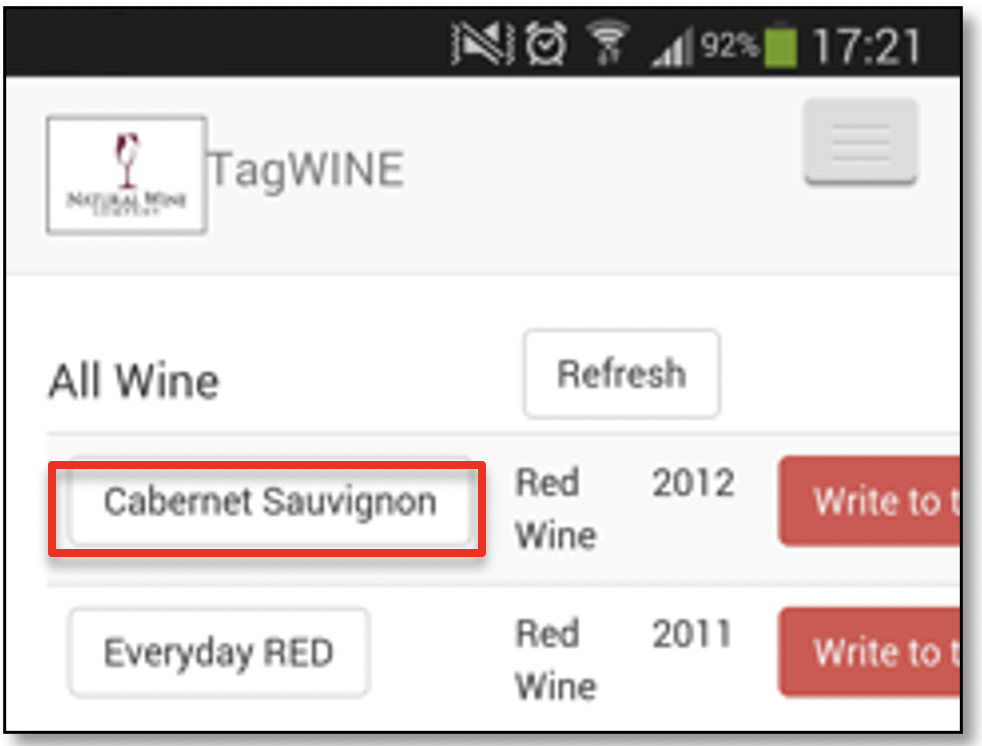}
    \caption{\textit{List View of Wine Records in TagWINE}}
    \label{fig:tagwineapp}
\end{figure}

Those related objects in the aforementioned APIs would be invoked by methods in "\emph{mainLogic.js}", since they are all “public” in nature. Whenever there is a set of specific methods required to be performed by the mobile applications, all required data retrieved by the objects will go through the operating logic in "\emph{mainLogic.js}" which will then forward to the user interface of TagWINE for displaying the wine records in listed view, as demonstrated in \textit{Fig.~\ref{fig:tagwineapp}}.

\begin{figure}[h]
    \centering
    \captionsetup{justification=centering}
    \includegraphics[width=0.5\textwidth]{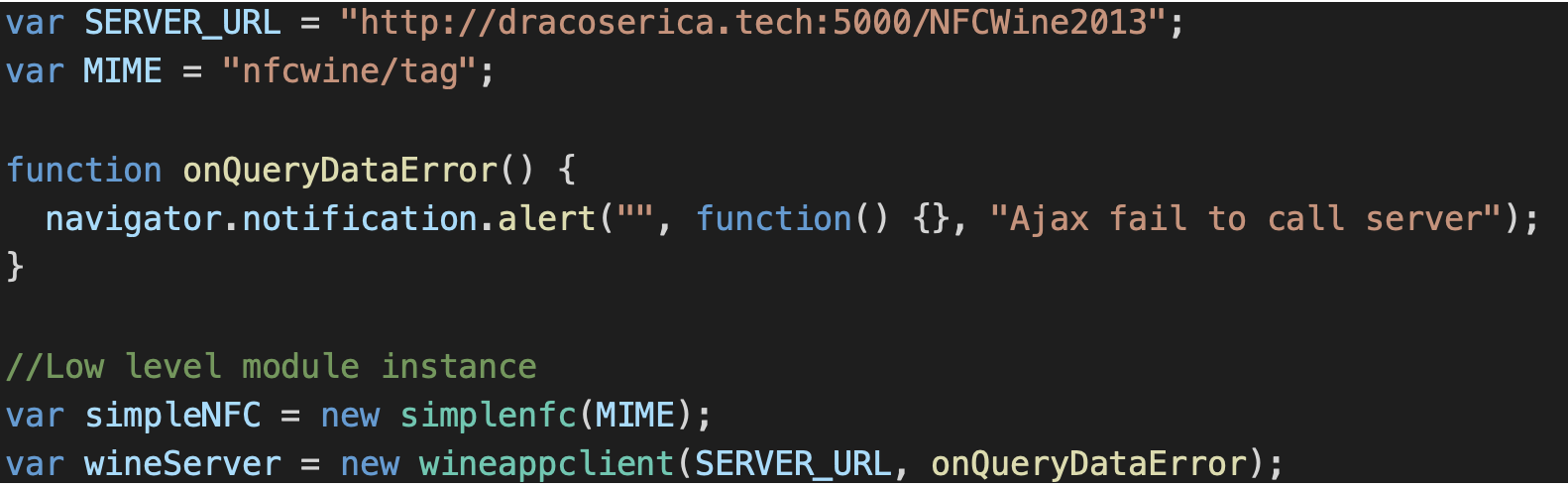}
    \caption{\textit{Definition of Environment Variables in mainLogic.js}}
    \label{fig:envvars}
\end{figure}

\begin{figure}[h]
    \centering
    \captionsetup{justification=centering}
    \includegraphics[width=0.5\textwidth]{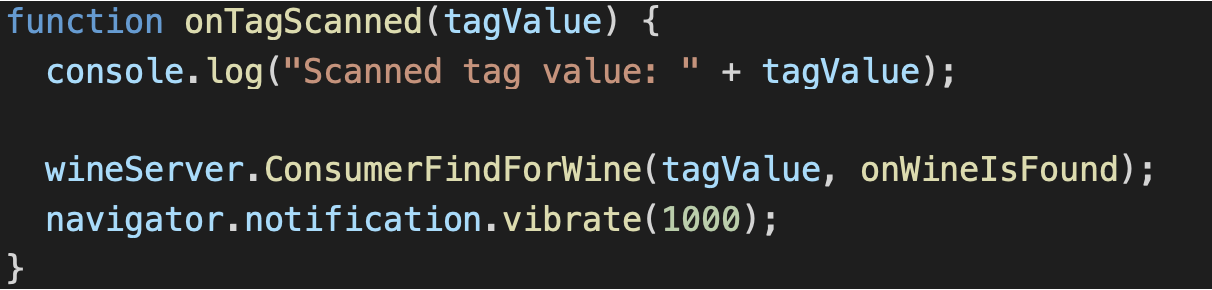}
    \caption{\textit{Sample Event Listener of NFC Tag Scanning Events}}
    \label{fig:ontag}
\end{figure}

Regarding "\emph{mainLogic.js}", environment variables, such as "\emph{SERVER_URL}", are defined at the beginning of the JavaScript file, via direct definition or injection from environment setup, as demonstrated in \textit{Fig.~\ref{fig:envvars}}, so that the application can connect to the server end whenever its objects, invoked by respective methods in "\emph{mainLogic.js}", requesting queries for wine data sent to the back-end server based on the URL registered. The APIs, such as the "\emph{wineappclient}" and "\emph{simpleNFC}", are also defined and instantiated, as demonstrated in \textit{Fig.~\ref{fig:envvars}}, so that its constituent objects - "\emph{wineServer}" for this case could be invoked by the respective methods, such as the event listener - "\emph{onTagScanned}" as demonstrated in \textit{Fig.~\ref{fig:ontag}}, of "\emph{mainLogic.js}".

For the event listener "\emph{onTagScanned}" of "\emph{mainLogic.js}", as demonstrated in \textit{Fig.~\ref{fig:ontag}}, the method, named "\emph{ConsumerFindForWine(tagValue, onWineIsFound)}", is being invoked by the corresponding instance of \emph{wineappclient} API which is defined as "\emph{wineServer}", so that a series of callbacks on requesting wine data, such as "\emph{nextNFCTag}" and "\emph{isInCommit}" could therefore be invoked with response payloads returned to the server end.

\begin{figure}[h]
    \centering
    \captionsetup{justification=centering}
    \includegraphics[width=0.5\textwidth]{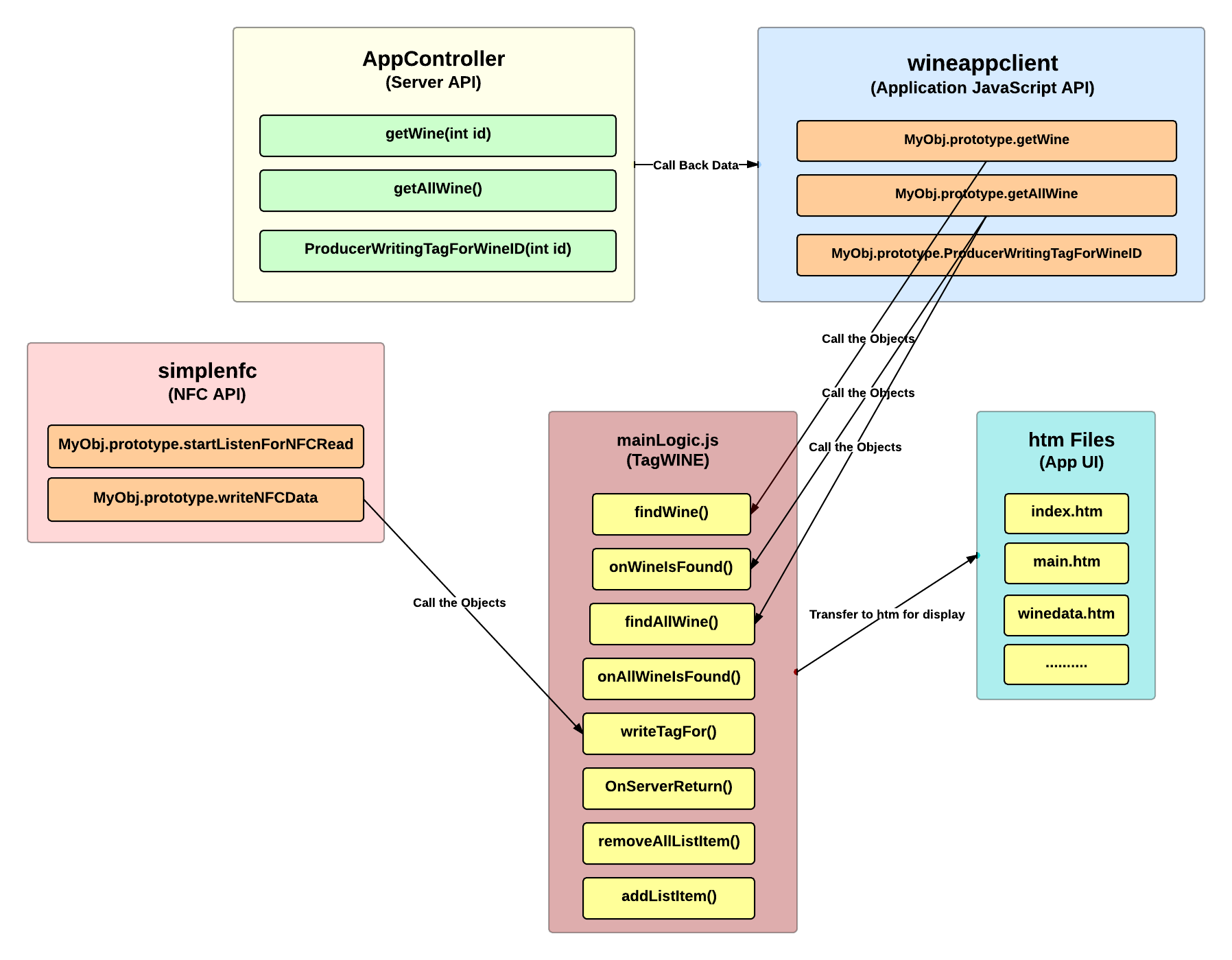}
    \caption{\textit{Module and Functional Relationship of TagWINE}}
    \label{fig:modulerelationtagwine}
\end{figure}

The module and functional relationships of \emph{TagWINE} with individual methods listed is actually demonstrated in \textit{Fig.~\ref{fig:modulerelationtagwine}}. Regarding the "\emph{mainLogic.js}" of \emph{TagWINE}, there are in total eight local methods included so that related mechanisms, such as how the required wine data could be located for displaying at user interfaces before the tag-writing process, how to write tag values into a NFC tag, and even how to list all the wine records stored in the wine database available for any tag-writing process, could all be achieved by invoking public objects of other APIs. For instance, the method "\emph{writeTagFor()}" could invoke the object "\emph{writeNFCdata}" in order to perform the tag-writing process using NFC technology by instructing the smartphone to perform NFC scanning based on events triggered by \emph{TagWINE}.

\subsection{The Tag-Reading Android Application – ScanWINE}
\emph{ScanWINE} is also an essential component of NAS, which enables both supply chain participants and wine consumers to authenticate any wine product circulating along the supply chain, during a wine transfer process or before purchasing the wine product at retail points. \emph{ScanWINE} allows its users to scan NFC tags attached on any wine bottleneck with NFC technology using NFC-enabled smartphones. Apart from the functionality of scanning specific NFC tags, \emph{ScanWINE} also allows its user to perform functions, such as accepting wine products from winemakers or even the previous node along the supply chain during the wine transferring process performed by supply chain participants, or buying a wine product while the authenticity of it could be confirmed. Features, such as sharing wine products to counterparts along the supply chain, manually checking the NFC tag ID against the database of winemakers, and checking any historical activity history, could all be available to users once the registration process of NAS is completed. 

Regarding \emph{ScanWINE}, the mobile application itself is also developed based on the OOP concept, under which all the elementary objects enabling the callback mechanism at the server end to trigger the NFC tag-writing and tag-reading features, are all developed and included in mainly two APIs, namely the "\emph{wineappclient}" and "\emph{phonegap-nfc-simplenfc}". The objects and methods of these APIs are invoked by local methods consisting the "\emph{mainLogic.js}" of \emph{ScanWINE}, whenever there is an event needed to be triggered relating to the object, such as the NFC-reading process.

\begin{figure}[h]
    \centering
    \captionsetup{justification=centering}
    \includegraphics[width=0.5\textwidth]{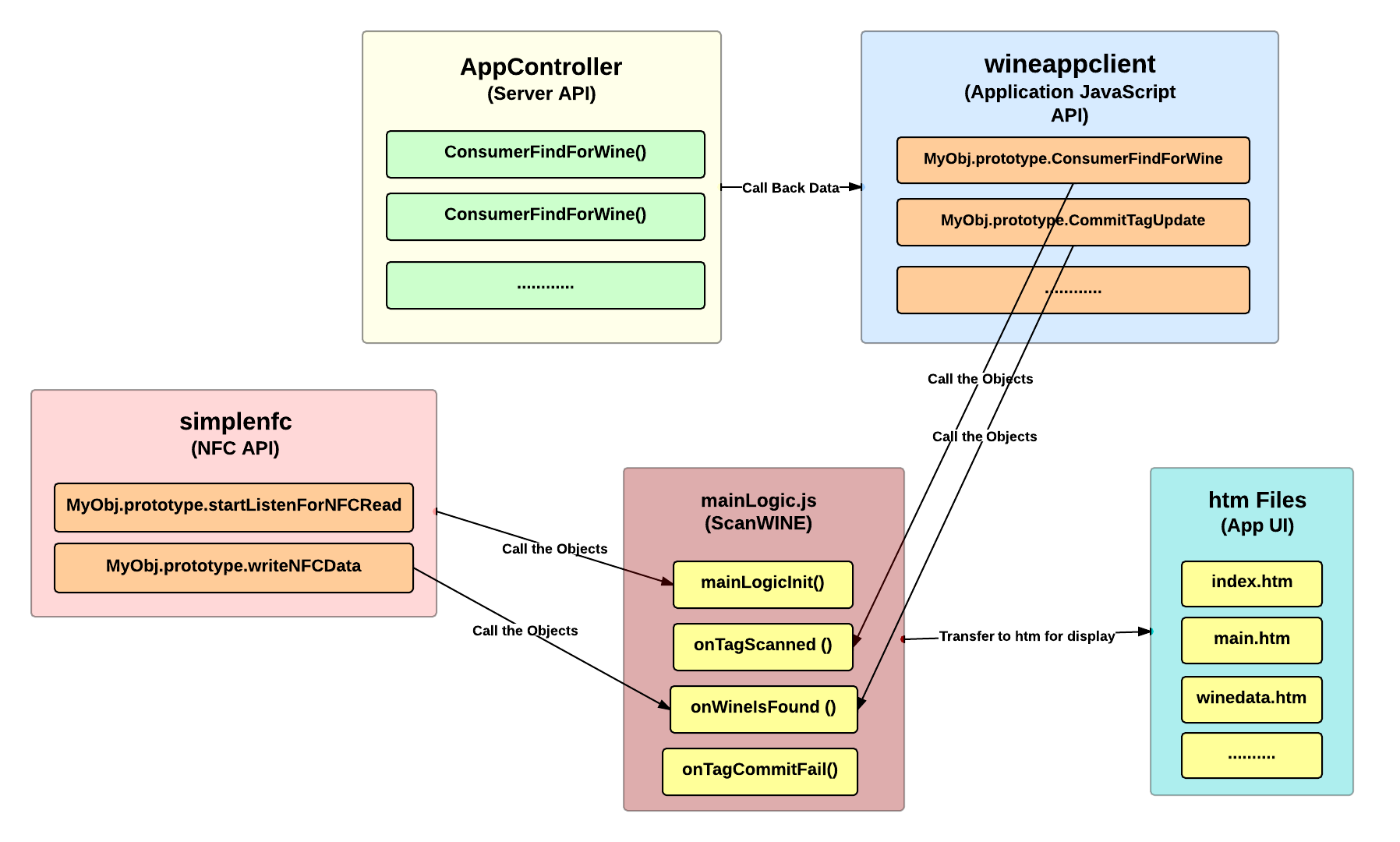}
    \caption{\textit{Module and Functional Relationship of ScanWINE}}
    \label{fig:modulerelationscanwine}
\end{figure}

According to the module and functional relationship diagram of \emph{ScanWINE} as depicted in \textit{Fig.~\ref{fig:modulerelationscanwine}} with individual methods listed, there are 4 local methods included in "\emph{mainLogic.js}". For instance, the method "\emph{mainLogicInit()}" would invoke the objects, such as "\emph{ConsumerFindForWine}", "\emph{startListenForNFCRead}" and "\emph{writeNFCData}", of "\emph{wineappclient}" so that the specific wine record could be located, and the tag-reading process as well as the tag-writing process could be performed in \emph{ScanWINE}.

\begin{figure}[h]
    \centering
    \captionsetup{justification=centering}
    \includegraphics[width=0.38\textwidth]{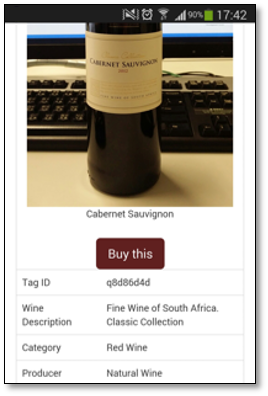}
    \caption{\textit{The Wine Product View on ScanWINE}}
    \label{fig:scanwineview}
\end{figure}

The reason why there should be an object "\emph{writeNFCData}", despite the fact that \emph{ScanWINE} is mainly for tag-reading process, is that the“Read Count” methodology is introduced as a step of security improvement, under which the tag value is needed to be refreshed with the specific NFC tag every time a tag-reading process is completed. All the wine data required will then be processed and passed to those HTML files based on design patterns defined in "\emph{mainLogic.js}", which will then return the detailed product view displayed on user interfaces of \emph{ScanWINE} running on the NFC-enabled smartphone as demonstrated in \textit{Fig.~\ref{fig:scanwineview}}.

\section{The Anti-Counterfeiting Mechanism of NAS}
It is too early and naive to say that NAS is the most secured anti-counterfeiting system ever adopted in wine industry to relieve the growing market of wine counterfeit given that there will always be a more innovative alternative, or rather, what NAS could transfer is the notion that it can add one more layer of security to those wine products in the market based on those anti-counterfeiting technologies already adopted on wine bottles. Nonetheless, what anti-counterfeiting and security values could NAS provide in today’s wine industry? The proposed NAS performs wine anti-counterfeiting by maintaining supply chain integrity, not manually but automatically, and the significance is twofold. First, the transaction path of a wine product is clear and its origin can be traced accordingly with the data stored in a series of servers with transaction path tracked and updated along the supply chain. Second, wine authenticity can be validated off-hand with one’s own NFC-enabled smartphone with mobile applications available to the specific winemakers. However, we all have to bear in mind with a fact that NAS is always possible to be attacked, and we should try our utmost to make it with least chance to be hacked and manipulated.

The back-end wine database owned by winemakers is the heart of the proposed NAS, connoting that only winemakers would be allowed to create, edit and delete any wine record, information about the trusted supply chain participants and the details about projects involving various combinations of wine products. There will be no right for any other players of different industry role to modify any column of a wine record stored in the system through HTTP/HTTPs, such as using their own wine supply chain database to connect with the one owned by specific winemakers of specific wine products, when using the applications. Nevertheless, there is only one situation that a wine record stored in the winemaker’s wine database, could be accessed and even be updated; for instance, the transaction record of a specific wine product could be updated when it was accepted by the next node along supply chain from that winemakers, such as the distributor of wine products, or being sold to wine consumers once the payment platform, integrated with NAS, with payment transaction is confirmed. The related servers of NAS will then update the wine status with text such as "\emph{wine being accepted by}" or "\emph{wine sold}" accordingly and automatically, with payloads in JSON via HTTP/HTTPs, without any human intervention during the whole updating process of the wine product for fear that some of the supply chain participants may manipulate the states of any wine record causing even more wine counterfeits in wine industry.

The product detail and full pedigree of wine products with the respective transaction records gathered along the supply chain, which are stored in the host server of winemakers, could only be accessed while the NFC-enabled smartphones and the controllers of host servers of winemakers, such as Wine Pedigree Controller and Authentication Controller, are being connected via the internet with payloads in JSON data format. Machine-to-machine communication between \emph{ScanWINE} running on NFC-enabled smartphones and Application Controllers of the back-end server, will be performed without any human intervention and errors, so as to prevent any manual modification performed by malicious supply chain participants or wine consumers from causing any system failure. Although one of the integrated features of NAS is that wine records could be shared to the next node of participant along the supply chain once the wine product and its authenticity are being confirmed, the data shared to other participants’ databases could only be viewed, without any access to any manual modification to make state-transitioned changes on any wine record. With the growing popularity of Internet connection with payloads in HTML and JSON, along with the emergence of more NFC-enabled smartphone or device in the market, NAS could be more mature to allow wine consumers to check wine pedigrees of specific wine products they would like to purchase, with the checkout channel integrated with electronic payment system and behavioral suggestions system integrated with NAS as well.

Regarding wine consumers, making sure that wine products are genuine could guarantee value for money and product safety. There are indeed sufficient incentives for wine participants and consumers to verify the authenticity of wine products with their own NFC-enabled smartphones, which is already a more convenient way to perform steps in order to claim authenticity of wine products. Similarly, since a wine product without a plausible history may not be saleable to the next carriers or wine consumers at retail points, the current carriers bear the responsibility of returning the suspicious wine products spotted with problems flagged while reading it with \emph{ScanWINE} and being processed with NAS. In anticipation of enhanced customer confidence, it would be justifiable for winemakers to host NAS. The winemakers, who are also the server-hosting entity of NAS along its product line with its own relations of supply chain participants and consumers, are responsible for tracking and recording suspicious transactions and tracing through the sources of any security breaches in various supply chains as wine products move along the supply chain of the wider wine industry.

Specifically, how do the components, such as both mobile applications named \emph{ScanWINE} and \emph{TagWINE} respectively, the NFC Tags and the back-end wine databases owned by winemakers, of NAS transfer the anti-counterfeiting value in detail? For instance, how are they set up in a way so that the counterfeiters could hardly attack or manipulate NAS and its components? The mechanisms which could hinder counterfeiters from manipulating NAS or cloning the NFC tags are based on the following four reasons: (1) the secured setup on NFC tag, (2) the setup of an automatic updating system on both applications and database systems, (3) the security features on NFC tags and its deployments, and (4) the procedural setup of NAS.

\subsection{Secured Setup on NFC Tags}
\begin{figure}[h]
    \centering
    \captionsetup{justification=centering}
    \includegraphics[width=0.5\textwidth]{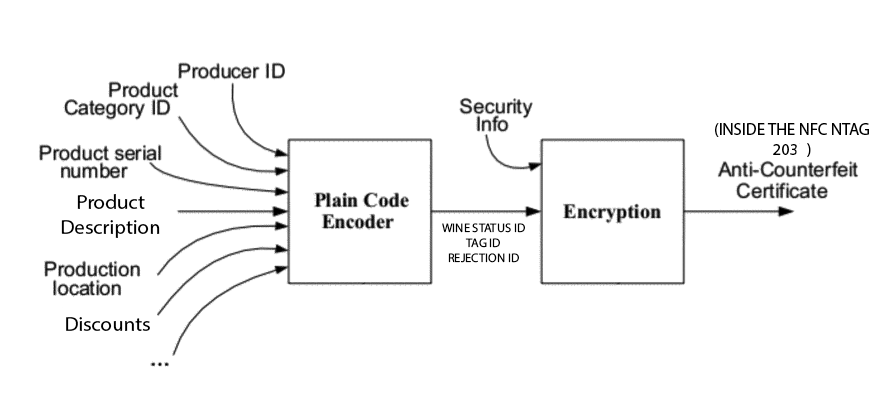}
    \caption{\textit{The Offline Data Tag Processing of Typical Anti-Counterfeiting System}}
    \label{fig:offlinedata}
\end{figure}

Regarding the detail of what the NFC tags actually store under NAS, only the data of tag value \emph{(WID+readCount)} of a wine product will be encrypted and stored in the NFC tag. In case the NFC tag is \emph{Ultralight C}, which is 3DES Encryption-protected, the data of tag value is stored in the NFC tag after being formatted with NDEF messages, instead of encoding, encrypting and storing every single column of wine record in the tag. Major industry players, who are using RFID technology for anti-counterfeiting purpose, are actually processing, encrypting and writing every detail of a product record into specific tags, in which it is being termed as \emph{offline data tag}, as depicted in \textit{Fig.~\ref{fig:offlinedata}}, that is highly susceptible to attacks or malicious manipulations such as spoofing, replay attacks, repudiation, etc.

\begin{figure}[h]
    \centering
    \captionsetup{justification=centering}
    \includegraphics[width=0.5\textwidth]{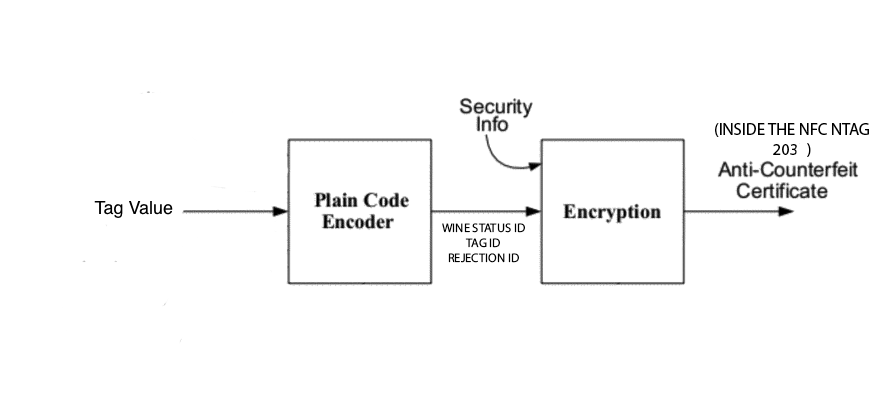}
    \caption{\textit{The Online Data Tag Processing of NAS}}
    \label{fig:onlinedata}
\end{figure}

The use of \emph{offline data tag} can easily lead to system failure or even collapse, making it even more susceptible to wine counterfeiting. However, for the situation of NAS, the tag value is the only data stored in the NFC tag acting as the only key to link up those applications and the corresponding application controllers of the back-end wine databases for communication. With only the tag value stored in NFC tags, the tag must be an \emph{online data tag} in nature, which is more secured for sure, with which updating a column of data of a specific wine record at the database’s end would in turn be demonstrated with an updated record on the user interfaces of both NFC-enabled mobile applications running on an NFC-enabled smartphone, whenever there is an NFC reading process triggered in \emph{ScanWINE}. Instead of putting every detail of a wine record into a RFID/NFC tag, which could be vulnerable to attacks and not convenient as updating processes could only be completed if the suspicious wine products are sent back to winemakers for further updating processes, the methodology of storing, processing and encrypting only tag value in the online data tag is appeared to be more secured, light-weighted and scalable as depicted in \textit{Fig.~\ref{fig:onlinedata}}. For instance, if the data of "\emph{winetitle}" defined as \emph{ABC} in database, and someone performed a modification on the wine record, the updated "\emph{winetitle}" would be shown on the \emph{ScanWINE} next time when the reading process on the NFC tag attached is triggered. It turns out that the tag value stored in the NFC tag is the only data the NAS needed to protect from malicious manipulations.

Undeniably, it is also unsafe if we only put the "\emph{WID}" and "\emph{readCount}" in the NFC tag and make them visible to the public, though the tag value will be hashed later on at the improvement stage, as it will in all likelihood be more susceptible to the attacks. So as to strengthen the security on the NFC tags and NAS as a whole, the data type of WID has been altered from "\emph{integer}" to "\emph{uniqueidentifier}" and it is believed that the "\emph{uniqueidentifier}" data type is far more secured than with purely the data type of "\emph{integer}", in which the \emph{uniqueidentifier} data type stores 16-byte binary values operating as globally unique identifiers (GUIDs). A GUID is a unique binary number, and no other computer in the world will generate a duplicate of that GUID value. The main use of a GUID is for assigning an identifier that must be unique in the whole computer network. Due to the fact that GUID values are long and obscure, they are not meaningful to the users. If randomly generated GUIDs are used for key values and someone inserts a lot of rows, random I/O would be included to the indexes, which can negatively impact computational performance. GUIDs are also relatively large when compared to other data types, such as "\emph{integer}". In short, deciding to adopt data type of "\emph{uniqueidentifier}" on NAS, there will no chance for finding duplication of \emph{WID} when the spoofing attack intended or happened on NAS which could enhance the security of NAS.

\begin{figure}[h]
    \centering
    \captionsetup{justification=centering}
    \includegraphics[width=0.4\textwidth]{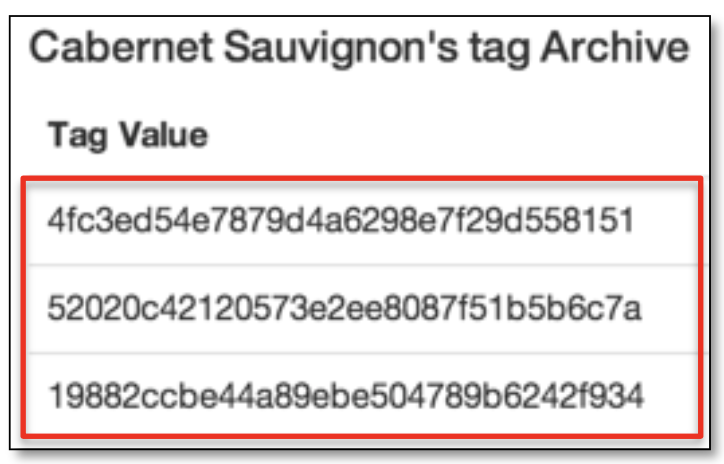}
    \caption{\textit{The List of Hashed WIDs Recorded under Specific Cabernet Sauvignon}}
    \label{fig:tagarchive}
\end{figure}

Nevertheless, NAS is also possible to be attacked or manipulated though the data type of \emph{WID} was being set with "\emph{uniqueidentifier}" if there is no cryptography technique applied to the tag value. The \emph{WID} is therefore considered to be hashed with the specific hashing algorithm, named MD5, and the masked \emph{WID} could be of no meaning to the malicious entities, as demonstrated in \textit{Fig.~\ref{fig:tagarchive}}, and less available for any attack involving spoofing or peeking the \emph{WID} and the "\emph{readCount}" stored in any NFC tag of NAS. Hashing algorithm is indeed a popular cryptography technique, but the principal difference between hashing and encryption is that encrypted values could be reversible via decryption, and hashing is one-way and not reversible. While it is not possible to take a hashed result and "\emph{de-hashed}" it to retrieve the original input, hashing functions are typically created to minimize the chance of collisions and make it hard to produce the same set of hashed values again. As such, only the winemaker will know the "\emph{secret}" which is the tag value \emph{(WID+readCount)} linking both the application and the back-end wine database together for performing functions, such as wine record request or retrieval.

There are also two columns of data set, namely "\emph{TagID}" and "\emph{TagStatusID}" under the database table of "\emph{ActiveWine}" making NAS a more secured system compared with its counterparts in the wider wine industry. For \emph{TagID}, it is actually the unique identifier of any NFC tag as every NFC tag comes with an unique identifier. With the data of \emph{TagID} returned from the database and visibly shown on the user interface of \emph{ScanWINE}, supply chain participants and wine consumers can key in the required \emph{TagID} and see if there is any matching result of the same \emph{TagID} returned from the database, to verify whether the NFC tag is the one originally attached to the wine product since the bottling process, so as to prevent counterfeiter from putting fabricated tag on the second-hand wine bottles. While "\emph{TagStatusID}" is actually with a data type of binary string and defaulted as "\emph{0}", under which when the data is written into a NFC tag with the use of \emph{TagWINE}, it will turn to be "\emph{1}" to denote the tag-writing process is completed. If "\emph{TagStatusID}" is turned to be "\emph{1}", there will be no such wine record shown on the user interface of \emph{TagWINE}, and it will be unable to be written any data into the NFC tag again. In short, the introduction of the \emph{TagStatusID} prevents the wine record linked with specific NFC tag from being updated again, and it could only be updated once only if "\emph{TagStatusID = 0}" denoting the tag-writing process is still available while "\emph{TagStatusID = 1}" denoting further tag-writing process is halted.

\subsection{Hashed Functions on Tag Value Stored in NFC Tags}
\begin{figure}[h]
    \centering
    \captionsetup{justification=centering}
    \includegraphics[width=0.5\textwidth]{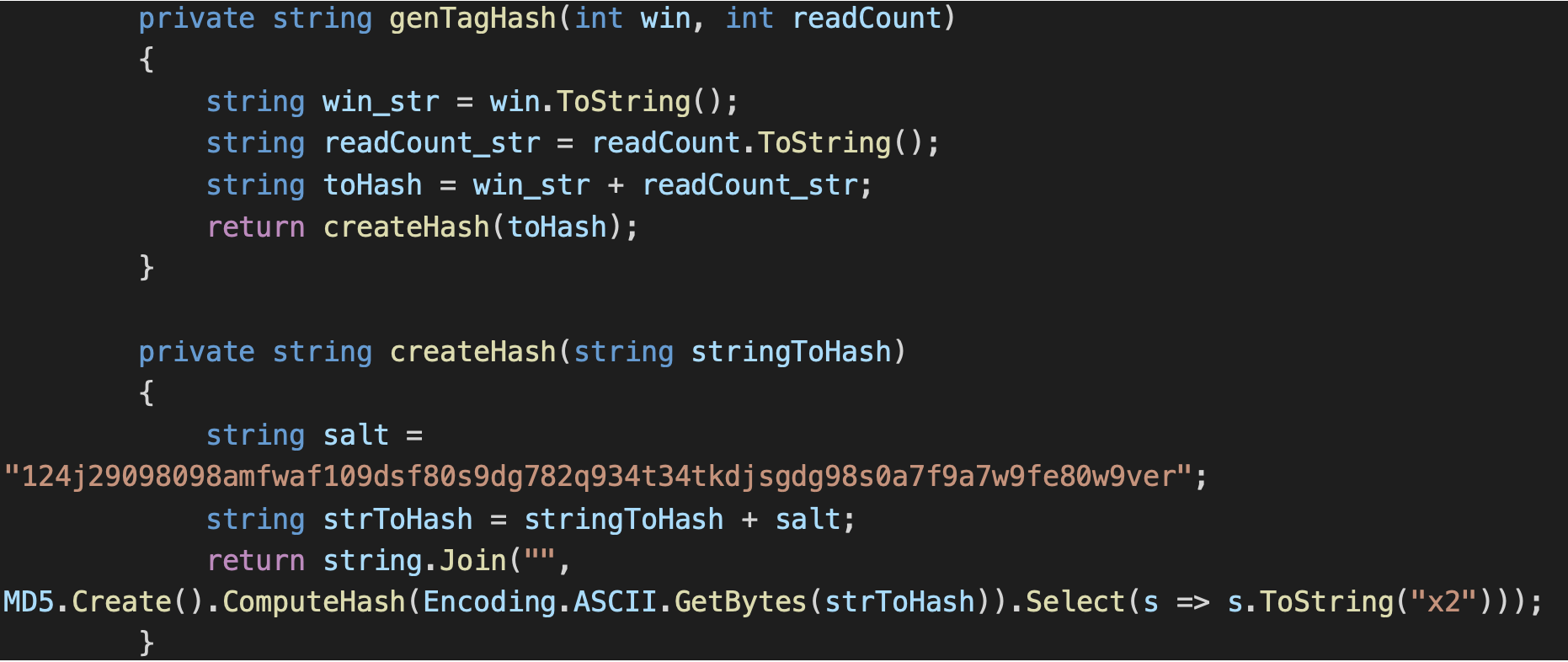}
    \caption{\textit{Design Pattern of MD5 Hash Function Applied in NAS}}
    \label{fig:gentaghash}
\end{figure}

As aforementioned there is only \emph{WID} written into NFC Tags which will in all likelihood be far more secured than storing more attributes of wine records beyond merely a \emph{WID}. The above improved security system with the concept of "\emph{Read Count}" should also be stored in NFC tags so that the corresponding methods could update the tag value every time it is being scanned by the NFC-enabled smartphone with tag-writing or tag-reading events triggered in \emph{ScanWINE}. There are actually 2 methods related to hash generation, with which the hash of specific tag values (\emph{WID+readCount}) could be generated using the Message-Digest Algorithm 5 (MD5). Regarding the method of "\emph{genTagHash()}", as demonstrated in \textit{Fig.~\ref{fig:gentaghash}}, both \emph{WID} and \emph{readCount} are standardized with data type of "\emph{string}" and being combined to form a string named "\emph{toHash}", representing the "\emph{compound}" stored in the NFC Tags, while the "\emph{toHash}" will then be passed to the method of \emph{createHash()}. 

Regarding the method of "\emph{createHash()}", there are steps performed so that every data generated in this method will be a 32-byte random string with no visible meaning to those who intended to copy the "\emph{compound}" stored in NFC tags, which are (1) adding a set of random number named "\emph{salt}" to further complicate the hashed chunks, obfuscating counterfeiters or any malicious entities to be even harder to manipulate, and (2) invoking the MD5 hashing algorithm to generate a hashed chunk of string based on the salt and the "\emph{compound}" to produce a hashed tag value which will eventually be stored in NFC tags, for which the hashed tag value stored will be updated whenever there is a tag-reading or tag-writing process performed on the same NFC tag.

\subsection{The "Read Count" Methodology}
\begin{figure}[h]
    \centering
    \captionsetup{justification=centering}
    \includegraphics[width=0.5\textwidth]{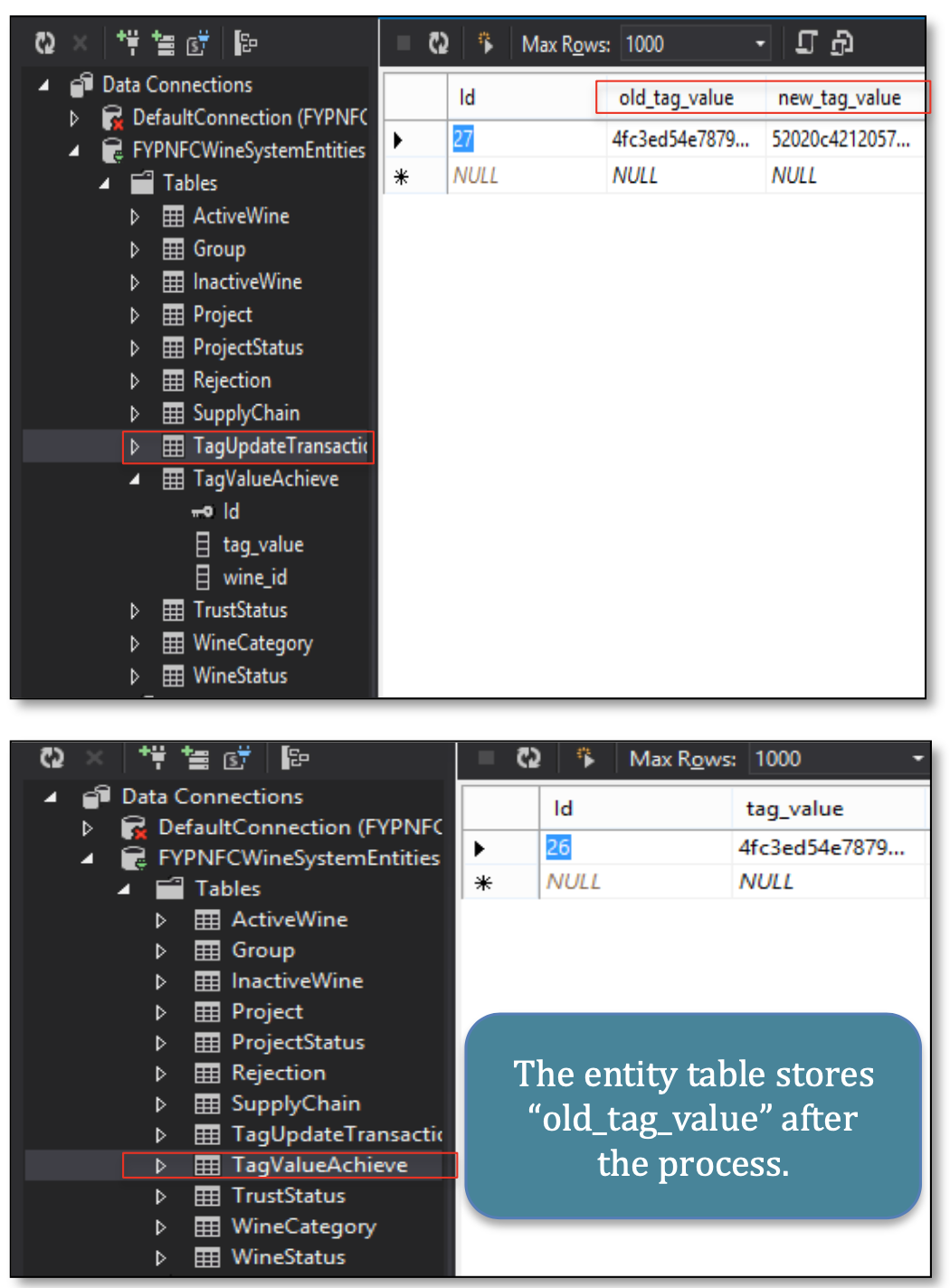}
    \caption{\textit{Entity Tables of the NAS Database}}
    \label{fig:databaseview}
\end{figure}

So as to enhance the security aspects of NAS regardless those security features already implemented in NAS, a mechanism termed as "\emph{Read Count}" is introduced with those hashed tag values generated in every tag-reading or tag-writing process to reach such purpose. To achieve this, there are two new entity tables added with the entity table of "\emph{ActiveWine}" altered as well based on the existing data models, namely "\emph{TagUpdateTransaction}" and "\emph{TagValueArchive}". Based on the database design as demonstrated in \textit{Fig.~\ref{fig:databaseview}}, the former is used to store all the "\emph{old_tag_value}" and "\emph{new_tag_value}" generated during every tag-reading process, while the latter is defined to store the "\emph{old_tag_value}" after every tag-reading process as a "\emph{tag_value}" for the purpose of future references.

With the advent of those hash-generating methods implemented in "\emph{AppController}" of the wine database, namely "\emph{createHash()}" and "\emph{genTagHash()}" respectively, a 32-byte hashed "\emph{NFCCurrentTag}", which is the tag value stored in NFC tags, is firstly generated and returned after every tag-writing process triggered in \emph{TagWINE}. There is also a new 32-byte hashed "\emph{new_tag_value}" generated and returned with the required wine record after every tag-reading process triggered in \emph{ScanWINE}. All the tag values, including "\emph{NFCCurrentTag}", "\emph{old_tag_value}" and "\emph{new_tag_value}", are actually a hashed combination of "\emph{WID}" and "\emph{readCount}", which is the only "\emph{compound}" stored in NFC tags. To a certain extent, the incremental \emph{readCount} can contribute to prompt a different "\emph{new_tag_value}" generated after every tag-reading process, as the \emph{readCount} itself is indeed a constituent of "\emph{new_tag_value}" in which a slight difference on the constituent \emph{readCount} could result in a completely different hashed tag value.

\begin{figure}[h]
    \centering
    \captionsetup{justification=centering}
    \includegraphics[width=0.5\textwidth]{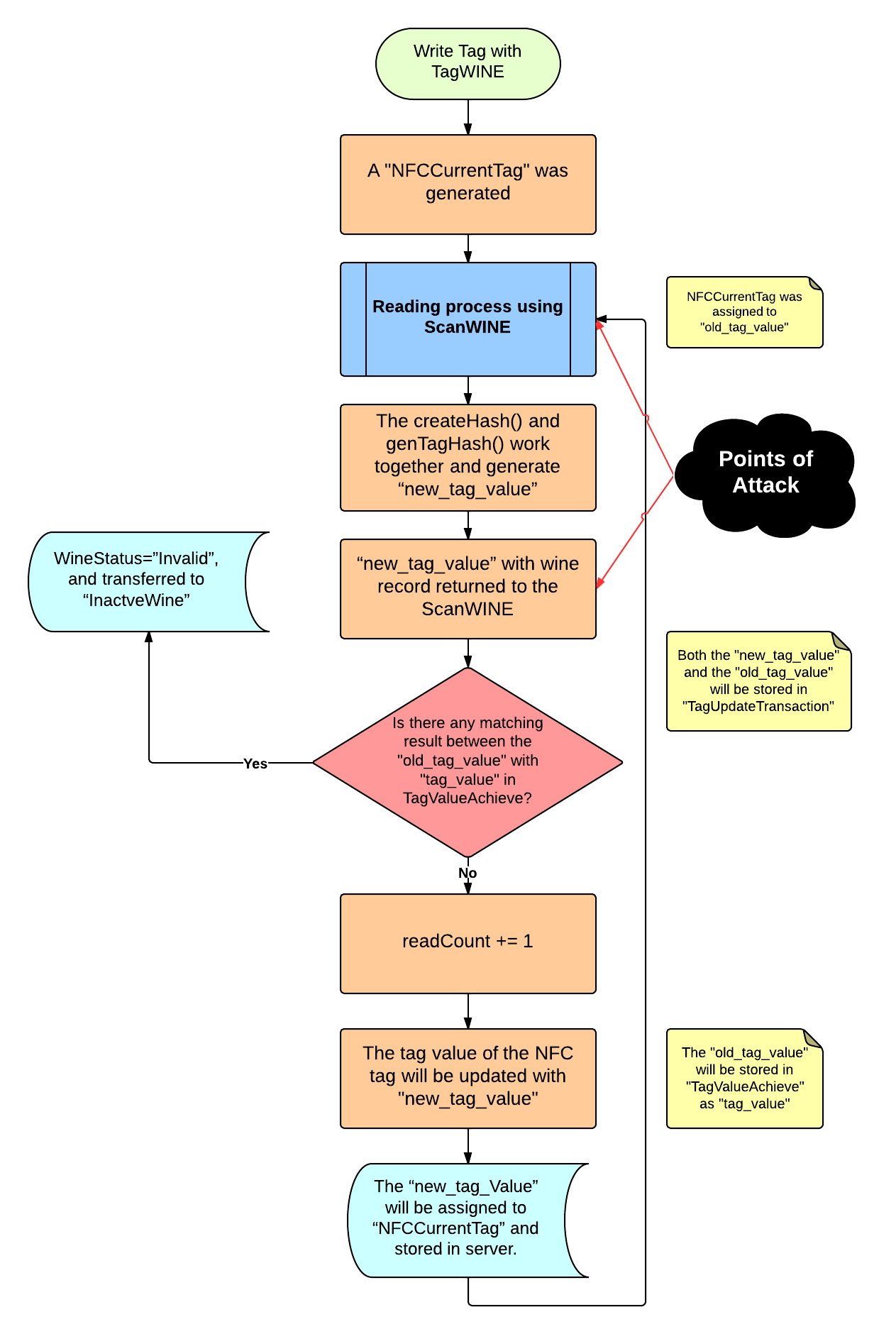}
    \caption{\textit{Flowchart of Security Enhancement Logic based on readCount and Different Data Models}}
    \label{fig:flowchart}
\end{figure}

Provided that there is no matching result of "\emph{new_tag_value}" with those "\emph{tag_value}" already stored in "\emph{TagValueArchive}", there will be a tag-updating command sent to the server, with which the "\emph{new_tag_value}" would therefore be assigned as "\emph{NFCCurrentTag}" and stored in the \emph{ActiveWine}. The "\emph{old_tag_value}" will be stored in the "\emph{TagValueArchive}" as "\emph{tag_value}" for the comparison for every oncoming tag-reading process, and the \emph{readCount} will be incremented after every tag-reading process. The security enhancement, introduced with the concept of \emph{readCount}, different data models and entity tables defined in the database, is detailed in the flowchart of \textit{Fig.~\ref{fig:flowchart}}.

\begin{figure}[h]
    \centering
    \captionsetup{justification=centering}
    \includegraphics[width=0.5\textwidth]{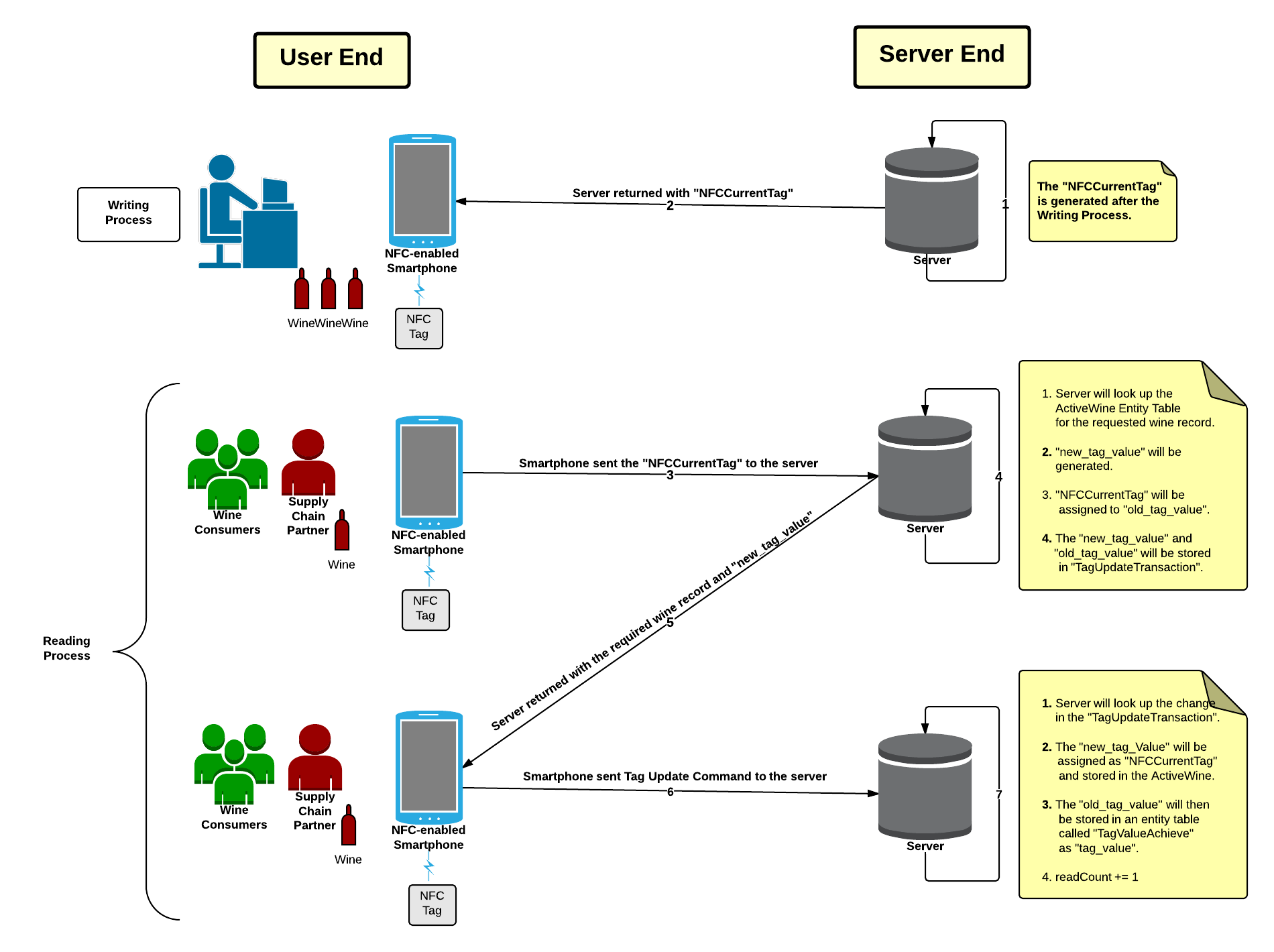}
    \caption{\textit{Tag-Reading Procedural Steps in Different Scenarios}}
    \label{fig:readingstep}
\end{figure}

However, we all knew that the tag value stored in NFC tags is updated after every tag-reading process, and there will then be two important cases to note when such process has progressed at between step 5 and step 6, of \textit{Fig.~\ref{fig:flowchart}}, in case there will be a disconnection or failure of communication somehow happened during the tag-reading process. For instance, the first case could be the situation that the "\emph{new_tag_value}" could not even be updated and overwritten to the "\emph{NFCCurrentTag}" stored in that NFC tag, leading to failure to commit the updated tag value back to the server and with a scenario that the "\emph{old_tag_value}" will still be sent to the server if there is any upcoming tag-reading processes. The second case would be the situation that the NFC tag itself is updated with the "\emph{new_tag_value}" which as well overwrites the "\emph{NFCCurrentTag}" stored, but somehow, due to a disconnection, the change of "\emph{NFCCurrentTag}" is failed to commit back to the server, leading to the tag value stored in the tag and that in the server are not synchronized for upcoming tag-reading processes. Both scenarios could well lead to chaos on tag values stored in NFC tags and the server, and could in turn cause security loopholes in NAS. There are methods implemented in "\emph{AppController}" of the database, so that \emph{ScanWINE} could continue on the NFC tag-updating process with "\emph{new_tag_value}" as if someone proceeds with the tag-reading process in first run to address the first case, while for the second case, both \emph{ScanWINE} and \emph{TagWINE} should be continuing in committing the updating process according to the tag-reading procedural steps as described in \textit{Fig.~\ref{fig:readingstep}}.

\begin{figure}[h]
    \centering
    \captionsetup{justification=centering}
    \includegraphics[width=0.5\textwidth]{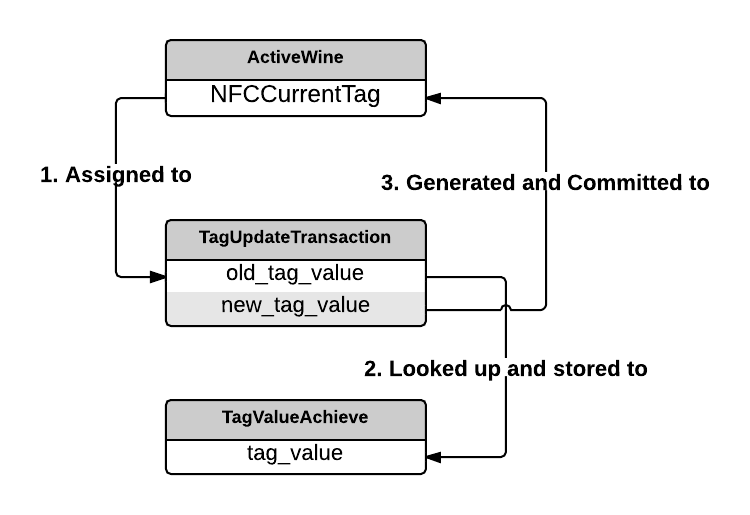}
    \caption{\textit{The Entity Table Relationship Regarding Tag Value of NAS}}
    \label{fig:tablerelation}
\end{figure}

Given the fact the hashed tag value, at different stage of same wine product, is the key to the anti-counterfeiting value of NAS, the database design has also included a separate table, namely "\emph{TagUpdateTransaction}", to store the intermediate tag values updated during every tag-reading process, such as "\emph{new_tag_value}" and "\emph{old_tag_value}" of every scanned wine product for fear that some entities try to scan that uncommitted NFC tag of a specific wine product again. With the advent of entity tables, such as "\emph{TagUpdateTransaction}" and "\emph{TagValueArchive}", the specific wine record will turn to be "\emph{Invalid}" in case there is a matching result of  "\emph{NFCCurrentTag}" (\emph{old_tag_value}) and every "\emph{tag_value}" (stored in the "\emph{TagValueArchive}"), as detailed in \textit{Fig.~\ref{fig:tablerelation}},under which such wine status of "\emph{Invalid}" would be attained should more than one NFC tag store the same tag value. It follows that the "\emph{readCount}" could prevent the wine supply chain with NAS adopted from having any NFC tag storing with the same tag value simultaneously.

\begin{figure}[h]
    \centering
    \captionsetup{justification=centering}
    \includegraphics[width=0.4\textwidth]{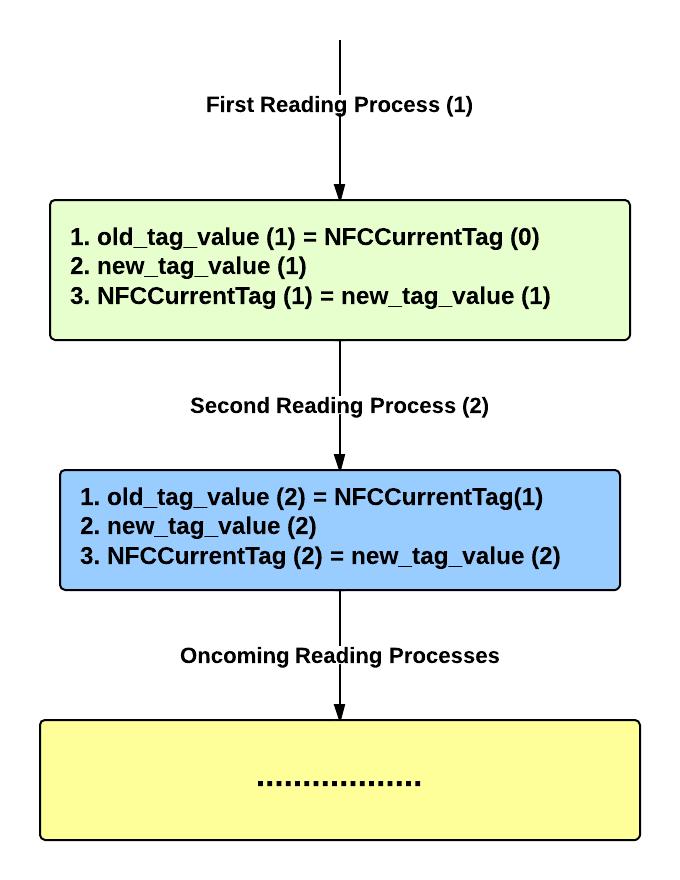}
    \caption{\textit{State Transitioned on Tag Value of Every Tag-Reading Process}}
    \label{fig:tagstep}
\end{figure}

There will be indeed an "\emph{new_tag_value}" generated, which will further be assigned to the  "\emph{NFCCurrentTag}" and stored in the entity table of "\emph{ActiveWine}" after every tag-reading process, and the "\emph{NFCCurrentTag}" from the most previous tag-reading process (such as the last tag-reading process) will be assigned as the "\emph{old_tag_value}" of the next tag-reading process as specified. The state-transitional changes on tag value after each tag-reading process are depicted step-wise in \textit{Fig.~\ref{fig:tagstep}}.

\begin{figure}[h]
    \centering
    \captionsetup{justification=centering}
    \includegraphics[width=0.5\textwidth]{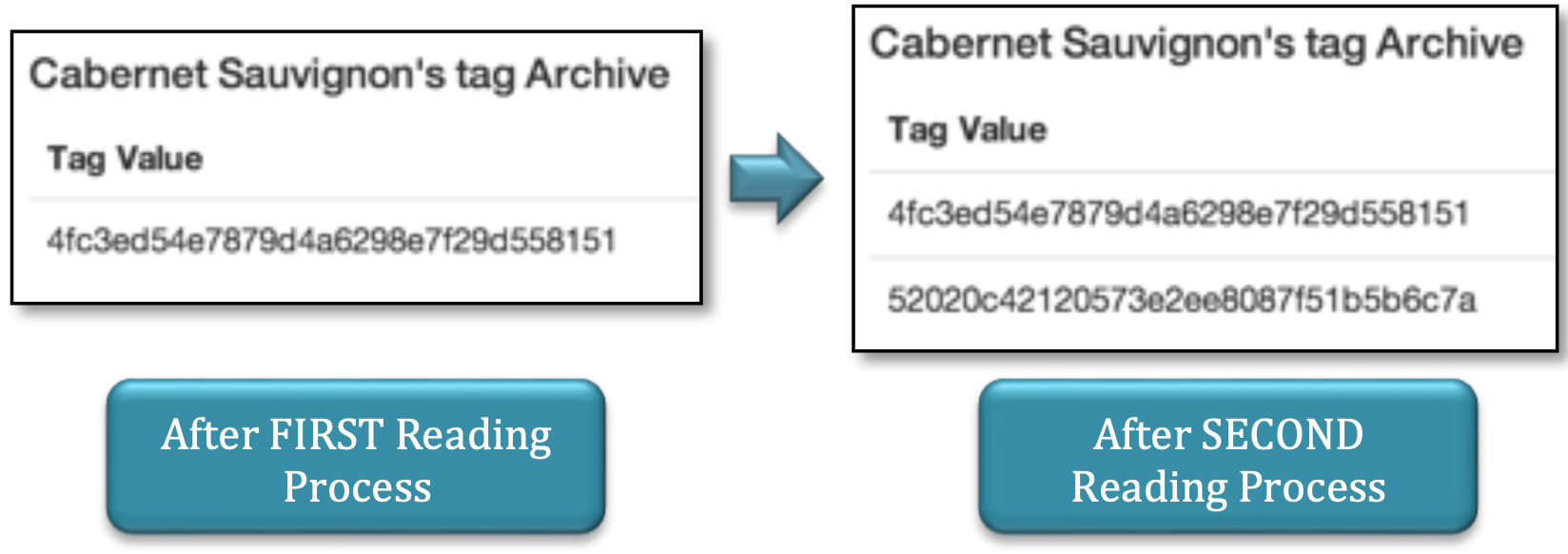}
    \caption{\textit{An Example of Tag Archive After Every Tag-Reading Process}}
    \label{fig:cabexam}
\end{figure}

In short, a hashed tag value (\emph{WID+readCount}) stored in an NFC tag, would be updated every time the tag is being read and returned with incremental \emph{readCount}, and those "\emph{tag_value}"("\emph{old_tag_value}" of all the previous tag-reading processes) stored in the entity table of "\emph{TagValueArchive}". An example on tag archives of Cabernet Sauvignon is demonstrated in \textit{Fig.~\ref{fig:cabexam}} after the first two tag-reading processes. The ever-changing nature of tag values stored in the NFC tag and the server could make the tag value and the NFC tag itself harder to be cloned, or make NAS a more secured solution which is less susceptible to be manipulated, by counterfeiters.

\subsection{The Automatic Updating Process on Wine Records}
The automatic updating process means some of the columns of specific wine records would be updated simultaneously and automatically whenever there is any NAS tag-reading process, performed by both supply chain participants and wine consumers, performed via Internet network with payloads in JSON under which the related logic is defined over \emph{ScanWINE} and "\emph{AppController}" of the back-end database. There are only four columns of a wine record related to such automatic updating feature. For instance, "\emph{WineStatus}" is a status of wine record with only two values – "\emph{Valid}" or "\emph{Invalid}" with default values set as "\emph{Valid}" when the wine record is first created. "\emph{WineStatus}" could turn from "\emph{Valid}" to "\emph{Invalid}" when two scenarios happened – (1) the wine is being sold and the payment is confirmed through pressing the button of "\emph{Buy it}" with \emph{ScanWINE}, or (2) the wine product moving along the supply chain is suspected as a wine counterfeit and cannot get through the Authentication Controller. Now that the wine product with its wine record is labeled with a wine status of "\emph{Invalid}", it will be put into the Unsuccessful Record Controller for storage and further reviews by the corresponding winemaker.

\begin{figure}[h]
    \centering
    \captionsetup{justification=centering}
    \includegraphics[width=0.5\textwidth]{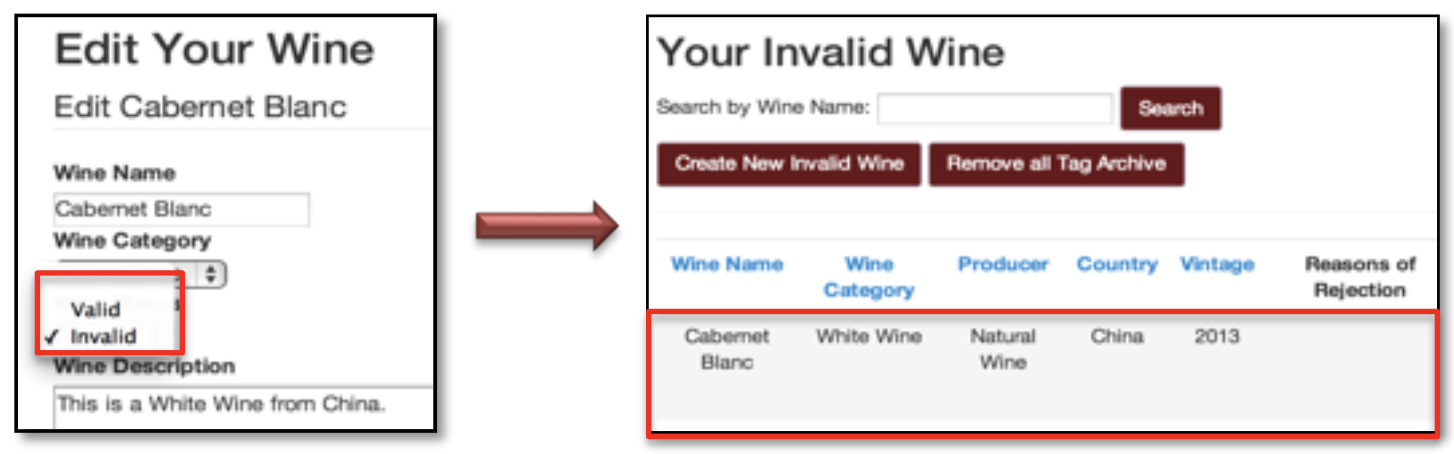}
    \caption{\textit{"Invalid" Wine Status to Label an Invalid Wine Product}}
    \label{fig:winestatuschange}
\end{figure}

Some of the programming methods defined in "\emph{AppController}" of the wine database and the logic performed in the mobile applications are amended with new elements added so that some parts could perform the automatic updating features and contribute to the security enhancement of NAS. For the wine database web application, if the wine status of a specific wine record is changed from “Valid” to “Invalid”, the wine record will be transferred to the tab of \emph{Inactive Wine} automatically after the change of wine status is completed, as the "\emph{InactiveWineController}" in the database is set to only list the wine record with a wine status of “Invalid”. The functionality is demonstrated in \textit{Fig.~\ref{fig:winestatuschange}} using the wine product of "\emph{Cabernet Blanc}" as an example.

\begin{figure}[h]
    \centering
    \captionsetup{justification=centering}
    \includegraphics[width=0.3\textwidth]{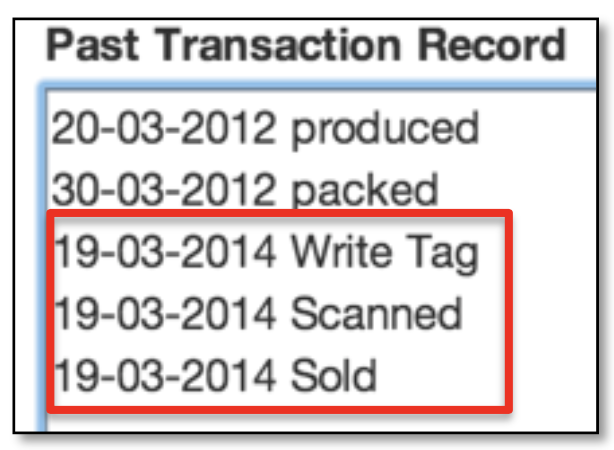}
    \caption{\textit{Automatic Updating Feature on Transaction Records of A Wine Product}}
    \label{fig:txrecord}
\end{figure}

There are some other automatic updating features introduced across different components throughout NAS. For instance, as demonstrated in \textit{Fig.~\ref{fig:txrecord}}, the automatic updating functions on transaction records of a specific wine record with "\emph{Date + "Write Tag"} whenever a specific wine record with its tag value is written into an NFC tag, with "\emph{Date + "scanned"} whenever the NFC tag containing with the right tag value of a specific wine record is scanned by a NFC-enabled smartphone with \emph{ScanWINE} running on it. The transaction record will also be updated with "\emph{Date + "sold"}, whenever the wine product is being sold, and the wine status of that wine record will therefore be labeled from "\emph{Valid}" to "\emph{Invalid}" with the corresponding wine record being transferred to the tab of "\emph{Inactive Wine}".

\subsection{Security Features of the NFC Tag and Its Deployment on Wine Bottle}
It is possible to deliver anti-counterfeiting value with the NFC technology being chosen as the primary and fundamental communication technology for NAS as mentioned in the previous chapters about NFC technology, so does the chosen NFC tag itself. Some NFC tags are actually manufactured with 3DES encryption, such as the \emph{Ultralight C}, protecting the NFC tag from being manipulated, editing or reading the data stored into it without any permission. As such, there is no other participant of specific supply chain, except the winemaker who formats the tag and is aware of the password protecting the NFC tags, could make any change and even get access into the editing procedures of the NFC tags.

Besides, \emph{TagWINE}, performing features such as writing tag value of specific wine record only for winemakers, will not be available for making it public and open for downloads on Google Store or any other application platforms. It follows that only winemakers could operate different wine manufacturing processes with \emph{TagWINE}, which originally designed and developed for them on writing tag values of a specific wine record, such as \emph{WID}, into the NFC tags via the communication between \emph{TagWINE} and Application Controllers of the back-end database.

Regarding the deployment of NFC tags on wine bottles of specific wine products, the best position of tag application is to put the NFC tags on the bottleneck with NFC tag encapsulated in the foil package. Such tag deployment is considered as a way to protect wine products from counterfeiting. For example, once the wine product is purchased and uncorked for consumption, the NFC tag storing the tag value, which is the only key connecting mobile applications and the back-end database of NAS, will be completely damaged alongside the foil package surrounding the bottleneck and the cap, while the wine product is uncorked for consumption. Provided that the \emph{WID} is invisible and not being manipulated by counterfeiters, the reused and recycled bottle of the wine product will not be scanned and gone through with NAS again, indicating the wine product was already consumed before.

\subsection{Procedural Policies of NAS}
With the advent of NAS, supply chain participants and wine consumers could now utilize their own NFC-enabled smartphone to authenticate and validate whether a certain wine product is genuine or not before accepting, sending it to the next nodes along supply chain and even purchasing it respectively. There are also some procedures necessary throughout NAS, in which supply chain participants have to pre-register before they can be granted to access the host server and applications offered by winemakers to enable NFC tag-scanning processes so as to update any wine transactions. In order to enable the NAS features to the related supply chain participants, the company particulars of that supply chain participant is indeed stored in Supply Chain Information Controller approved by the winemaker, and they also need to login before being able to enjoy the anti-counterfeiting features offered in \emph{ScanWINE}, thus excluding those counterfeiters from attempting to commit wine counterfeiting via using \emph{ScanWINE} without registration to capture wine pedigree of the rare wine products for cloning of a fabricated NFC tag, or even applying the updated wine record to a second-hand wine bottle of a similar wine product.

The suspicious transactions will also be screened accordingly if it happens with a situation that any wine record is with a wine status of "\emph{Invalid}", the invalid wine record will then be transferred to Unsuccessful Record Controller as it cannot pass through Authentication Controller during the tag-reading process due to its invalid wine status. The supply chain participants and wine consumers will eventually be notified of the invalid wine product and its wine record, which would then lead to a failed tag-reading process with no wine record returned. The supply chain participants and wine consumers should refuse to accept, send and even make purchase for any suspicious wine product detected in NAS. It is essential that the related supply chain participants should be responsible for returning the suspicious wine products back to the corresponding winemakers for fear that the suspicious wine products will still be circulating in the market exacerbating the product counterfeiting of wine industry.

One of the reasons why NAS could be more popular in conveying anti-counterfeiting value is that winemakers, supply chain participants and wine consumers could simply utilize their own NFC-enabled smartphone with both \emph{TagWINE} and \emph{ScanWINE} to authenticate a wine product or update any state transition of certain wine products at anytime along the supply chain before the next node accepting or purchasing it.

\section{The Project Conclusion}
Based on the design concepts and implementation details of NAS explained in this project alongside how NAS could bring innovations to contribute with anti-counterfeiting value to, and improve the exacerbating wine counterfeiting problems of, the wine industry, the future works on how NAS could be improved with anti-counterfeiting features diversified and system security enhanced are identified with project findings also concluded.

\subsection{Future Works on NAS}
\subsubsection{Further Possible Improvements on NAS}
Indeed, the development of NAS is actually based merely on winemakers or wineries, not to mention there will be tons of other mobile applications similar to ScanWINE on the application platforms, loaded onto consumers’ NFC-enabled smartphones, if there are more individual winemakers aiming to deliver anti-counterfeiting value to the wider wine industry. It turns out that the situation will well be more chaotic if NAS is not secured, efficient and scalable. Therefore, it would be better if there will be any centralized wine database and even a centralized NAS-like architecture developed and implemented by any governmental or well-known international organizations of global wine industry, such as The International Organization of Vine and Wine (OIV) promoting development on quality fine wine products, to store all the standardized wine records of wine products imported and exported back and forth the city with centralized database architecture, and allowing all the nodes along the supply chain to update wine pedigrees of all the wine products circulating in the wine market concerning the trustworthiness, effectiveness, security and accuracy of every wine record stored, the centralized wine database, and even those mobile applications developed providing to wine consumers or distributors for usage before any transaction is taken place. The NAS-like architecture concept could also be delivered with a decentralized and distributed fashion where all the supply chain participants in wine industry can take part into sending and validating standardized cryptography transactions based on a pre-defined consensus algorithm of a chosen peer-to-peer network, with the advent of blockchain technology and emergence of enterprise blockchain development. 

Nonetheless, the idea of NAS should be expanded and not based on only a winemaker, but a cluster of them or a centralized NAS should be developed by the government, some professional wine organizations or a strategic alliance of a cluster of wine brands, so as to better deliver the efficiency, effectiveness and anti-counterfeiting value of NAS instead of making the wine consumers and the wine supply chain participants to download a new mobile application every time they need to deal with another brand of wine product. A decentralized NAS based on blockchain technology could also be a more pragmatic way to enhance the anti-counterfeiting value of NAS for the wider wine industry.

Undeniably, it is too early to regard NAS is the most secured and innovative anti-counterfeiting system for the wine industry. However, the idea is that NAS could add one more layer of security to the wine industry. It is also known to everyone that a combination of labeling technologies would definitely be more secured and complicated than only applying either of them to the wine industry. It is always good to say NAS is only based on NFC technology, which is also one of the reason why it is NFC-enabled, but the reality is that integrating with more labeling or other anti-counterfeiting technologies could make NAS an even more secured and trustworthy option to the users, such as integrating QR code or barcode scanning features to the \emph{ScanWINE}. It is also recommended that wine brands should introduce manufacturing steps to include barcodes and communication tags, in order to reach the greatest number of customers as well as integrating with many Internet-of-Things system for future opportunities on data analytics applied to different aspects of the business and efficient inventory management with lean manufacturing plans implemented.

\subsubsection{Development of NFC-Enabled Wine Closure System}
\begin{figure}[h]
    \centering
    \captionsetup{justification=centering}
    \includegraphics[width=0.17\textwidth]{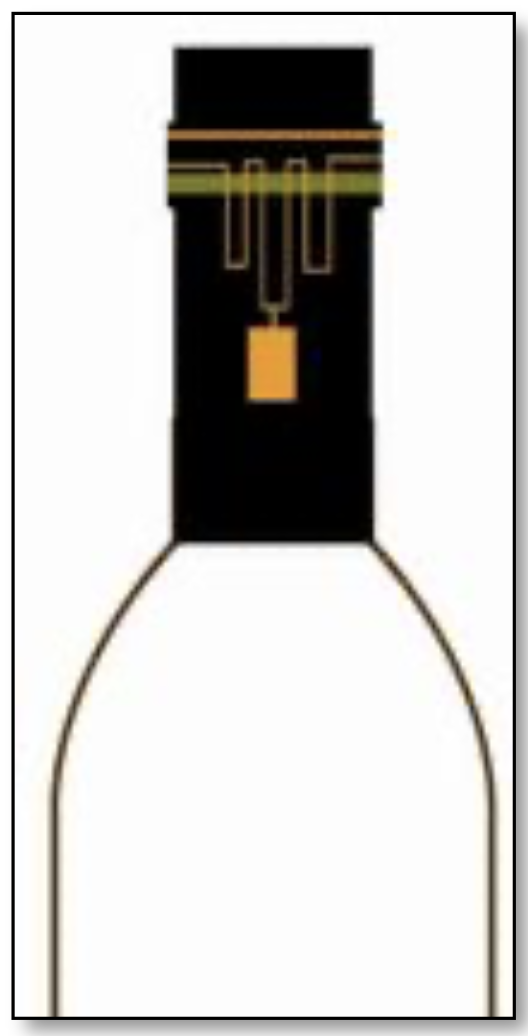}
    \caption{\textit{The Concept of NFC-Enabled Wine Closure System}}
    \label{fig:wineclosure}
\end{figure}

According to the chapter describing the deployment of NFC tags on wine bottles, the best position to place the NFC tag would be right on the bottleneck and fully encapsulated into the packaging metal foil of the wine product. However, it was proved that encapsulating an individual NFC Ferrate tag, storing tag value, into a metal foil would greatly lower the performance of NAS, influencing both the response rate and effectiveness of NAS. It was then suggested that NFC Ferrate tags should be physically intact with the packaging foil, either being part of the packaging foil to seal the wine bottle or only make contact with the foil with an example as depicted in \textit{Fig.~\ref{fig:wineclosure}}.

We all knew that wine bottles are traditionally sealed with a wooden cork or a screw-top cap, and there are several other methods adopted to seal a wine bottle. There is a commonality of all these sealing techniques, in which all of them require packaging foil for sealing the wine bottlenecks. As such, an idea termed as NFC-enabled Wine Closure System should suit both the aforementioned sealing methods, namely the screw-top cap and sealing cork. The NFC-enabled Wine Closure System is appeared as an one-piece foil packaging with NFC-enabled \emph{printed circuit board} (PCB) antenna connected. The whole packaging is functioned and of no difference with any ordinary NFC tag, just imagine the NFC tag is now transformed from circular surface to cylindrical surface in shape.

The NFC-enabled Wine Closure System itself could be an overt counterfeiting deterrent if pairing with NAS since its unique NFC-enabled seal is affixed to an individual wine bottle along with tag value stored into it, fundamentally integrated the NFC-enabled Wine Closure System with other components of NAS, such as \emph{ScanWINE} and \emph{TagWINE}, reading and writing tag values into the NFC-enabled Wine Closure System respectively for the purpose of authenticating and validating wine products. Most importantly, the NFC-enabled Wine Closure System shoud be best compatible with NAS as it remains NAS as an open system for winemakers, supply chain participants and wine consumers, as well as transferring one of the most significant anti-counterfeiting value in which the NFC-enabled Wine Closure System, connected with NFC PCB antenna must be damaged before the wine product could be uncorked for consumption, making the wine closure system itself and the recycled wine bottle less feasible for wine counterfeits, and ensuring no more update on unique wine record of a consumed wine product would be made possible along the supply chain of wine industry anymore.

\subsubsection{Integration of Digital Marketing and Data Analytics Features with NAS}
Indeed, NAS could actually further combine its anti-counterfeiting technology with brand-marketing messages so as to boost the sales performance of wine products and enhance brand reputation. NAS could include marketing features to \emph{ScanWINE} so that it not only authenticates wine products but also enables winemakers and related supply chain participants, such as wine brand promoters, to disseminate wine consumers with more customized information, via a tap on their NFC-enabled smartphones. With a label of further sales information of the selected wine products, it could further direct application users to another user interface with more wine information on it, such as sales information, brand updates or even information about upcoming wine products. 

Interest is growing amongst advertisers of wine products in utilizing NFC tags to promote products and gather behavioral data of consumers, such as consumers' preferences or behaviors, so as to customize advertising resources based on targeted advertising. For instance, with NFC tags and the digital marketing interface on \emph{ScanWINE}, brands and winemakers could collect data about which types of wine products are popular amongst wine consumers, which categories of user interfaces of \emph{ScanWINE} are most viewed, and which types of wine are the most purchased ones, geographically or based on behavioral data identified for individual wine consumers.

It is undeniable to say that wine consumers will only allow advertisements which are able to add value to their preference and interest, such as information about suggested types of cheese pairing for specific wine products, rather than those "\emph{random}" Google-type adverts for potential customers to purchase wine products without any value added to wine consumers throughout the process of selecting and purchasing wine products. Wine advertisers could also send more narrowly targeted follow-up promotions to any smartphone user demonstrating the most interest in NFC-enabled advertisements of specific wine products, even if they do not know the brands and models. The tracking should be made possible only if wine consumers are agreed to opt in to participate.

\subsubsection{Integration of NFC-Enabled Track-and-Trace System with Advanced Features}
Instead of only performing wine-authenticating features to deliver anti-counterfeiting value on wine products, NAS could also equip with track-and-trace features on shipments of wine products and real-time updates along the supply chain of wine products, especially for supply chain participants of specific wine products.

Track-and-trace feature could also be introduced to NAS, with which each NFC-enabled smartphone utilized by supply chain participants could be assigned with a unique serial number by the winemaker. The NFC-enabled smartphone could also be registered before any wine acceptance is approved by the supply chain participants when using \emph{ScanWINE}, under which location data of a wine product could be captured, utilizing GPS features of the NFC-enabled smartphone. The location data captured could then be written and stored into the wine database as a data attribute of the corresponding wine product, such as "\emph{wineLocation}" to store the corresponding location data at the point of ownership transfer, which would in turn display on the user interface of \emph{ScanWINE}. In this case, whenever there is a completed tag-reading process along the supply chain, an update will always be performed on the wine record based on the location data of the NFC-enabled smartphone scanning the NFC-tagged wine product, returned from the back-end database with the GPS data sent by the NFC-enabled smartphone. This could be used to track wine products throughout the supply chain if winemakers is in favor of enabling the possible track-and-trace feature of NAS. The registration of smartphone's serial number is not compulsory but could help contribute to a more secured track-and-trace system as every NFC-enabled smartphone is required to register with the winemaker before any transfer of ownership could take place along the supply chain.

\begin{figure}[h]
    \centering
    \captionsetup{justification=centering}
    \includegraphics[width=0.30\textwidth]{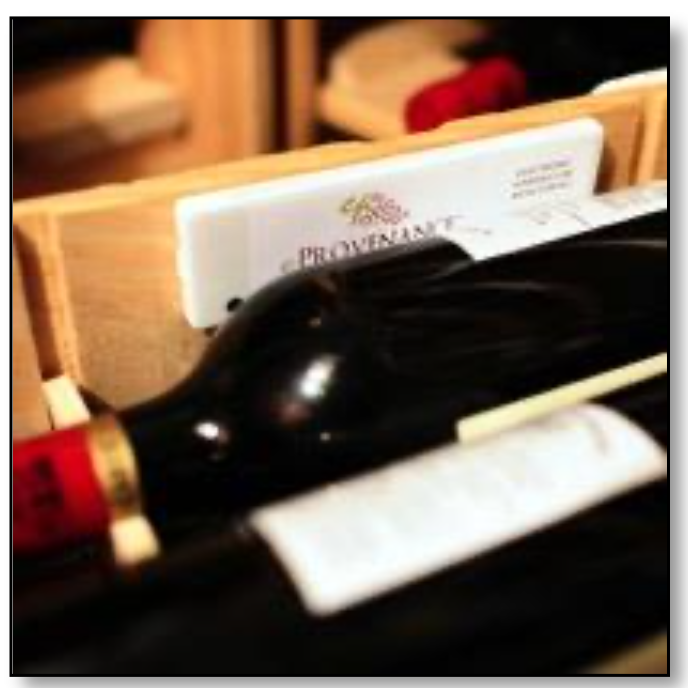}
    \caption{\textit{Adoption of NFC-Compatible Temperature Sensors System in Wine Product Packaging}}
    \label{fig:tempbox}
\end{figure}

There are also some advanced features which could be developed based on the track-and-trace system, such as the NFC-compatible temperature sensors system, with which there will be a new NFC-compatible temperature sensors with durable battery life, which could also be scanned through wooden packaging of the wine products with required temperature data returned to the NFC-enabled reader, as demonstrated in \textit{Fig.~\ref{fig:tempbox}}. The cases of packed wine products or simply a bottle of wine product with individual packaging box could motivate the supply chain participants and wine consumers to verify the quality conditions especially the temperature data along the supply chain before accepting or purchasing it. It could in turn provide retailers with more information to share at the point of transfer or point of sale, which could encourage wine consumers to check the quality of what they will be purchasing. Similar to the track-and-trace system, some attributes, such as "\emph{wineTemperature}" storing the temperature data whenever there is a tag-reading process triggered along the supply chain, could be introduced to the corresponding tables of the back-end database system.

\subsection{Summary and Conclusion}
To sum up, the NAS project has actually achieved all the objectives, described under the “\emph{Objectives of Constructing NAS}”, in which the objectives, such as understanding the background of the rampant wine-counterfeiting situation in global wine industry, designing an overall NAS structure and communication mechanism between different components of NAS and building solid models for components like the wine database, \emph{ScanWINE} and \emph{TagWINE} with the compatible NFC-enabled smartphones and NFC tag selected. It would be fair to comment that the project has been successful as the solid model of NAS could eventually be defined, designed, developed and tested with different sets of scenario, such as the situation when NFC tags are being cloned by counterfeiters. NAS was even demonstrated to the public for further improvements on security consideration and aesthetic values on the general layout of both wine database and mobile applications. 

The NAS project has indeed developed, exactly based on a standard software development life cycle, namely the NAS Development Life-cycle as detailed in \textit{Fig.~\ref{fig:naslifecycle}}, in which the whole NAS development process is actually commenced at a stage of problem definition where the global wine counterfeiting situation, the limited effectiveness of existing labeling technologies are identified according to the literature reviews, giving strong incentives to design and develop an anti-counterfeiting system, using NFC technology, for which such labeling technology has unprecedentedly been applied for this purpose. The components, such as the mobile applications and database, are then designed and developed once every requirement definition and analysis of different components are completed. All the components are then connected together via APIs designed and developed with application development tools and programming methods included in "\emph{AppController}" of the wine database web application. The NAS prototype is then put to test with different sets of scenario with different tests cases produced in a wide variety of unit tests, integration tests and system tests, so that the most suitable NFC tag working best with NAS could therefore be identified and adopted for a series of demonstrations to potential users of the wine industry. It is not surprised to say that every improvement of NAS would require feedback or reactions from actual users of the wine industry.

\begin{figure}[h]
    \centering
    \captionsetup{justification=centering}
    \includegraphics[width=0.5\textwidth]{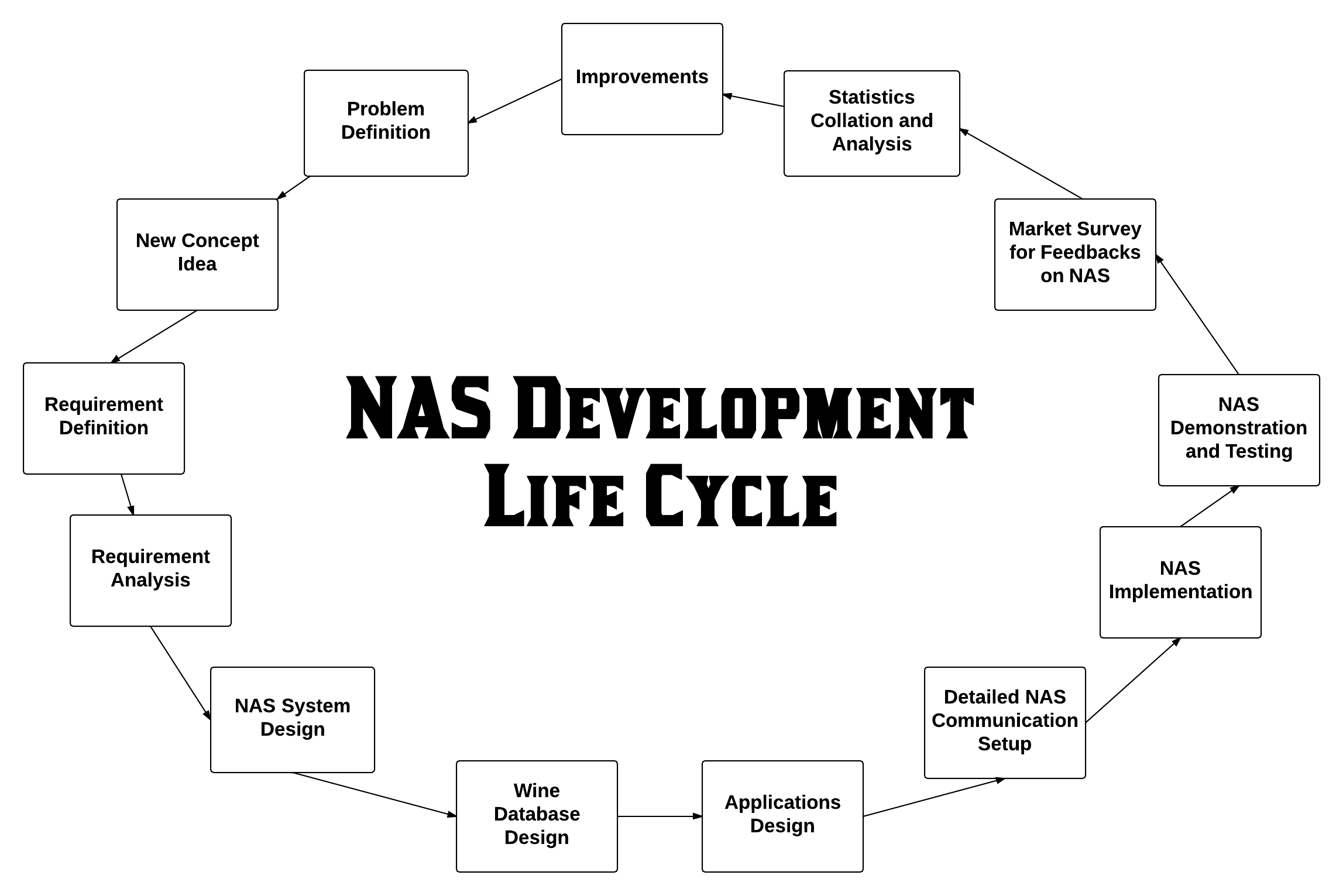}
    \caption{\textit{The NAS Development Life Cycle}}
    \label{fig:naslifecycle}
\end{figure}

As such, there are surveys to collect general feedback towards the NAS prototype, and it is further improved based on the findings from these surveys, although both wine consumers and supply chain participants are delighted with such a new application of NFC technology on anti-counterfeiting purpose and satisfied with the project outcomes of NAS development. Apart from the improvements made, there are also a series of future works suggested to further improve NAS, such as developing the extension version of wine database and \emph{ScanWINE} for supply chain participants so that they could view the shared wine records on the same platform, which will well be making the system more user-friendly and integrated.

Though NAS has now successfully been developed, implemented and available for further industrial testing activities and demonstrations, it will never be possible to conclude that such NFC-enabled anti-counterfeiting system would be the most secured anti-counterfeiting technique being adopted especially in wine industry. It is, however, right to a certain extent that NAS is actually able to introduce an additional layer of security to wine products circulating in the current wine industry. NAS is the first anti-counterfeiting system utilizing labeling technology – NFC, to apply anti-counterfeiting techniques on wine bottles, specially on the bottleneck so that NFC tags would be damaged before any wine product is uncorked for consumption, and this innovative and unprecedented application of NFC for wine industry, using any ordinary NFC-enabled smartphone, will in all likelihood contribute, set a solid example and even lead to further improvements so as to develop a more innovative and sophisticated anti-counterfeiting system for wine industry or even for other industries of luxury goods in the near future.

\bibliographystyle{ieeetr}

\begin{IEEEbiography}[{\includegraphics[width=1.1in,height=1.25in,clip,keepaspectratio]{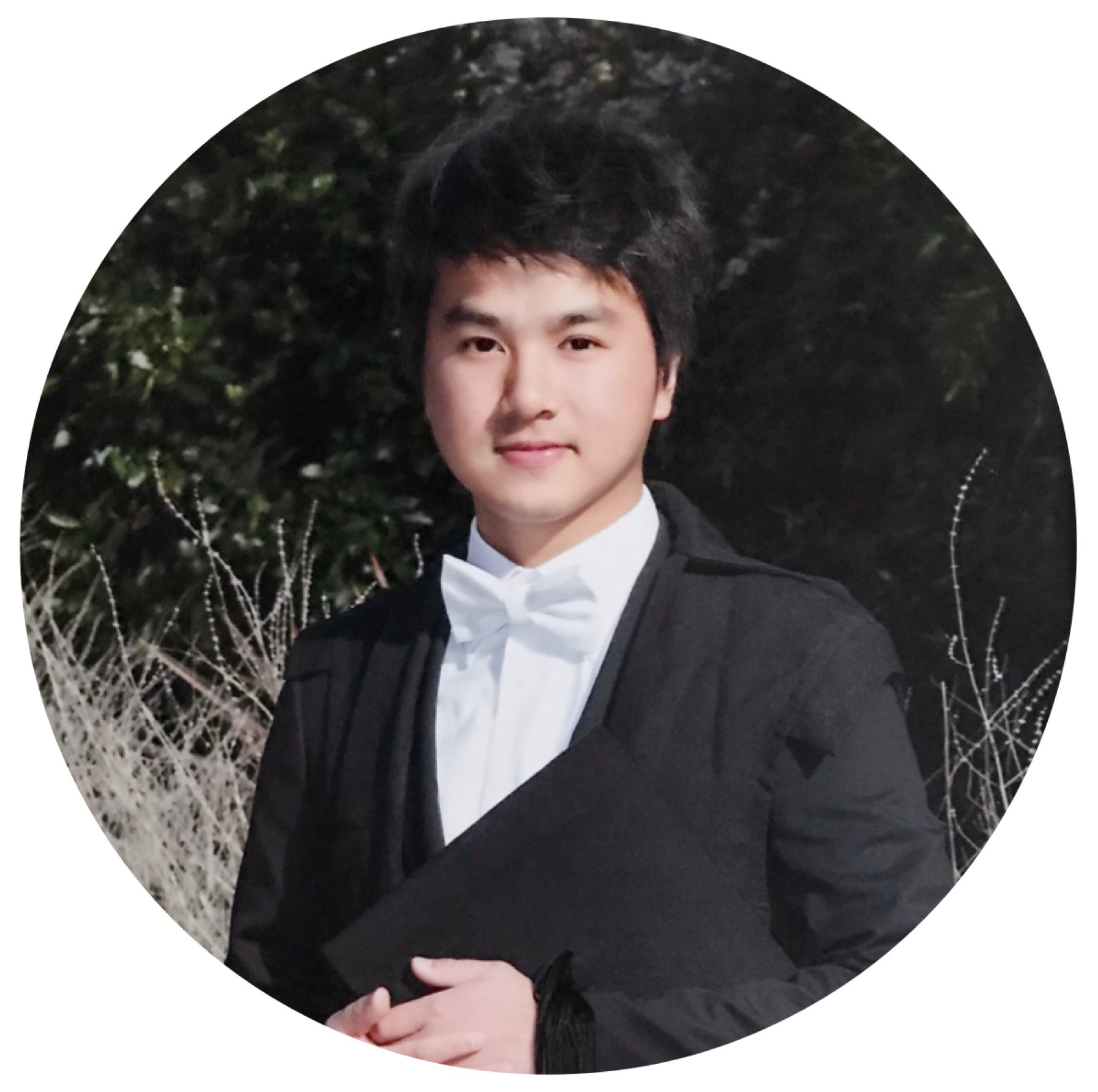}}]{Mr. Neo C.K. Yiu IEEE}
is a computer scientist and software architect specialized in developing decentralized and distributed software solutions for industries. Neo is currently the Lead Software Architect of Blockchain and Cryptography Development at De Beers Group on their end-to-end traceability projects across different value chains with the Tracr™ initiative. Formerly acting as the Director of Technology Development at Oxford Blockchain Society, Neo is currently a board member of the global blockchain advisory board at EC-Council. Neo received his MSc in Computer Science from University of Oxford and BEng in Logistics Engineering and Global Supply Chain Management from The University of Hong Kong.
\end{IEEEbiography}
\vfill

\newpage
\appendices
\section{Selection of NAS Hardware Components}
\subsection{List of NFC-Enabled Smartphones} \label{a1}
The updated list of NFC-enabled smartphones is available at:
\url{https://en.wikipedia.org/wiki/List_of_NFC-enabled_mobile_devices}
\subsection{NFC Forum Standard on Choosing NFC Tag} \label{a2}
The updated map of the NFC Standards, Products and Specifications is available at: \url{http://open-nfc.sourceforge.net/wp/}
\subsection{Technical Specification of Hardware Tools} \label{a3}
\begin{figure}[h]
    \centering
    \captionsetup{justification=centering}
    \includegraphics[width=0.4\textwidth]{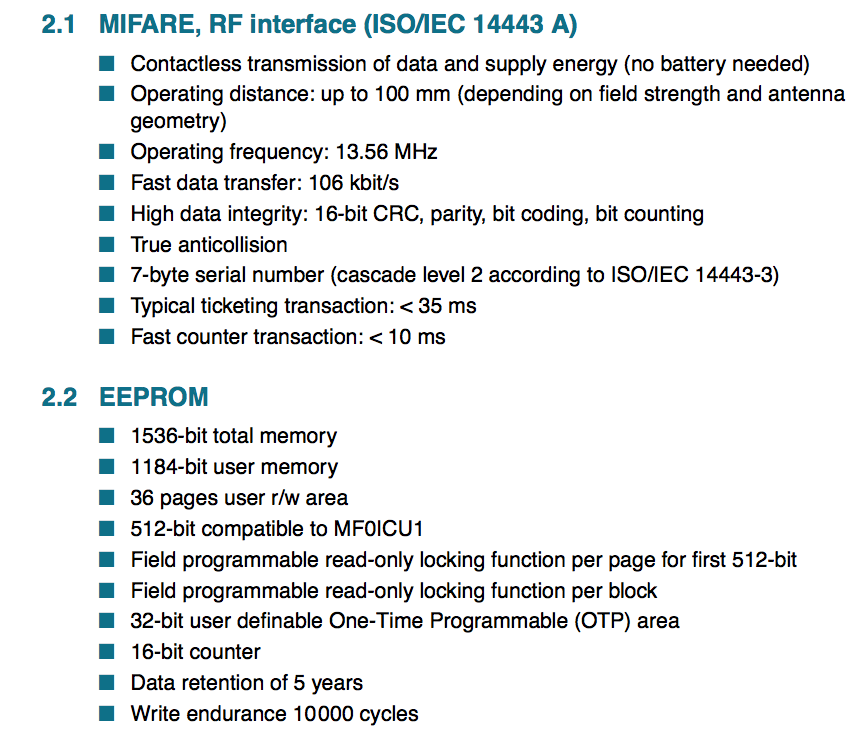}
    \caption{\textit{Technical Specification - Mifare Ultralight C NFC Tag}}
\end{figure}
\begin{figure}[h]
    \centering
    \captionsetup{justification=centering}
    \includegraphics[width=0.4\textwidth]{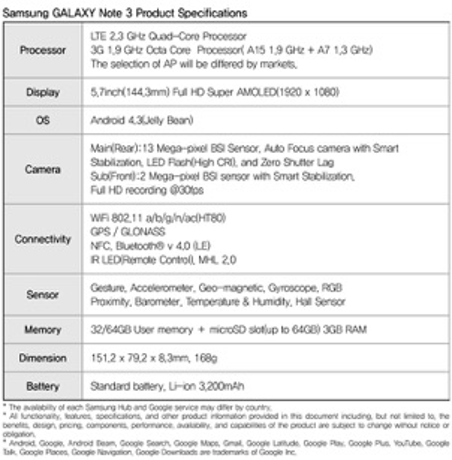}
    \caption{\textit{Technical Specification – Samsung Galaxy Note 3}}
\end{figure}
\begin{figure}[h]
    \centering
    \captionsetup{justification=centering}
    \includegraphics[width=0.4\textwidth]{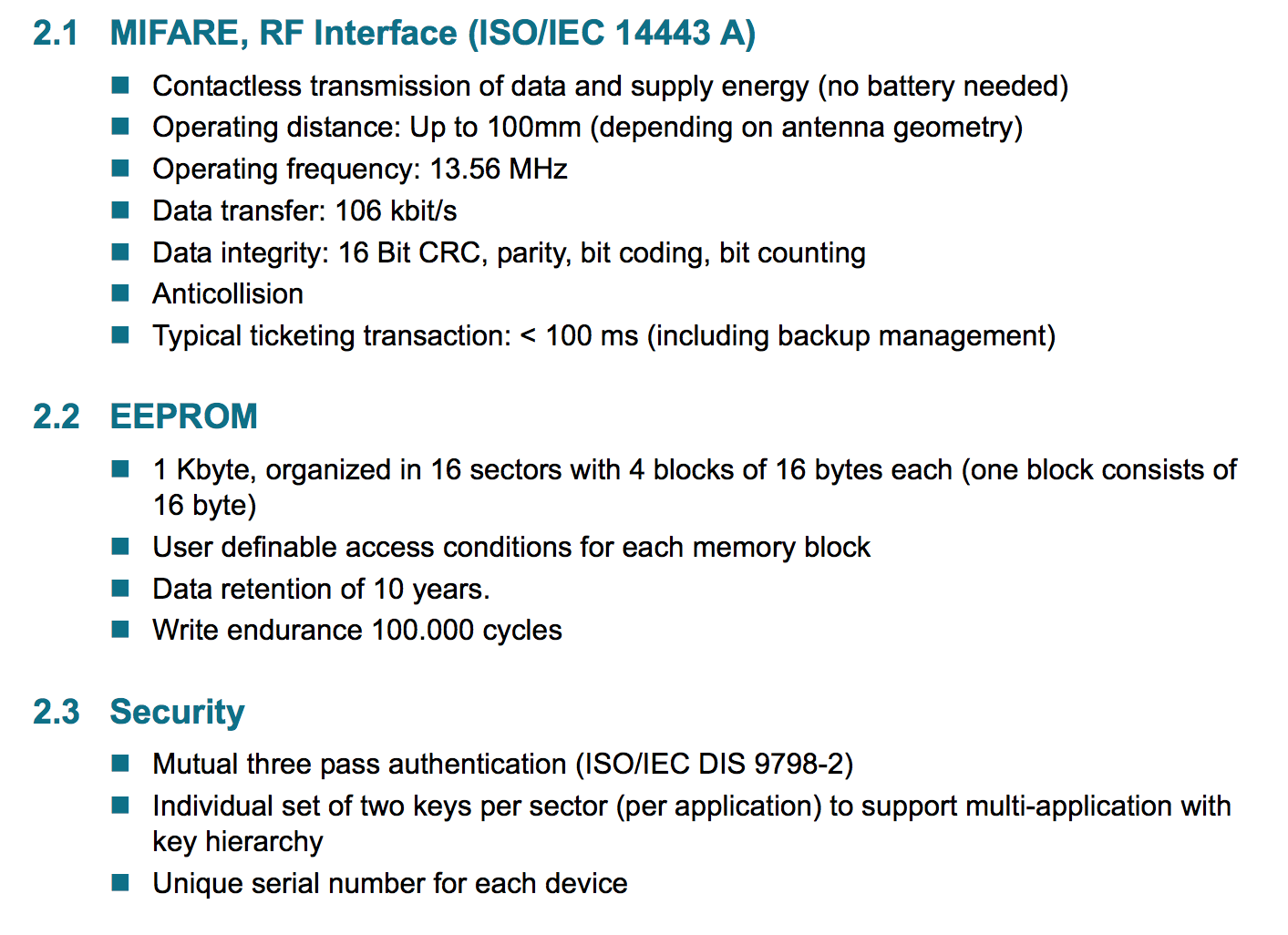}
    \caption{\textit{Technical Specification – Mifare Classic 1K NFC Tag}}
\end{figure}
\begin{figure}[h]
    \centering
    \captionsetup{justification=centering}
    \includegraphics[width=0.4\textwidth]{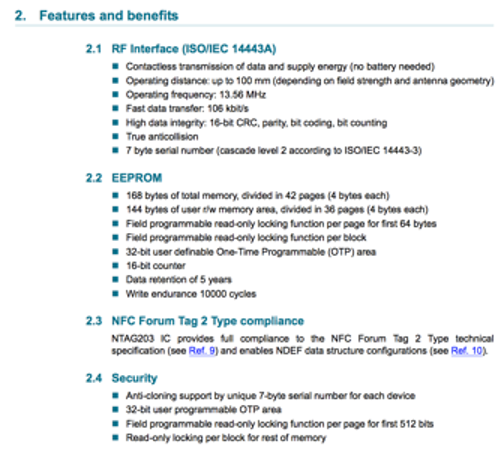}
    \caption{\textit{Technical Specification – NTAG203 NFC Tag}}
\end{figure}
\subsection{The NFC Data Exchange Format (NDEF)} \label{a4}
The NFC Forum has created NFC Data Exchange Format (NDEF) and NFC Forum Type Tag Operation. NDEF is a data format to encapsulate and identify application data that is exchanged between NFC-enabled tags and devices. A type of such device is the NFC Forum Type Tag. The Type Tags are contactless tags based on currently available products capable to store NDEF formatted data and operate one of the many NFC Forum Tag Platform. Just imagine a situation when we just bought an USB, we have to format before using it so does the NFC tag, and NDEF is the format of every NFC tag, just like the customized EPC numbering scheme and memory structure of those UHF RFID C1G2 tag demonstrated in \cite{7} – "Implementation issues in RFID-based anti-counterfeiting systems", the NFC tags have their own default numbering scheme set by the NDEF format.

\begin{figure}[h]
    \centering
    \captionsetup{justification=centering}
    \includegraphics[width=0.36\textwidth]{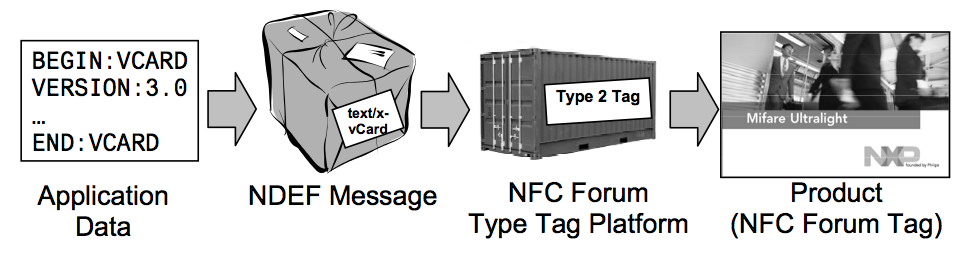}
    \caption{\textit{The NFC Forum Type Tag Operation}}
    \label{fig:ndefprocess}
\end{figure}

The application data stored inside an NFC Forum Tag is encapsulated firstly into an NDEF message and secondly into the data structure specified by the NFC Forum Type Tag Platform. The NDEF message and the NFC Forum Type Tag Platform encapsulations are used to identify the type of application data stored in the NFC Forum Tag e.g. an URL, a v-Card, a JPEG image, signature or text and to guarantee the interoperability and the co-existence between applications. An example of Type Tag Operation is depicted in \textit{Fig.~\ref{fig:ndefprocess}}.

NDEF is strictly a message format. It is a lightweight, binary message encapsulation format to exchange information between NFC Forum Device and another NFC Forum Device or an NFC Forum Tag. Below are some of the key specifications of NDEF - The application payloads may be of arbitrary type and size. The entire payload will be encapsulated in one NDEF message. The type identifiers may be URIs, MIME Types or NFC specific types. The length of the payload is an unsigned integer indicating the number of octets in the payload. There is an optional payload identifier field, which will help in the association of multiple payloads and cross-referencing between them. NDEF payloads may include NDEF messages or set of data chunks whose length is unknown at the time data is generated.

The design goal of NDEF is to provide an efficient and simple message format that can accommodate the following - encapsulating arbitrary documents and entities, including encrypted data, XML documents, XML fragments, image data like GIF and JPEG files, etc. Encapsulating documents and entities initially of unknown size. This capability can be used to encapsulate dynamically generated content or very large entities as a series of chunks. NFC applications can take advantage of such formats by encapsulating them in NDEF messages. For example, NDEF can be used to encapsulate an NFC-specific message and a set of attachments of standardized types referenced from that NFC-specific message. Compact encapsulation of small payloads should be accommodated without introducing unnecessary complexity to the parser.

\subsubsection{NDEF Message}
\begin{figure}[h]
    \centering
    \captionsetup{justification=centering}
    \includegraphics[width=0.42\textwidth]{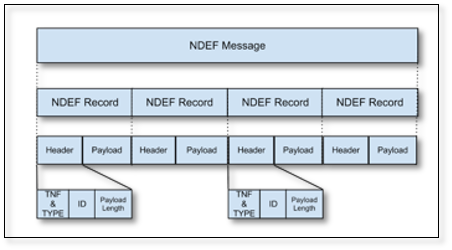}
    \caption{\textit{The Data Structure of NDEF Message}}
    \label{fig:ndefstructure}
\end{figure}

NFC Data Exchange Format is a lightweight binary message format .It is designed to encapsulate one or more application payloads in to single message construct. The single message construct is called NDEF Message. Each NDEF message consists of one or more NDEF Records, as structured in \textit{Fig.~\ref{fig:ndefstructure}}. Each NDEF Record can carry a payload of an arbitrary type and up to $2^{31} - 1$ octets in size. If the payload is larger, then the records can be chained to support bigger data.

\begin{figure}[h]
    \centering
    \captionsetup{justification=centering}
    \includegraphics[width=0.30\textwidth]{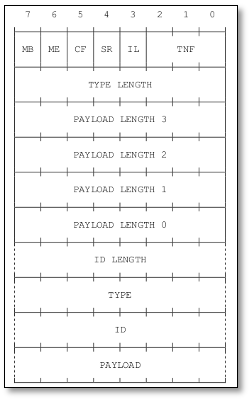}
    \caption{\textit{The Standardized Format of NDEF Message}}
    \label{fig:ndefformat}
\end{figure}

As depicted in \textit{Fig.~\ref{fig:ndefformat}}, the first record in the NDEF Message is marked with MB (Message Begin) flag set and the last record is marked with ME (Message End) set. The minimum record length is "\emph{one}" which can be constructed by setting MB and ME in the first record. The maximum number of NDEF records that can be carried in an NDEF message is unbounded. NDEF messages must not overlap. If you want to send more than one NDEF message, then each NDEF message must encapsulated in the form of a NDEF record. NDEF records do not carry any index. The ordering of the records is given by the way they are serialized. If an intermediate application repacks the data, then it must take care of the ordering too.

NDEF Records are variable length records with common format. Below is a NDEF Record layout –

•	MB: It is a 1-bit field. When this flag is set, it indicates the start of NDEF Message.

•	ME: It is a 1-bit field. When this flag is set, it indicates the end of NDEF Message.

•	CF: It is a 1-bit field. It indicates whether the record is either the first record chunk or a middle record chunk of a chunked payload.

An NDEF message can contain zero or more chunked payloads. A record chunk carries a chunk of payload. Chunked payloads can be used to partition dynamically generated content or very large entities into multiple subsequent record chunks serialized within the same NDEF message. In each chunked payload is encoded as an initial chunk followed by zero or more middle chunks and finally by terminating chunk. Each record chunk is encoded as an NDEF record using the following guidelines - The initial record chunk is an NDEF record with the CF (Chunk Flag) flag set. The type of the entire chunked payload MUST be indicated in the TYPE field regardless of whether the PAYLOAD_LENGTH field value is zero or not. The PAYLOAD_LENGTH field of this initial record indicates the size of the data carried in the PAYLOAD field of this record only, not the entire payload size. The ID field MAY be used to carry an identifier of the entire chunked payload.

Each middle record chunk is an NDEF record with the CF flag set indicating that this record chunk contains the next chunk of data of the same type and with the same identifier as the initial record chunk. The value of the TYPE_LENGTH and the IL fields MUST be zero and the TNF (Type Name Format) field value MUST be 0x06 (Unchanged). The PAYLOAD_LENGTH field indicates the size of the data carried in the PAYLOAD field of this single middle record only. The terminating record chunk is an NDEF record with the CF flag cleared, indicating that this record chunk contains the last chunk of data of the same type and with the same identifier as the initial record chunk. The value of the TYPE_LENGTH and the IL fields MUST be zero and the TNF (Type Name Format) field value MUST be 0x06 (Unchanged). The PAYLOAD_LENGTH field indicates the size of the data carried in the PAYLOAD field of this single terminating record only.

•	SR: It is a 1-bit field. If this flag is set, then the PAYLOAD_LENGTH field is a single octet. The short record layout is intended to compact encapsulation of small payloads, which will fit within PAYLOAD fields of size ranging between 0 and 255.

•	In the above layout SR = 1. The PAYLOAD_LENGTH filed is of only 1 octet. The max value is $2^8 - 1$.

•	IL: It is a 1-bit field. If this field is set, then the ID_LENGTH field is present in the header as a single octet. If the IL flag is zero, the ID_LENGTH field is omitted from the record header and the ID field is also omitted from the record.

•	TNF: This is a 3-bit value. It indicates the structure of the value of TYPE field.

•	TYPE_LENGTH: The TYPE_LENGTH field is an unsigned 8-bit integer that specifies the length in octets of the TYPE field.

•	ID_LENGTH: The ID_LENGTH field is an unsigned 8-bit integer that specifies the length in octets of the ID field. This field is present only if the IL flag is set to 1 in the record header. An ID_LENGTH of zero octets is allowed and, in such cases, the ID field is omitted from the NDEF record.

•	PAYLOAD_LENGTH: The PAYLOAD_LENGTH field is an unsigned integer that specifies the length in octets of the PAYLOAD field (the application payload). If the SR flag is set, the PAYLOAD_LENGTH field is a single octet representing an 8-bit unsigned integer. The max size will be $2^8 - 1$ octets. If the SR flag is clear, the PAYLOAD_LENGTH field is four octets representing a 32-bit unsigned integer. The max size will be $2^{32} - 1$ octets.

•	TYPE: The value of the TYPE field is an identifier describing the type of the payload. The value of the TYPE field MUST follow the structure, encoding, and format implied by the value of the TNF field as described in TNF section above.

•	PAYLOAD: The PAYLOAD field carries the payload intended for the NDEF user application. Any internal structure of the data carried within the PAYLOAD field is opaque to NDEF.

\section{NAS System Design Components}
\subsection{Sitemap Diagram of NAS Wine Database Web Application} \label{b1}
\begin{figure}[h]
    \centering
    \captionsetup{justification=centering}
    \includegraphics[width=0.5\textwidth]{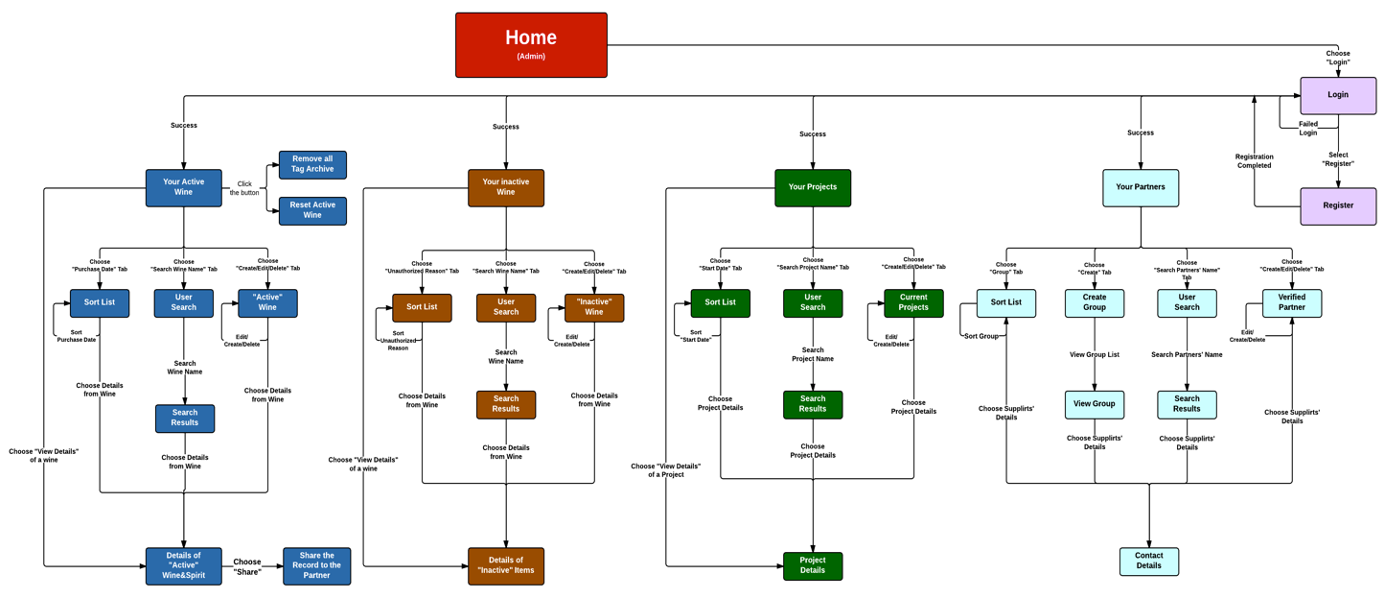}
    \caption{\textit{Sitemap Diagram of Wine Database Web Application}}
    \label{fig:naslifecycle}
\end{figure}
\subsection{Entity and Attribute Diagram of NAS Wine Database} \label{b2}
\begin{figure}[h]
    \centering
    \captionsetup{justification=centering}
    \includegraphics[width=0.5\textwidth]{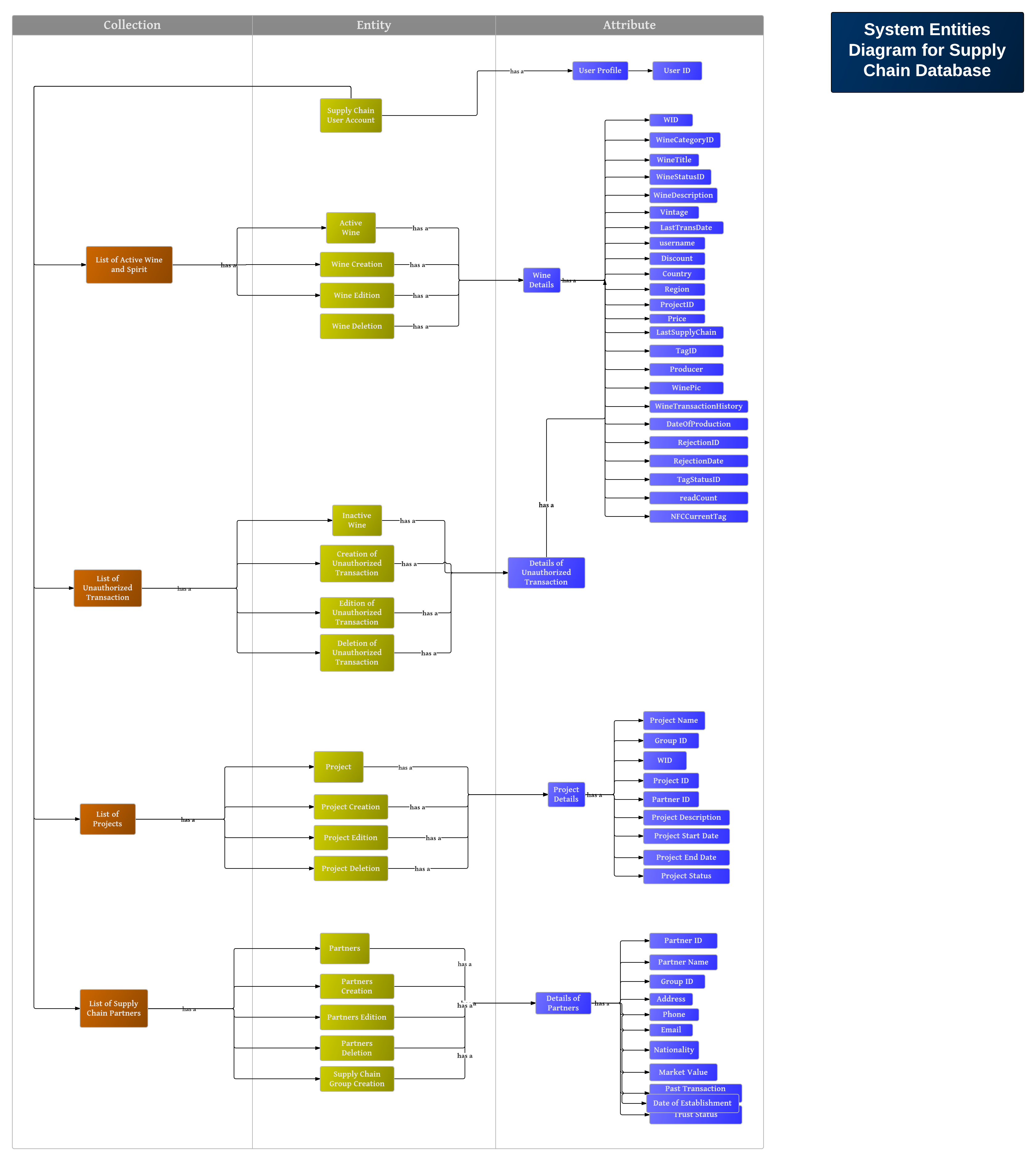}
    \caption{\textit{Entity and Attribute Diagram of Wine Database}}
    \label{fig:naslifecycle}
\end{figure}
\subsection{Entities Relationship Diagram of NAS Wine Database} \label{b3}
\begin{figure*}[h]
    \centering
    \captionsetup{justification=centering}
    \includegraphics[width=1\textwidth]{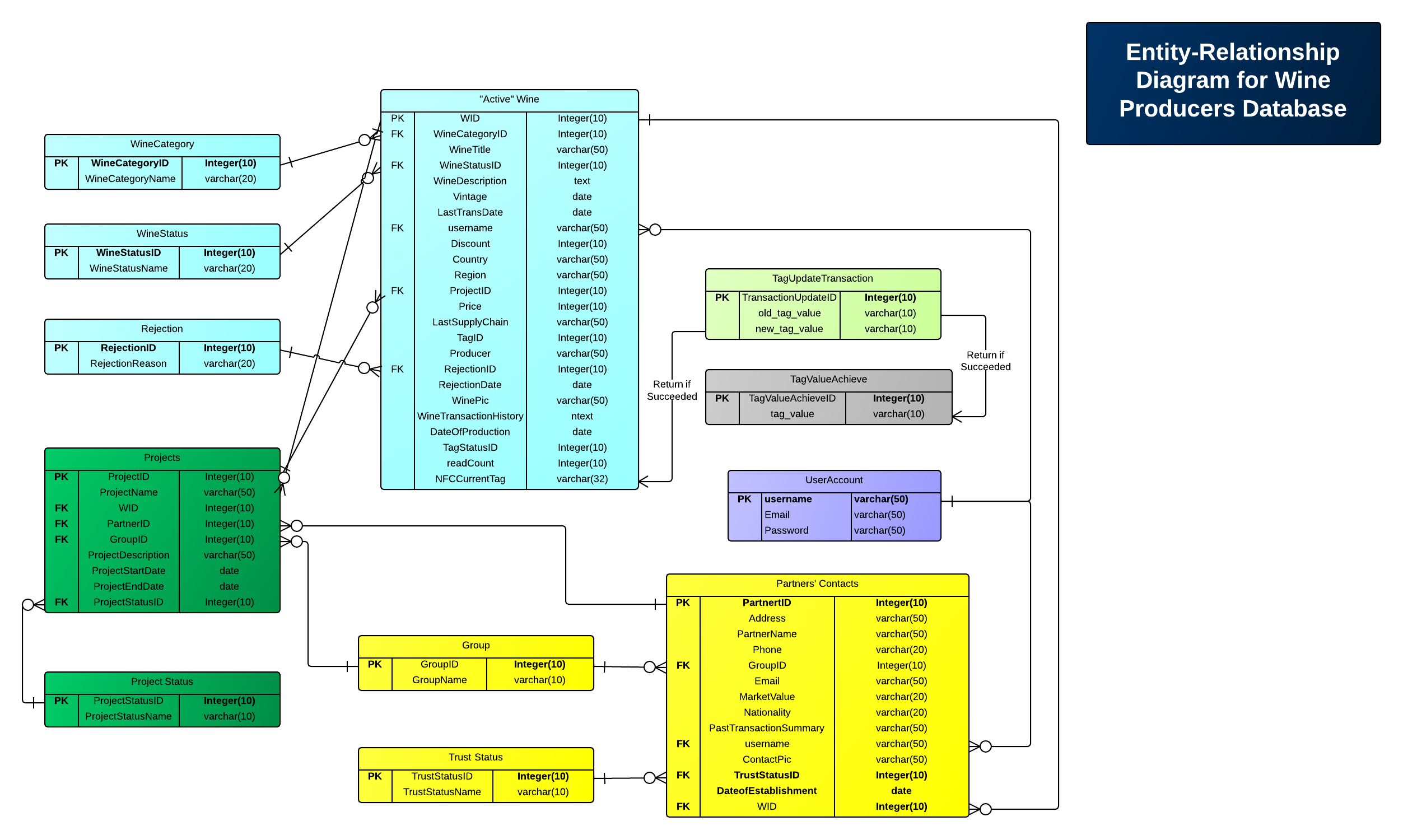}
    \caption{\textit{Entities Relationship Diagram of Wine Database}}
    \label{fig:erd}
\end{figure*}

The Entities Relationship Diagram is depicted in \textit{Fig.~\ref{fig:erd}}.
\subsection{The Database Structure Design} \label{b4}
For the Identity Entities Diagram, it is defined into three parts, which are Collection (Red), Entity (Yellow) and Attribute (Blue) respectively. The lists of Active wine, Inactive wine, Project and the supply chain participants are put under the layer of "\emph{Collection}" meaning the whole data will be stored and appeared in the database, and there are associated functions, also known as events, such as \emph{User Login}, \emph{Edition/Creation} of every wine record, project and supply chain participant, as well as some other basic features such as sorting, searching and sharing are also put under the layer of "\emph{Entity}", so that the database of NAS with aforementioned functions could better enhance the efficiency on data operation and management, and interaction between different entities. In fact, there are FIVE basic entities shown in this Identity Entities Diagram, namely "\emph{User Account}", "\emph{Active Wine}", "\emph{Inactive Wine}", "\emph{Projects}" and "\emph{Partners}" respectively. 

Regarding to the layer of "\emph{Entity}" and "\emph{Attribute}", the "\emph{has a}" relationships are shown appropriately based on the inclusive relationship between different items on those layers, such as each Active Wine and its association will "\emph{have a}" Wine Detail, the project and its associations such as creation and edition will "\emph{have a}" project detail, so far and so on, indicating that each entity "\emph{has a}" attribute to operate. For the detail of attributes each entity stored, for instance, the wine detail has a lot of necessary attribute, such as wine title, wine status, last transaction date, the transaction record, so that the wine record is indicative and informational to perform those functions and even convey the anti-counterfeiting values of the whole NAS. 

Furthermore, there are also some important attributes set for different purposes of the NAS. For instance, the "\emph{WineStatusID}" is the wine status to control where the wine record will be under the tab of Active Wine or being transferred to the tab of Inactive Wine for denoting it as a suspicious wine product found along the supply chain when there is nothing returned while the mobile applications made request to assess the back-end database, or a sold wine product when the status and the transaction record are updated while the payment is confirmed. There are also some noted attributes, such as the "\emph{RejectionID}", in which it relates to the reason of rejection while there is a wine product being rejected by supply chain participants, and "\emph{TrustStatusID}" in which it refers to the Trust Status of a supply chain participant which will also be updated when pre-registration is confirmed from that participant.

For the Entities Relationship Diagram, there are totally 11 entities, for those additional six sub-entities, which are from the result of normalization. Normalization can make the data model more flexible and reliable. However, it does generate some overheads because it usually requires to get more tables, but it enables to do many things with the data model without having to adjust it. To achieve normalization, the database should confine with the following 2 properties, which are no repeating elements or groups of the elements and all foreign keys attributes fully functionally depending on the whole primary key. For instance, there are repeating elements, such as wine category and wine status. Therefore, there are two more additional sub-entity tables, namely "\emph{WineCategory}" and "\emph{WineStatus}" developed with \emph{WineCategoryID} and \emph{WineStatusID} to link back to the original entity table of "\emph{Active Wine}" so as to prevent from repeating the elements, leading to low efficiency of data response and bulkiness of the general database.

As aforementioned, there are FIVE basic entities tables, which are "\emph{User Account}", "\emph{Active Wine}", "\emph{Inactive Wine}", "\emph{Projects}" and "\emph{Participant Information}". Each of them stores the necessary attributes with appropriate data type for the purpose of the entity table. For example, the attribute of "\emph{TransationRecord}" is needed to store those status records of wine products either manually input by winemakers, including the status of production, bottling and even the NFC writing processes, or atomically updated based on NFC scanning activities throughout the supply chain of wine products.

\begin{figure}[h]
    \centering
    \captionsetup{justification=centering}
    \includegraphics[width=0.5\textwidth]{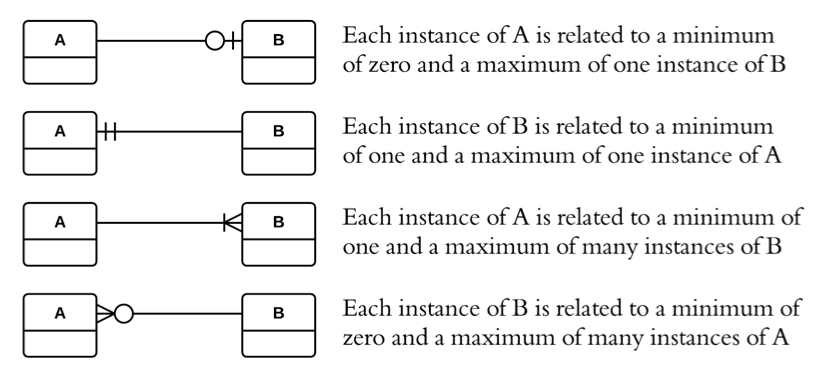}
    \caption{\textit{The Relationship of Entity Instance}}
    \label{fig:keyrelationship}
\end{figure}

To enable the entity relationship, the relationship of primary key and foreign key should be set, as depicted in \textit{Fig.~\ref{fig:keyrelationship}}, so that a relative entity tables could be linked together in which there will be a primary key assigned for each entity table. There are 11 primary keys (denoted as "\emph{PK}") and 17 foreign keys (denoted as "\emph{FK}") enabling 19 relationships throughout the whole information structure of the database.

Indeed, according to the Entity Relationship Diagram in \textit{Appendix~\ref{b3}}, there are different categories of relationship. For the entity table of "\emph{Project}", there could be more than one project status with data type of "\emph{varchar(10)}" stored in the entity of "\emph{Project Status}", and there would be more than one project status able to be chosen by winemakers when creating or editing a project record. An one-to-many relationship would be set between the entity tables of Project and Project Status, in which the relationship with each instance of Project Status is related to a minimum of zero and a maximum of many instances of Project is set, as the entry of key - "\emph{ProjectStatusID}" could be "\emph{Null}" in which the columns of input of this data will not be compulsory and could be with no data at all under which the minimum of zero instance could be possible between this relationship. According to the same methodology, the similar relationships are set between the entity tables of "\emph{Active Wine}" with that of "\emph{WineCategory}" and "\emph{WineStatus}", between entity tables of "\emph{Partner Information}" with tables of "\emph{Group}" and "\emph{Trust Status}", and between tables of "\emph{Unauthorized Transaction}" with that of "\emph{Rejection}". This kind of relationship could be visualized whenever there will be a drop-down list for "\emph{Wine Status}" to choose the desired status during the process of creating a new wine record at the wine database web application.

While for the entity table of "\emph{User Account}", another kind of one-to-many relationship is being constructed, between that entity table with other relating tables, such as the table of "\emph{Active Wine}", the table of "\emph{Unauthorized Transaction}" and "\emph{Partner Information}". However, the one-to-many relationship is different from that shown above, under which for this case each instance of those tables connected with the table of "\emph{UserAccount}" will be related to a minimum of one and a maximum of many instance of the table of "\emph{UserAccount}".

\subsection{The Adoption of PhoneGap Framework} \label{b5}
PhoneGap (Cordova) is an open-source mobile development framework for quickly building cross-platform mobile applications using HTML5, JavaScript and CSS, instead of being bounded by those device-specific languages, such as Java, in which, normally, different operating systems will need different languages and platforms to build mobile applications. For instance, Android and BlackBerry require Java, while the iOS requires Objective-C as the primary programming language.  With the advent of PhoneGap, all the functions could be designed and built using JavaScript, which is way easier than using mainly Java and the Bootstrap CSS could be adopted.

\begin{figure}[h]
    \centering
    \captionsetup{justification=centering}
    \includegraphics[width=0.5\textwidth]{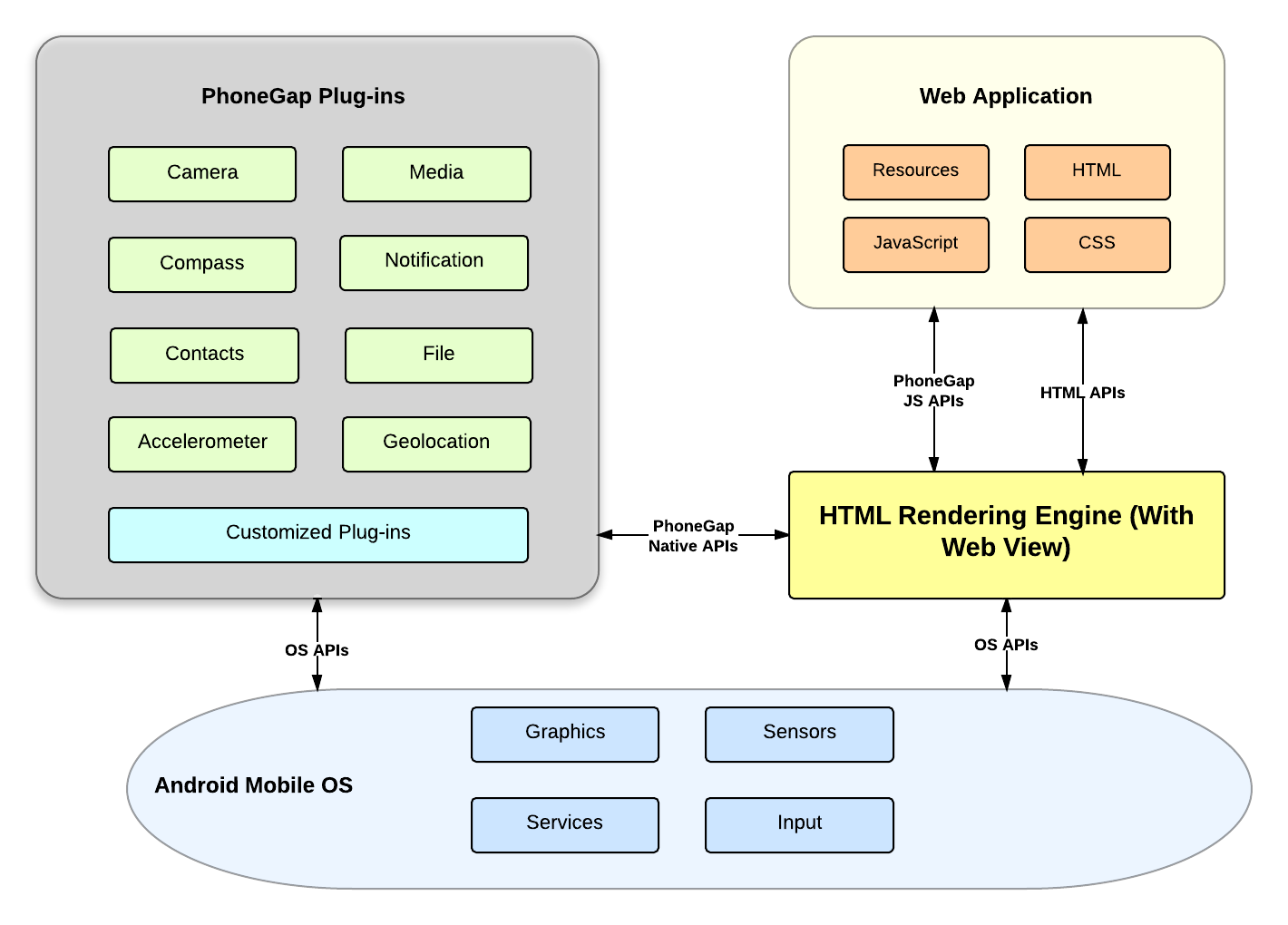}
    \caption{\textit{The Communication Mechanism of PhoneGap Framework}}
    \label{fig:phonegapframework}
\end{figure}

The PhoneGap solves this by using standards-based web technologies to bridge web applications and mobile devices, making PhoneGap application is cross-platform in nature, with the communication mechanism as depicted in \textit{Fig.~\ref{fig:phonegapframework}}. PhoneGap is indeed a set of device APIs that allow a mobile application developer to access native device function, such as the camera or accelerometer from JavaScript. Combined with an user interface framework, such as jQuery Mobile or Dojo Mobile or Sencha Touch, this would allow a smartphone application to be developed with just HTML, CSS, and JavaScript. When using the Cordova APIs, a mobile application can be built without any native code (Java and Objective-C) from the application developer. Instead, web technologies are used, and they are hosted in the application itself locally. And because these JavaScript APIs are consistent across multiple device platforms and built based on web standards, the application should be portable to other device platforms with minimal to no changes. 

\end{document}